\documentclass[11pt]{article}
\pdfoutput=1

\usepackage{jheppub}
\usepackage{amssymb,amsfonts,amsmath,amsthm, enumerate}
\usepackage{mathrsfs}
\usepackage{bbold}
\usepackage{tikz}
\usetikzlibrary{shapes.geometric, arrows}
\usepackage{hyperref}
\usepackage[normalem]{ulem}

\usepackage{arydshln}

\usepackage{calc}

\newcommand{\g}{nonrelativistic }
\newcommand{\gc}{Nonrelativistic }

\usepackage{accents}

\makeatletter
\newcommand{\Vast}{\bBigg@{4.75}}
\makeatother

\newcommand{\be}{\begin{equation}}
\newcommand{\ee}{\end{equation}}
\newcommand{\bea}{\begin{eqnarray}}
\newcommand{\eea}{\end{eqnarray}}

\newcommand{\CA}{\mathcal{A}}

\newcommand{\CC}{\mathcal{C}}

\newcommand{\CF}{\mathcal{F}}
\newcommand{\CG}{\mathcal{G}}
\newcommand{\CH}{\mathcal{H}}
\newcommand{\CI}{\mathcal{I}}
\newcommand{\CJ}{\mathcal{J}}
\newcommand{\CK}{\mathcal{K}}
\newcommand{\CL}{\mathcal{L}}
\newcommand{\CN}{\mathcal{N}}
\newcommand{\CM}{\mathcal{M}}

\newcommand{\CP}{\mathcal{P}}

\newcommand{\CV}{\mathcal{V}}

\newcommand{\lr}{\left (}
\newcommand{\rr}{\right )}
\newcommand{\ls}{\left [}
\newcommand{\rs}{\right ]}

\newcommand\qt\tau

\newcommand{\p}{\partial}

\renewcommand{\tilde}[1]{\widetilde{#1}}
\newcommand{\tr}{\text{tr}}

\makeatletter
\renewcommand{\@seccntformat}[1]{\csname the#1\endcsname.\,\,}
\makeatother

\let \savenumberline \numberline
\def \numberline#1{\savenumberline{#1.}}

\makeatletter
\def\@fpheader{\relax}
\makeatother

\def\bea{\begin{eqnarray}}
\def\eea{\end{eqnarray}}

\usetikzlibrary{shapes,arrows,chains}
\usetikzlibrary{decorations.markings}
\usetikzlibrary{decorations.pathmorphing}
\usetikzlibrary{shapes.multipart}
\tikzset{snake it/.style={decorate, decoration=snake}}

\title{\ \vspace{1.6cm} \\
\scalebox{1}{Dual D-Brane Actions in Nonrelativistic String Theory}}
\author[a]{Stephen Ebert,}
\author[b]{Hao-Yu Sun}
\author[c]{and Ziqi Yan}
\affiliation[a]{Mani L. Bhaumik Institute for Theoretical Physics, University of California,\\ Los Angeles, CA 90095-1547, USA\smallskip}
\affiliation[b]{Theory Group, Department of Physics,
University of Texas, \\Austin, TX 78712-1192, USA\smallskip}
\affiliation[c]{Nordita, KTH Royal Institute of Technology and Stockholm University,\\
Hannes Alfv\'{e}ns v\"{a}g 12, SE-106 91 Stockholm, Sweden\medskip}
\emailAdd{stephenebert@physics.ucla.edu}
\emailAdd{hkdavidsun@utexas.edu}
\emailAdd{ziqi.yan@su.se}

\abstract{We study worldvolume actions for D-branes coupled to the worldvolume $U(1)$ gauge field and Ramond-Ramond (RR) potentials in nonrelativistic string theory. This theory is a self-contained corner of relativistic string theory and has a string spectrum with a Galilean-invariant dispersion relation. We therefore refer to such D-branes in nonrelativistic string theory as nonrelativistic D-branes. We focus on the bosonic fields in spacetime and also couple the D-branes to  general closed string geometry, Kalb-Ramond, and dilaton background fields. We dualize nonrelativistic D-branes by performing a duality transformation on the worldvolume $U(1)$ gauge field and uncover novel dual D-brane actions. This generalizes familiar properties, such as the SL($2,\mathbb{Z}$) duality in Type IIB superstring theory and the relation between Type IIA superstring and M-theory, to nonrelativistic string and M-theory.
Moreover, we generalize the limit of string theory, in which nonrelativistic string theory arises, to include RR potentials. This stringy limit induces a codimension-two foliation structure in spacetime. This spacetime geometry is non-Riemannian and known as string Newton-Cartan geometry.
In contrast, nonrelativistic M-theory that we probe by dualizing D2- and D4-branes in nonrelativistic string theory arises as a membrane limit of M-theory, and it is coupled to a membrane Newton-Cartan geometry with a codimension-three foliation structure.  
We also discuss T-duality in nonrelativistic string theory and generalize Buscher rules from earlier work to include RR potentials.}

\begin{document}

\maketitle
\vfill\eject

\section{Introduction}

One of the most fascinating discoveries in string theory is that different superstring theories are corners of M-theory in eleven dimensions. Theories that arise as various limits of M-theory and describe rather different physics are frequently related to each other by duality transformations. These dualities provide powerful techniques for probing different corners in M-theory and bring useful intuitions about various nonperturbative regimes in string theory. Even though a comprehensive understanding of M-theory is still lacking, it is widely believed that the full dynamics of M-theory is captured by a simple quantum mechanical system of D0-branes, known as the Matrix theory conjecture \cite{Banks:1996vh, Susskind:1997cw, Seiberg:1997ad, Sen:1997we}. This description of M-theory is approached from considering a discrete light cone quantization (DLCQ) of M-theory, typically defined by taking an infinite-boost limit of compactification on a spacelike circle. Moreover, the Matrix theory description of the DLCQ of string theory was later studied in \cite{Motl:1997th, Banks:1996my, Dijkgraaf:1997vv}. Compared to the infinite-momentum frame and the large-$N$ limit considered in the seminal work on Matrix theory (with $N$ D0-branes), the DLCQ of string/M-theory has the advantage of making various dualities manifest at finite $N$. 

Not relying on the subtle infinite-boost limit, the DLCQ of string theory is known to be related to a unitary and ultra-violet (UV) complete theory called nonrelativistic string theory, which in flat spacetime is defined by a two-dimensional quantum field theory (QFT) with a (string-)Galilean global symmetry \cite{Gomis:2000bd}. The string excitations in nonrelativistic string theory have a Galilean-invariant dispersion relation and a nonrelativistic spacetime S-matrix; the spectrum there contains \emph{no} massless gravitons.~\footnote{The closed string spectrum was first obtained from a limit of relativistic string theory \cite{Klebanov:2000pp}. Also see \cite{Danielsson:2000gi}.} Instead, the asymptotic closed string states necessarily carry nonzero windings and they can exchange instantaneous gravitational forces that are Newtonian-like \cite{Gomis:2000bd, Danielsson:2000mu}. In curved spacetime, this theory describes strings propagating in the so-called string Newton-Cartan geometry \cite{Andringa:2012uz, Bergshoeff:2018yvt},~\footnote{Background geometries with different torsional constraints that nonrelativistic strings are coupled to have been discussed in the literature, depending on what global symmetries are imposed on the worldsheet. See, \emph{e.g.}, \cite{Harmark:2018cdl, Bergshoeff:2019pij, Harmark:2019upf, Bergshoeff:2021bmc, Yan:2021lbe, Bidussi:2021ujm, Bergshoeff:2021tfn} for studies of strings in torsional string Newton-Cartan geometries. These subtle differences will not affect our studies of D-brane actions in this paper. For simplicity, we will collectively refer to this class of background geometries as ``string Newton-Cartan geometry."} which is a non-Riemannian geometry equipped with a codimension-two foliation structure. This geometry naturally generalizes Newton-Cartan geometry, the geometrization of Newtonian gravity equipped with a codimension-one foliation. Moreover, T-dualities of nonrelativistic string theory with arbitrary background fields have been studied in \cite{Bergshoeff:2018yvt}. It is shown that the DLCQ of relativistic string theory arises from performing a T-duality transformation along a spacelike circle in nonrelativistic string theory. This relation provides a first principle definition of the DLCQ of relativistic string theory, which is otherwise only defined by the subtle infinite-boost limit. 

In recent years, significant progress has been made towards the formulation of nonrelativistic string theory in general backgrounds.~\footnote{See \cite{Oling:2022fft} for a review of recent developments in nonrelativistic string theory.}  These studies have been generating new excitements about exploring a landscape of non-Lorentzian gravity and field theories, as well as their applications to the AdS/CFT correspondence; see \emph{e.g.}, \cite{Gomis:2005pg, Andringa:2012uz, Ko:2015rha, Morand:2017fnv, Harmark:2017rpg, Bergshoeff:2018yvt, Harmark:2018cdl, Gomis:2019zyu, Gallegos:2019icg, Harmark:2019upf, Bergshoeff:2019pij, Blair:2019qwi, Yan:2019xsf, Harmark:2020vll, Gallegos:2020egk, Blair:2020gng, Blair:2021ycc, Bergshoeff:2021bmc, Bidussi:2021ujm, Yan:2021lbe, Bergshoeff:2021tfn, Fontanella:2021hcb, Fontanella:2021btt, Kluson:2021tub}.~\footnote{Also see \cite{Yan:2021hte} for a companion paper on KLT relations for amplitudes in nonrelativistic string theory.} Lately, the quantum conformal invariance of the worldsheet theory with boundary terms describing nonrelativistic string theory in arbitrary open string backgrounds is analyzed in \cite{Gomis:2020fui}. This study of worldsheet conformal anomalies leads to a set of spacetime equations of motion that are nonlinear and describe the low-energy dynamics of the open string fields. For a single D-brane, these equations of motion can be derived from a nonlinear action of the curvature of the $U(1)$ connection, which describes a local field theory with Galilean symmetry. This theory is referred to as Galilean Dirac-Born-Infeld (DBI) theory in \cite{Gomis:2020fui}.  
Such a Galilean DBI action can be readily generalized to D$p$-brane actions that describe the low-energy dynamics of $(p+1)$-dimensional membranes on which nonrelativistic open strings end.~\footnote{In \cite{Gomis:2020fui}, strictly speaking, only D$25$-branes in bosonic string theory are considered.} These Galilean DBI actions describing D-branes in nonrelativistic string theory significantly differ from their relativistic counterparts, and only arise as a nontrivial limit of relativistic D-branes \cite{Gomis:2020fui}. The realization of such worldvolume actions enable concrete studies of D-branes in nonrelativistic string theory. In particular, D-branes are useful probes for mapping out physics in different (strongly-coupled) regimes of string/M-theory, which are accessed by performing duality transformations on the associated worldvolume actions of D-branes (see \emph{e.g.}, \cite{Tseytlin:1996it, Aganagic:1997zk}).

Based on the previous advancements in nonrelativistic string theory, it is timely to systematically investigate its duality web. This endeavor will not only improve our understanding of extended objects in nonrelativistic string theory, but also allow us to probe ``nonrelativistic M-theory" that arises in a strongly coupled regime. It will also be interesting to investigate how the expectations from the well-studied dualities of D-brane actions in relativistic string theory can be applied to the DBI actions in nonrelativistic string theory, and whether novel nonrelativistic twists arise. 

Such a duality web in nonrelativistic string theory can be accessed using two complementary methods. First, it is known that nonrelativistic string theory arises as an intriguing limit of relativistic string theory that requires a subtle cancellation between the string tension and a critical Kalb-Ramond field \cite{Gomis:2000bd}. It must be possible to generalize such a limiting procedure to derive the desired duality web in nonrelativistic string theory using the known ingredients from relativistic string theory. As we will construct later in this paper, such a limiting procedure in general involves highly nontrivial cancellations among various divergent terms and has to be treated with care. Second, since nonrelativistic string theory is self-contained and can be studied independently of relativistic string theory, the desired duality web can also be accessed from first principles. This program imposes new challenges as D-branes in nonrelativistic string theory are coupled to string Newton-Cartan geometry, which is non-Riemannian and equipped with a two-dimensional foliation structure. Such a geometry is fundamentally distinct from Einstein's gravity emerging in relativistic string theory. Consequently, duality transformations of the associated D-brane action (whose DBI part is only recently realized in \cite{Gomis:2020fui}) do not find any direct analogue in relativistic string theory and are expected to reveal novel brane configurations in nonrelativistic string theory. Even though a systematic analysis of the duality web in nonrelativistic string theory is missing, some related aspects in both approaches have been explored to certain extent in the literature, which we briefly review below.

Hitherto, S-duality transformations of the Galilean-invariant D-brane actions from \cite{Gomis:2020fui} have been explored almost only in flat spacetime, mostly in the context of noncommutative open string (NCOS) theory.~\footnote{For examples, see \cite{Berman:2000jw,Berman:2001rka} for NCOS-type limits of various extended objects in string/M-theory and relevant applications to holography.}  The NCOS is known to be an open string sector in the framework of nonrelativistic string theory \cite{Klebanov:2000pp, Gomis:2000bd, Danielsson:2000gi}. 
NCOS was originally introduced by taking a stringy limit of relativistic string theory in the presence of a Kalb-Ramond field tuned to cancel the string tension. This limit naturally induces a codimension-two foliation structure that also appears in string Newton-Cartan geometry to which nonrelativistic closed strings are coupled. It is known that the four-dimensional NCOS gives an S-dual description of the strongly coupled, spatially noncommutative Yang-Mills theory with $\CN = 4$ supersymmetry \cite{Gopakumar:2000na}. It is also shown in \cite{Gopakumar:2000ep, Bergshoeff:2000ai} that the strongly-coupled five-dimensional NCOS is described by a theory of light open membranes (OM) on an M5-brane at a near critical three-form field strength. Such an OM theory arises as a membrane generalization of the NCOS limit of M-theory. This limit induces a codimension-three foliation structure in eleven-dimensional spacetime. This is in contrast to the codimension-two foliation in string Newton-Cartan geometry that nonrelativistic strings are coupled to.

The NCOS limit has been generalized in \cite{Andringa:2012uz, Bergshoeff:2019pij, Gomis:2020fui} to relativistic strings coupled to arbitrary geometry, Kalb-Ramond and dilaton background field, together with a worldvolume gauge field. This limit leads to nonrelativistic string theory that can be equivalently defined using the worldsheet theory introduced in \cite{Gomis:2000bd}. Moreover, this limit has been generalized to the so-called $p$-brane limits involving a near critical $(p+1)$-form field strength \cite{Gomis:2000bd}. In particular, the two-brane limit coincides with the limit of M-theory that leads to the OM theory. S-dualities of theories that arise as various $p$-brane limits of relativistic string/M-theory have also been discussed, for examples, in \cite{Gomis:2000bd, Kamimura:2005rz}. It is also known that a dimensional reduction of the two-brane limit of M-theory leads to nonrelativistic string theory \cite{Blair:2021ycc, Kluson:2019uza}. However, relations between general $p$-brane limits and nonrelativistic string theory are unclear. 

On the other hand, T-duality transformations in nonrelativistic string theory have been studied in \cite{Bergshoeff:2018yvt}, where general Buscher rules for geometry, Kalb-Ramond and dilaton background fields have been derived from first principles by using the worldsheet theory.~\footnote{Also see \cite{Gomis:2000bd, Danielsson:2000gi, Kluson:2018vfd, Bergshoeff:2019pij}. Moreover, see \cite{Harmark:2017rpg, Kluson:2018egd, Harmark:2019upf, Harmark:2018cdl} for related works on null reductions.} These Buscher rules have been applied to the DBI-like part of the D-brane actions in nonrelativistic string theory in \cite{Gomis:2020izd}, where it is shown how relativistic, nonrelativistic and noncommutative open strings are related to each other. 

\vspace{2mm}

In this paper, we will first generalize the D-brane actions found in \cite{Gomis:2020fui} to include all relevant bosonic terms, which will be our starting point for analyzing both S- and T-duality transformations of such D-brane actions in nonrelativistic string theory.~\footnote{There are a series of standard simplifications we make throughout the paper. We will restrict to the case where the geometric curvatures are very small and hence omit any curvature contributions to the effective action of a D-brane (see, \emph{e.g.}, \cite{Bershadsky:1995qy, Green:1996dd, Morales:1998ux, Bachas:1999um} for inclusion of corrections from the geometric curvature in the worldvolume actions of D-branes in relativistic string theory). Moreover, for simplicity, we will only consider a single D-brane with a worldvolume $U(1)$ connection, but our construction can be in principle generalized to non-abelian cases with multiple coinciding D-branes. 
We further assume the worldvolume gauge field strength to be slowly varying at the string length scale, 
which allows us to drop derivatives of the field strength (see, \emph{e.g.}, \cite{Andreev:1988cb, Wyllard:2000qe} for the corrections beyond the slowly-varying-field approximation in relativistic D-brane actions).} We will refer to such D-branes as \emph{\g D-branes}.
In relativistic string theory, since Polchinski's realization \cite{Polchinski:1995mt}, it is well established that D-branes are charge carriers for RR potentials, which extends the D-brane action to include a Chern-Simons (CS) term, in addition to the DBI term. Analogously, introducing similar CS terms in nonrelativistic D-brane actions allows the inclusion of RR potentials in nonrelativistic string theory. We will also derive how the nonrelativistic string limit is applied to RR potentials, and show that a careful cancellation among different RR potentials and the Kalb-Ramond field is required for reproducing the finite D-brane action in nonrelativistic string theory.

Based on these new developments of D-brane actions in nonrelativistic string theory, we study in detail the duality transformations of D$p$-branes in Type II nonrelativistic superstring theories with $p=1,\cdots,4$\,, by dualizing a $U(1)$ gauge field on the branes' worldvolumes. Firstly, the S-duals of nonrelativistic D1- and D3-branes reveal the SL(2,$\mathbb{Z}$) symmetry as in Type IIB relativistic superstring theory. Secondly, the dual actions of nonrelativistic D2- and D4-branes describe nonrelativistic M2- and M5-branes, respectively. These duality transformations are significantly different from those in relativistic string theory, and lead to novel dual D-brane actions coupled to various non-Riemannian spacetime geometries. While nonrelativistic string theory and the associated D-branes are coupled to ten-dimensional string Newton-Cartan geometry with a codimension-two foliation, nonrelativistic M2- and M5-branes are coupled to eleven-dimensional membrane Newton-Cartan geometry with a codimension-three foliation. Moreover, we also generalize the previous studies on T-duality transformations in nonrelativistic string theory \cite{Bergshoeff:2018yvt, Bergshoeff:2019pij, Gomis:2020izd} to include RR potentials. These results generalize the previous works to a larger duality web for nonrelativistic string/M-theory coupled to RR potentials, in addition to other bosonic background fields that have been considered before. 

\vspace{2mm}

The paper is organized as follows. In \S \ref{sec:D-branes in NRST}, we construct the bosonic part of the worldvolume action describing D-branes in nonrelativistic string theory coupled to string Newton-Cartan geometry, Kalb-Ramond, dialton, $U(1)$ gauge and RR potential background fields. In \S \ref{sec:s-duality}, we study S-duality transformations of various D$p$-branes by dualizing the worldvolume $U(1)$ gauge field. In \S \ref{sec:T-duality}, we extend the previous works on T-duality transformations in nonrelativistic string theory to include RR potentials. In particular, we realize in \S\ref{sec:lltd} a novel double-scaling limit to derive the Buscher rules associated with the longitudinal lightlike T-duality transformation that relate nonrelativistic string theory to NCOS. In \S \ref{sec:conclusions}, we conclude our paper. In Appendix \ref{app:dpbpbncg}, we construct the D$p$-brane action coupled to $p$-brane Newton-Cartan geometry.~\footnote{The D2-brane action that arises as a three-brane limit of relativistic D2-brane action has been formulated in \cite{Kluson:2019uza} by dimensionally reducing the nonrelativistic M2-brane action along a transverse direction. Moreover, it is also shown in \cite{Kluson:2019uza} that a double dimensional reduction in a longitudinal spatial direction of nonrelativistic M2-brane gives rise to the Nambu-Goto action describing nonrelativistic strings in string Newton-Cartan geometry, which arises in the $p$-brane limit of relativistic string theory. See relevant discussions in Appendix \ref{app:dpbpbncg} and \S\ref{sec:drm2}, respectively.}  In Appendix \ref{app:stbr}, we provide an extra check using a symmetry argument for the Buscher rules derived in \S\ref{sec:T-duality}.

\section{D-Branes in Nonrelativistic String Theory}
\label{sec:D-branes in NRST}

In this section, we construct the effective worldvolume actions describing D-branes in nonrelativistic string theory coupled to arbitrary bosonic background fields. In particular, we turn on arbitrary RR potentials here. We will focus on the case of a single D-brane with an abelian worldvolume gauge potential. Moreover, we only keep leading-order terms in the derivative expansion, such that derivatives of the worldvolume field strength are neglected. We also assume the geometric curvature to be small and omit any dependence on the geometric curvature. This will form the foundation for our later studies of duality transformations in nonrelativistic string theory. We will also discuss how to reproduce these D-brane actions from a stringy limit of relativistic string theory that induces a codimension-two foliation structure in the target-space geometry. Such a stringy limit can be generalized to a $p$-brane limit of relativistic string/M-theory that imposes a codimension-$(p+1)$ foliation in spacetime, which we review at the end of this section in \S\ref{sec:pbl}. Among these $p$-brane limits, the two-brane limit will play an important role when we dualize the worldvolume $U(1)$ gauge field for D2- and D4-branes later in \S\ref{sec:s-duality}, where nonrelativistic M-theory will be probed at the strongly-coupled regime of Type IIA nonrelativistic superstring theory. 

\subsection{Nonrelativistic String Limit of D-Brane Actions} \label{sec:zrsl}

We start with the Dirac-Born-Infeld (DBI) action that describes the dynamics of a D$p$-brane in relativistic string theory. We denote the coordinates on the spacetime manifold $\CM$ as $X^I$, $I = 0, 1, \cdots, 9$\,. Such spacetime coordinates $X^I$ are associated with the worldsheet fields that map the worldsheet $\Sigma$ to the target space $\CM$ in relativistic string theory. On the D-brane submanifold $\CN$\,, we denote the coordinates as $Y^\mu$, $\mu = 0,1,\cdots,p$\,. In curved spacetime, we write
$X^M \big|_{\p\Sigma} = f^M (Y^\mu)$\,, where $f^M$ describes how the D$p$-brane is embedded in spacetime. Consider the closed string background described by a metric field $\hat{G}_{\mu\nu}$\,, a Kalb-Ramond field $\hat{B}_{\mu\nu}$ and a dilaton field $\hat{\Phi}$\,. We also introduce a $U(1)$ gauge field $\hat{A}_\mu$ together with its field strength $\hat{F} = d\hat{A}$ on the D-brane. It is useful to define the pullbacks $\hat{G}_{\mu\nu} = \p_\mu f^M \, \p_\nu f^N \, \hat{G}_{MN}$ and $\hat{\CF}_{\mu\nu} = \p_\mu f^M \, \p_\nu f^N \, \hat{B}_{MN} + \hat{F}_{\mu\nu}$\,. Then, the DBI action of a D$p$-brane in relativistic string theory takes the form,
\be \label{eq:hsp}
    \hat{S}_p = - \hat{T}_p \int d^{p+1} Y \, e^{-\phi} \sqrt{-\det \lr \hat{G}_{\mu\nu} + \hat{\CF}_{\mu\nu} \rr}\,.
\ee 
The brane tension $\hat{T}_p$ is related to the string coupling $\hat{g}_s = e^{\hat{\Phi}_0}$ and the Regge slope $\hat{\alpha}'$ via 
\be
    \hat{T}_p = \frac{1}{(2\pi)^p \, \hat{g}_s \, (\hat{\alpha}')^{(p+1)/2}}\,.
\ee
Here, $\hat{\Phi}_0$ is constant and the dilaton field is now split to be $\hat{\Phi} = \Phi_0 + \phi$\,.
%
Consider the zero Regge slope limit $\hat{\alpha}' \rightarrow 0$ with the following field configurations \cite{Klebanov:2000pp, Gomis:2000bd, Danielsson:2000gi}:
\begin{align} \label{eq:ncos}
    \hat{G}_{MN} = 
        \begin{pmatrix}
            \eta_{AB}^{} & \,\, 0 \\[2pt]
            0 & \,\, \frac{\hat{\alpha}'}{\alpha'} \, \delta_{A'B'}
        \end{pmatrix},
        \quad%
    \hat{B}_{MN} = 
        \begin{pmatrix}
            - \epsilon_{AB}^{} & \,\, 0 \\[2pt]
            0 & \,\, 0
        \end{pmatrix} + \frac{\hat{\alpha}'}{\alpha'} \, B_{MN}\,,
        \quad%
    \hat{F}_{\mu\nu} = \frac{\hat{\alpha}{}'}{\alpha'} \, F_{\mu\nu}\,.
\end{align}
We also set $\phi = 0$\,.
Here, the index $M$ is split into the longitudinal part $A = 0,1$ and the transverse part $A' = 2, \cdots, 9$\,. Moreover, $F_{\mu\nu} = \p_\mu A_\nu - \p_\nu A_\mu$\,, where $A_\mu$ is the rescaled $U(1)$ gauge field. We defined $\eta_{AB} = \text{diag} (-1, 1)$ and the Levi-Civita symbol $\epsilon_{AB}$ by $\epsilon_{01} = - \epsilon_{10} = 1$\,. We have introduced $\alpha'$ that will later become the effective Regge slope in nonrelativistic string theory.
For the meantime, we hold fixed the radius $R_{10} = \hat{g}_s \sqrt{\hat{\alpha}'}$ of the circle compactified over the eleventh dimension $X^{10}$ in 
M-theory. The effective string coupling is
\be \label{eq:gsrel}
    g_s = \hat{g}_s \sqrt{\frac{\hat{\alpha}'}{\alpha'}}\,.
\ee
We kept a remainder $B$-field $B_{MN}$ in \eqref{eq:ncos} for later use. Taking the limit $\hat{\alpha}{}'\rightarrow 0$ in \eqref{eq:hsp} then leads to a finite action referred to as Galilean DBI in \cite{Gomis:2020fui},
\be  \label{eq:spnr0}
    S_p = - T_p \int d^{p+1} Y \, \sqrt{- \det 
        \begin{pmatrix}
            0 & \,\,\,\, \p_\nu \! \lr f^0 + f^1 \rr \\[4pt]
            \p_\mu \! \lr f^0 - f^1 \rr & \,\,\,\, \p_\mu f^{A'} \, \p_\nu f^{A'} + \CF_{\mu\nu}
        \end{pmatrix}}\,,
\ee
where $\CF_{\mu\nu} = \p_\mu f^M \, \p_\nu f^N \, B_{MN} + F_{\mu\nu}$ and
\be
    T_p = \frac{1}{(2\pi)^p \, g_s \, (\alpha')^{(p+1)/2}}\,.
\ee
We will collectively refer to D$p$-branes described by the action \eqref{eq:spnr0} as \emph{\g D$p$-brane}. 
The near critical field limit treats the longitudinal and transverse sectors unequally. 
As a result, there are different configurations that we can consider, depending on whether the D-brane is transverse or extending in the longitudinal spatial $X^1$ direction. For historical reasons, this $\hat{\alpha}' \rightarrow 0$ limit is referred to as the noncommutative open string limit~\cite{Seiberg:2000ms, Gopakumar:2000na, Klebanov:2000pp}. In this paper, we will view this limit in a larger framework of nonrelativistic string theory and therefore refer to such a limit as the \emph{nonrelativistic string limit}.

We first consider a D-brane that is transverse to the longitudinal spatial $X^1$ direction. In this case, $f^0 = Y^0$ and $f^1 = x^1 + \pi$\,, where $x^1$ is the location of the D-brane in $X^1$. The field $\pi$ is the Nambu-Goldstone boson that emerges from the spontaneous symmetry breaking of the translational isometry in $X^1$\,. Similarly, $f^{A'}$ now splits into 
\begin{subequations}
\begin{align}
    f^{i+1} = Y^{i}\,, 
        & \qquad%
    i = 1, \cdots, \, p\,; \\[2pt] 
    f^{a} = x^{a} + \pi^a\,, 
        & \qquad%
    a = p+2\,, \cdots, 9\,.
\end{align}
\end{subequations}
Here, $x^a$ 
is the location of the D$p$-brane along $X^a$, and $\pi^a$ 
are the Nambu-Goldstone bosons that perturb perpendicularly to the D-brane in $X^a$. In the absence of $B$-field with $B_{MN} = 0$\,, the low-energy D$p$-brane action \eqref{eq:spnr0} 
becomes
\be \label{eq:nrosbi}
    S_p = - T_p \int d^{p+1} Y \, \sqrt{- \det 
        \begin{pmatrix}
            0 & \,\,\,\, \delta_\nu^0 + \p_\nu \pi \\[4pt]
            \delta_\mu^0 - \p_\mu \pi & \,\,\,\, \delta_\mu^{i} \, \delta_\nu^{i} + \p_\mu \pi^a \, \p_\nu \pi^a + F_{\mu\nu}
        \end{pmatrix}}\,.
\ee
This theory is the effective field theory on D-branes in nonrelativistic open string (NROS) theory.
At the quadratic order in field configurations, we find the effective action,
\be \label{eq:ged}
    S^{(2)}_p = \frac{T_p}{2} \int d^{p+1} Y \Bigr( \dot{\pi}^2 - 2 \, E_i \, \p_i \pi - \tfrac{1}{2} \, F_{ij} \, F^{ij} - \p_i \pi^a \, \p_i \pi^a \Bigr)\,.    
\ee
Here, $E_i = F_{0i}$ is the electric field.
This quadratic action \eqref{eq:ged} is invariant under a Galilean boost symmetry \cite{Festuccia:2016caf},
\be
    \tilde{Y}^0 = Y^0\,,
        \qquad
    \tilde{Y}^i = Y^i + v^i \, Y^0,
\ee
if supplemented with the following field transformations:
\begin{subequations}
\begin{align}
    \tilde{A}_0 \bigl(\tilde{Y}\bigr) & = A_0 (Y) - v^i A_i (Y) + \tfrac{1}{2} \, v_i \, v^i \, \pi\,, 
        &%
    \tilde{\pi} (\tilde{Y}) & = \pi (Y)\,, \\[2pt]
    \tilde{A}_i \bigl(\tilde{Y}\bigr) & = A_i (Y) - v_i \, \pi\,, 
        &%
    \tilde{\pi}^a (\tilde{Y}) & = \pi^a (Y)\,.
\end{align}
\end{subequations}
The action \eqref{eq:ged} without the last term that depends on $\pi^a$ is referred to as Galilean electrodynamics (GED) in the literature \cite{Santos:2004pq, Festuccia:2016caf, Bergshoeff:2015sic}. There are no propagating degrees of freedom in \eqref{eq:ged}. However, it is shown in \cite{Chapman:2020vtn} that, in 2+1 dimensions, coupling GED to propagating Schr\"{o}dinger scalars generates non-trivial modifications to renormalization group flows, which give rise to a family of conformal fixed points.

Next, we consider D-branes extending in the longitudinal spatial $X^1$ direction. We introduce a purely electric $B$-field with $B_{AB} = e\,\epsilon_{AB}/ 2$ and all other components in $B_{\mu\nu}$ are taken to be zero. Moreover, we take $f^A = Y^A$ and split $f^{A'}$ into
\begin{subequations}
\begin{align}
    f^i = Y^i \,, &\qquad i = 2, \cdots, p\,; \\[2pt]
    f^a = x^a + \pi^a\,, &\qquad a = p+1\,, \cdots, 9\,. 
\end{align}
\end{subequations}
The D$p$-brane action \eqref{eq:spnr0} 
now becomes
\be  \label{eq:ncosdbifs}
    S_p = - T_p \int d^{p+1} Y \, \sqrt{- \det 
        \begin{pmatrix}
            0 & \,\,\,\, \delta_\nu^0 + \delta_\nu^1 \\[4pt]
            \delta_\mu^0 - \delta_\mu^1 & \,\,\,\, \delta_\mu^i \, \delta_\nu^i + \p_\mu \pi^a \, \p_\nu \pi^a + \CF_{\mu\nu}
        \end{pmatrix}}\,.
\ee
%
%
At the quadratic order in field configurations of \eqref{eq:ncosdbifs}, we have
\begin{align} \label{eq:eftncos}
\begin{split}
    S^{(2)}_p & = - \frac{T_p}{4 \, e^{3/2}} \! \int d^{p+1} Y \, \Bigl( F_{AB} \, F^{AB} + 2 \, e \, F_{Ai} \, F^{A}{}_i + e^2 \, F_{ij} \, F_{ij} \Bigr) \\[4pt]
    & \quad - \frac{T_p}{2 \, e^{1/2}} \int d^{p+1} Y \, \Bigl( \p_A \pi^a \, \p^A \pi^a + e \, \p_i \pi^a \, \p_i \pi^a \Bigr)\,.
\end{split}
\end{align}
Taking the rescaling $Y^i \rightarrow e^{1/2} \, Y^i$ and $A_\mu \rightarrow e^{1/2} \, A_\mu$\,, the effective action \eqref{eq:eftncos} becomes
\be \label{eq:ff}
    S^{(2)}_p = - \frac{T_p \, e^{p/2}}{2} \int d^{p+1} Y \, \Bigl( \tfrac{1}{4} \, F_{\mu\nu} \, F^{\mu\nu} + \tfrac{1}{2} \, \p_\mu \pi^a \, \p^\mu \pi^a \Bigr)\,,
\ee
which is manifestly relativistic. Open strings that end on such a D-brane configuration described at low energies by \eqref{eq:ncosdbifs} is known as noncommutative open strings (NCOS) in the literature, where space/time noncommutativity arises due to the present of the nonzero Kalb-Ramond field $B_{AB}$ \cite{Gopakumar:2000na}. 
This noncommutative behavior becomes manifest after using the Seiberg-Witten map to rewrite the worldvolume theory in terms of the effective background fields seen by the open strings \cite{Seiberg:1999vs}, which we briefly describe below. We start with the DBI action \eqref{eq:hsp} that describes relativistic D-branes. The inverse effective metric $\hat{\CG}^{\mu\nu}$\,, the noncommutativity tensor $\Theta^{\mu\nu}$\,, and the effective open string coupling $\hat{\CG}_\text{o}$ seen by the open strings are given by the following Seiberg-Witten map between closed and open string background fields \cite{Seiberg:1999vs}:
\begin{subequations} \label{eq:swm}
\begin{align}
    \hat{\CG}^{\mu\nu} & = \frac{\hat{\alpha}'}{\alpha'} \lr \frac{1}{\hat{G} + \hat{B}} \, \hat{G} \, \frac{1}{\hat{G} - \hat{B}} \rr^{\mu\nu},
        \qquad%
    \hat{\CG}_\text{o} = \hat{g}_s \sqrt{\frac{\det \bigl( \hat{G} + \hat{B} \bigr)}{\det \hat{G}}}\,, \\[4pt]
    \hat{\Theta}^{\mu\nu} & = - \frac{\hat{\alpha}'}{\alpha'} \lr \frac{1}{\hat{G} + \hat{B}} \, \hat{B} \, \frac{1}{\hat{G} - \hat{B}} \rr^{\mu\nu}.
\end{align}
\end{subequations}
Here, the equal time commutator $[Y^\mu\,, Y^\nu] \sim \hat{\Theta}^{\mu\nu}$ measures the noncommutativity between different worldvolume coordinates.
In terms of these open string background fields, the associated field theory can be written in terms of the noncommutative Yang-Mills fields using the Moyal bracket \cite{Seiberg:1999vs}.~\footnote{In the noncommutative Yang-Mills action, the terms that are quadratic in $F_{\mu\nu}$ are still the same as in \eqref{eq:ff}. However, the noncommutative field strength involves higher order terms in the gauge potential $A_\mu$\,, with $F'_{\mu\nu} = \p_\mu A'_{\nu} - \p_\mu A'_\nu - i \, A'_\mu \star A'_\nu + i \, A'_\nu \star A'_\mu$\,, where ``$\star$" denotes the Moyal product, and the prime in $A'_\mu$ indicates that a point-splitting regularization has been introduced \cite{Seiberg:1999vs}.}   
Plugging \eqref{eq:ncos} and \eqref{eq:gsrel} into the Seiberg-Witten map \eqref{eq:swm}, and with the Kalb-Ramond field being purely electric as we have specified earlier, we derive the following NCOS variables that arise in the $\hat{\alpha}' \rightarrow 0$ limit: 
\begin{align} \label{eq:nstsw}
    \hat{\CG}^{\mu\nu} \rightarrow \CG^{\mu\nu} = 
    \begin{pmatrix}
         e^{-1} \, \eta_{AB} & 0 \\[2pt]
        0 & \delta^{}_{ij}
    \end{pmatrix}, 
        \quad%
    \hat{\Theta}^{\mu\nu} \rightarrow \Theta^{\mu\nu} = - \begin{pmatrix}
       e^{-1} \, \epsilon_{AB} & 0 \\[2pt]
        0 & 0
    \end{pmatrix},
        \quad%
    \hat{\CG}_\text{o} \rightarrow \CG_\text{o} = g_s \sqrt{e}\,.
\end{align}
The structure of the noncommutativity tensor $\Theta^{\mu\nu}$ implies that the longitudinal space and time coordinates $Y^0$ and $Y^1$ do not commute with each other, with $[Y^0\,, Y^1] \propto e^{-1}$\,.

As a final remark, we note that there is a formal T-dual relation between NCOS D$p$-branes and NROS D$(p\!-\!1)$-branes \cite{Gomis:2020izd}: Start with the NCOS D$p$-brane, we take an infinite boost in the longitudinal sector of the $X^1$ circle, along which the D-brane extends. Then, we are led to the DLCQ of NCOS on a compactified lightlike circle. Performing a T-duality transformation along this lightlike circle leads to a D$(p\!-\!1)$-brane in the DLCQ of NROS on a dual lightlike circle. 

\subsection{\gc DBI Action} \label{sec:gdbia}

There is a natural curved-spacetime generalization of the above zero Regge slope limit \cite{Andringa:2012uz}. This limit treats the longitudinal directions differently from the transverse directions, and thus induces a codimension-two foliation structure in spacetime. In curved spacetime, we introduce the longitudinal vielbein field $\tau_\mu{}^A$ and the transverse vielbein field $E_\mu{}^{A'}$. 
We consider the reparametrizations of relativistic background fields in terms of $\hat{\alpha}{}'$,
\begin{subequations} \label{eq:aexp}
\begin{align}
    \hat{G}^{}_{MN} & = \tau^{}_{MN} + \frac{\hat{\alpha}{}'}{\alpha'} \, E^{}_{MN}\,, 
        \qquad%
    e^{\hat{\Phi}} = e^\Phi \sqrt{\frac{\alpha'}{\hat{\alpha}{}'}}\,, \\[2pt]
    \hat{B}^{}_{MN} & = - \tau^{}_M{}^A \, \tau^{}_N{}^B \, \epsilon^{}_{AB} + \frac{\hat{\alpha}{}'}{\alpha'} \, M^{}_{MN}\,,
\end{align}
\end{subequations}
where $\tau^{}_{MN} = \tau^{}_M{}^A \, \tau^{}_N{}^B \, \eta_{AB}$\,, $E^{}_{MN} = E^{}_M{}^{A'} E^{}_N{}^{A'}$ and $M_{MN}$ is an antisymmetric two-tensor. The ansatz \eqref{eq:aexp} is a direct covariantization of \eqref{eq:ncos}. Applying the $\hat{\alpha}{}' \rightarrow 0$ limit to the relativistic DBI action \eqref{eq:hsp},  we find \cite{Gomis:2020fui} (also see \cite{Kluson:2019avy, Kluson:2020aoq, Gomis:2020izd}),
\be \label{eq:spnr}
    S_{\text{D}p} = - T_p \int d^{p+1} Y \, e^{-\phi} \sqrt{- \det 
        \begin{pmatrix}
            0 & \,\,\,\, \tau_\nu \\[4pt]
            \bar{\tau}_\mu & \,\,\,\, E_{\mu\nu} + M_{\mu\nu} + F_{\mu\nu}
        \end{pmatrix}}\,.
\ee
Here, we have written the dilaton field $\Phi$ as $\Phi = \log g_s + \phi$\,, such that the expectation value of $\phi$ is zero. Moreover, $F_{\mu\nu} = \p_\mu A_\nu - \p_\nu A_\mu$ for a $U(1)$ gauge potential $A_\mu$ on the D-brane. We also defined the following pullbacks to the worldvolume:
\be
    \tau_\mu{}^A = \tau^{}_M{}^A \, \p_\mu f^M \,,
        \qquad%
    E_{\mu}{}^{A'} = E^{}_M{}^{A'} \, \p_\mu f^M\,,
        \qquad%
    M_{\mu\nu} = M^{}_{MN} \, \p_\mu f^M \, \p_\nu f^N\,.
\ee
We also defined $\tau_\mu = \tau_\mu{}^0 + \tau_\mu{}^1$ and $\bar{\tau}_\mu = \tau_\mu{}^0 - \tau_\mu{}^1$. The action \eqref{eq:spnr} is invariant under the local string Galilean boost transformations parametrized by $\Lambda^A{}_{A'}$\,,
\be
    \delta^{}_\text{G} \tau^{}_M{}^A = 0\,,
        \qquad%
    \delta^{}_\text{G} E^{}_M{}^{A'} = \Lambda^{A'}{}_A \, \tau^{}_M{}^A\,,
\ee
supplemented with 
\be
    \delta^{}_\text{G} M^{}_{MN} = \Lambda^{A'}{}_{\!A} \, \epsilon^{A}{}_{B} \lr \tau^{}_M{}^B E^{}_N{}^{A'} \!- \tau^{}_N{}^B E^{}_M{}^{A'} \rr.
\ee
Together with diffeomorphisms, the Lorentz boost in the longitudinal sector, and rotations in the transverse sector, all these transformations form the string Galilei algebra.
It is useful to introduce an additional gauge field $m_\mu{}^A$ that transforms nontrivially under the string Galilean boosts,
\be
    \delta m^{}_M{}^A = - \Lambda_{A'}{}^A \, E_\mu{}^{A'}, 
\ee
and parametrize $M^{}_{MN}$ as
\be \label{eq:mbrelation}
    M^{}_{MN} = B^{}_{MN} + \lr m^{}_M{}^A \, \tau^{}_N{}^B - m^{}_N{}^A \, \tau^{}_M{}^B \rr \epsilon_{AB}\,.
\ee
Then, $B_{MN}$ is invariant under the string Galilei boosts. This Kalb-Ramond field transforms under the Neveu-Schwarz (NS) gauge symmetry parametrized by $\epsilon_M$ as
\be \label{eq:tgse}
    \delta_\epsilon B_{MN} = \p_M \epsilon_N - \p_N \epsilon_M\,.
\ee 
The action \eqref{eq:spnr} is invariant under \eqref{eq:tgse}, if supplemented with the following transformation of the gauge potential $A_\mu$\,:
\be \label{eq:deam}
    \delta_\epsilon A_M = - \epsilon_M\,,
\ee
apart from the $U(1)$ gauge transformation
\be \label{eq:dampeta}
    \delta_\eta A_\mu = \p_\mu \eta\,.
\ee
The target-space gauge fields $\tau^{}_M{}^A$, $E^{}_M{}^{A'}$ and $m^{}_M{}^A$ constitute the so-called \emph{torsional string Newton-Cartan geometry} \cite{Harmark:2018cdl, Harmark:2019upf,  Bergshoeff:2021bmc, Bidussi:2021ujm}.

It is known that the string Galilei symmetry is not sufficient for the sigma model describing nonrelativistic string theory to be self-consistent at the quantum level; instead, one has to introduce an extra counterterm that essentially deforms the theory towards relativistic string theory \cite{Gomis:2019zyu, Yan:2019xsf, Yan:2021lbe}. In order to define a renormalizable two-dimensional worldsheet QFT without this deformation, the string Galilei symmetry is extended to include a noncentral extension associated with the gauge field $m_M{}^A$ \cite{Gomis:2019zyu, Yan:2019xsf} (also see \cite{Andringa:2012uz, Bergshoeff:2018yvt, Bergshoeff:2019pij}). This extension modifies the Lie bracket between the string Galilei boost and transverse translation generator such that they commute into a new generator $Z_A$\,, which only acts nontrivially (and infinitesimally) on $m_M{}^A$ and $A_\mu$ as \cite{Gomis:2020fui}
\be \label{eq:dzam}
    \delta^{}_Z m_M{}^A = D_M \sigma^A\,,
        \qquad%
    \delta^{}_Z A_\mu = - \epsilon^{}_{AB} \, \sigma^A \, \tau_\mu{}^B\,.
\ee
Here, $\sigma^A$ is the Lie group parameter associated with $Z_A$ and $D_M \sigma^A = \p_M \sigma^A - \Omega_M \, \epsilon^{A}{}_B \, \sigma^B$, with $\Omega_M$ the spin connection for the longitudinal Lorentz boost. Extending the string Galilei algebra with the $Z_A$ generator (and together with other generators required for the closedness of the algebra) leads to the so-called string Bargmann algebra \cite{Brugues:2004an, Brugues:2006yd, Andringa:2012uz, Bergshoeff:2019pij}, which underlies the \emph{string Newton-Cartan geometry}. In the gauging procedure of the string Bargmann algebra, $m^{}_M{}^A$ turns out to be the gauge field associated with the $Z_A$ generator.
Moreover, the $Z_A$ symmetry imposes a torsional constraint on the longitudinal Vielbein field $\tau^{}_M{}^A$, with \cite{Andringa:2012uz, Bergshoeff:2018yvt}
\be \label{eq:dt0}
    D^{}_{[M} \tau^{}_{N]}{}^A = 0\,.
\ee
Using the torsional constraint \eqref{eq:dt0}, we find that the field strength $F = dA$ transforms under the $Z_A$ symmetry as
\be
    \delta_Z F_{\mu\nu} = - \epsilon^{}_{AB} \, D^{}_{\mu} \sigma^A \, \tau^{}_\nu{}^B.  
\ee
It is shown in \cite{Gomis:2020fui} that \eqref{eq:spnr} is invariant under the above $Z_A$ transformations and thus the string Bargmann symmetry. In \cite{Yan:2021lbe}, it is shown that preserving half of the lightlike components in the $Z_A$ symmetry, \emph{e.g.}, $Z_+ = Z_0 + Z_1$\,, also leads to a self-consistent algebra. This $Z_+$ symmetry only imposes half of the torsional constraint in \eqref{eq:dt0} that coincides with the one found in \cite{Bergshoeff:2021tfn} from supersymmetrizing string Newton-Cartan geometry. This relaxed torsional constraint also leads to a renormalizable worldsheet QFT that describes nonrelativistic strings \cite{Yan:2021lbe}. Since the above distinctions between these non-Lorentzian geometries with different torsional constraints will not play any role in this paper, we will refer to the target-space geometry as ``string Newton-Cartan geometry" in the generic sense, without specifying what torsional constraint is imposed on $\tau^{}_M{}^A$\,. 

In terms of the above geometric data, we write $S_{\text{D}p}$ in the form that is manifestly boost invariant, with
\be \label{eq:sphb}
    S_{\text{D}p} = - T_p \int d^{p+1} Y \, e^{-\phi} \sqrt{- \det 
        \begin{pmatrix}
            0 & \,\,\,\, \tau_\nu \\[4pt]
            \bar{\tau}_\mu & \,\,\,\, H_{\mu\nu} + B_{\mu\nu} + F_{\mu\nu}
        \end{pmatrix}}\,,
\ee
where both $H_{\mu\nu} \equiv E_{\mu\nu} + \lr \tau_\mu{}^A \, m_\nu{}^B + \tau_\nu{}^A \, m_\nu{}^B \rr \eta_{AB}$ and $B_{\mu\nu}$ are invariant under the string Galilei boost. This action \eqref{eq:sphb} introduces extra field contents that give rise to Stueckelberg-type symmetries \cite{Bergshoeff:2018yvt}: 
\begin{subequations} \label{eq:ss}
\begin{align} 
    H^{}_{MN} & \rightarrow H^{}_{MN} - \lr \tau^{}_M{}^A \, \Xi^{}_N{}^B + \tau^{}_N{}^A \, \Xi^{}_M{}^B \rr \eta^{}_{AB}\,, \\[2pt]
    B^{}_{MN} & \rightarrow B^{}_{MN} + \lr \tau^{}_M{}^A \, \Xi^{}_N{}^B - \tau^{}_N{}^A \, \Xi^{}_M{}^B \rr \epsilon^{}_{AB}\,. \label{eq:dxib}
\end{align}
\end{subequations}
Note that \eqref{eq:ss} are finite transformations.
This formalism in terms of $H_{MN}$ and $B_{MN}$ has the advantage that the geometric data is separated from the $B$-field. This (infinitesimal) Stueckelberg-type symmetry will also serve as a useful check later in the paper. 

In \cite{Gomis:2020fui}, the nonlinear equations of motion that govern the consistent open string background fields in nonrelativistic string theory are derived by demanding the quantum conformal invariance on the string worldsheet. Furthermore, it is shown that these equations coincide with the ones from varying \eqref{eq:sphb}. Therefore, \eqref{eq:sphb} defines the D-brane action that describes the low-energy dynamics of open string background fields in nonrelativistic string theory coupled to a string Newton-Cartan geometry, Kalb-Ramond and dilaton field. 



\subsection{Coupling to RR Potentials} \label{sec:nrsl}

We now couple the \g D$p$-brane action \eqref{eq:sphb} to the RR potentials via a CS term. We denote a differential $q$-form RR potential by $C^{(q)}$\,. The RR potentials arise in nonrelativistic superstring theory similarly as in relativistic superstring theory \cite{Kim:2007pc}. 
In addition, the $B$-field and the gauge field strength $F_{\mu\nu}$ also contribute to the CS term. This requires that the CS action also includes RR potentials of ranks no greater than $p+1$ for a D$p$-brane. The complete CS term is given by \cite{Li:1995pq, Douglas:1995bn}
\be \label{eq:sphbrr}
    S_\text{CS} = \mu_p \int \sum_{q} C^{(q)} \wedge e^{\mathcal{F}} \, \Bigr|_{p+1}\,,
        \qquad%
    \CF = B + F\,.
\ee
where $\mu_p$ is the D$p$-brane charge and the subscript $p+1$ indicates that only the $(p+1)$-forms are kept in the expression. In this paper, we assume that $q \geq 0$ for $C^{(q)}$\,.~\footnote{It is also interesting to consider $q = -1$\,, in which case there is a D$(-1)$-brane that plays the role of a spacetime instanton.} The factor $e^\CF$ denotes an infinite sum over wedge products of $\CF = B + F$\,. In the $B=0$\, case, and in a more standard normalization, this factor can be written as $e^{i F / (2\pi)}$\,, which is the Chern character that generates polynomials of Chern classes in the (generically non-abelian) gauge bundle, with $F$ being the associated curvature. As in relativistic string theory, the CS term in \eqref{eq:sphbrr} is related to various topological features of the gauge bundle on the D-brane's worldvolume.~\footnote{In relativistic string theory, such terms are also important for cancelling the anomalies in the Green-Schwarz mechanism and necessary for having the correct T-duality transformations. We will discuss T-duality transformations that involve the RR potentials in \S\ref{sec:T-duality} for nonrelativistic string theory.}   

The CS action \eqref{eq:sphbrr} is invariant under the NS gauge symmetry \eqref{eq:tgse} and \eqref{eq:deam}, and the worldvolume $U(1)$ gauge symmetry \eqref{eq:dampeta}. Additionally, it is also invariant (up to a boundary term) under the RR gauge transformation,
\be \label{eq:rrgt}
    \delta_\zeta {\sum}_{\substack{q}} C^{(q)} = d\zeta^{(q-1)} + dB \wedge \zeta^{(q-3)}\,.
\ee
We assume that $\zeta^{(q)} = 0$ for $q < 0$\,.
One may also express the CS action \eqref{eq:sphbrr} as an integral over a ($p+2$)-dimensional worldvolume with a ($p+1$)-dimensional boundary
\be 
    S_\text{CS} = \mu_p \int \sum_{q}R^{(q)} \wedge e^{\mathcal{F}} \, \Bigr|_{p+2}\,,
\ee
where the RR field strength
\be
R^{(q)} = dC^{(q-1)} + dB \wedge C^{(q-3)}\
\ee
is invariant under the RR gauge transformation \eqref{eq:rrgt} and the NS gauge transformation defined in \eqref{eq:tgse} and \eqref{eq:deam}.
%
%
%
%
Moreover, the RR potential $C^{(q)}$ is boost invariant, but it transforms infinitesimally under the $Z_A$ symmetry (if imposed) as
\be
    \delta_Z C^{(q)} = C^{(q-2)} \wedge D\sigma^A \wedge \tau^B \, \epsilon_{AB}\,,
\ee
and accordingly for the halved $Z_A$ symmetry in \cite{Yan:2021lbe}. Here, $D = dY^\mu \, D_\mu$ and $\tau^A = dY^\mu \, \tau_\mu{}^A$.

The finite Stueckelberg symmetry \eqref{eq:ss} can also be extended to act on the RR potentials:
\begin{align} \label{eq:ssC}
    C^{(q)} & \rightarrow C^{(q)} - C^{(q-2)} \wedge \tau^A \wedge \Xi^B \, \epsilon_{AB} + \tfrac{1}{2} \, C^{(q-4)} \wedge \tau^A \wedge \Xi^B \wedge \tau^C \wedge \Xi^D \, \epsilon_{AB} \, \epsilon_{CD}\,.
\end{align} 
Fixing the Stueckelberg symmetry by setting $\Xi^{}_M{}^A = m^{}_M{}^A$, we find that the Galilean DBI action becomes \eqref{eq:spnr} and the CS action becomes
\be \label{eq:scsm}
    S_\text{CS} = \mu_p \int \sum_{q} N^{(q)} \wedge e^{M + F} \Big|_{p+1}\,,
\ee
where $M = B + m^{(2)}$ as in \eqref{eq:mbrelation} and
\be \label{eq:mq}
    N^{(q)} = C^{(q)} - C^{(q-2)} \wedge m^{(2)} + \tfrac{1}{2} \, C^{(q-4)} \wedge m^{(2)} \wedge m^{(2)}\,.
\ee
The two-form $m^{(2)}$ is defined in components as
\be
    m^{(2)}_{MN} = \lr \tau^{\phantom{(}}_{M}{}^{\!A} \, m^{\phantom{(}}_{N}{}^B - \tau^{\phantom{(}}_{N}{}^{\!A} \, m^{\phantom{(}}_{M}{}^B \rr \epsilon^{}_{\!AB}\,. 
\ee
We have used the identity $m^{(2)} \wedge m^{(2)} \wedge m^{(2)} = 0$ to derive \eqref{eq:scsm}. Note that $N^{(q)}$ transforms nontrivially under the string Galilei boosts but trivially under the $Z_A$ symmetry.

It is useful to understand how to obtain the Galilean DBI action coupled to the RR potentials from the zero Regge slope limit of the Galilean DBI action as in \S\ref{sec:gdbia}. Starting with the relativistic DBI action \eqref{eq:hsp}, to which we add the following CS action: 
%
%
%
\be \label{eq:hs}
    \hat{S}_\text{CS} = \hat{\mu}_p \int \sum_{q} \hat{C}^{(q)} \wedge e^{\hat{\CF}} \, \Bigr|_{p+1}\,,
        \qquad%
    \hat{\mu}_p = \frac{1}{(2\pi)^p \, (\hat{\alpha}')^{(p+1)/2}}\,,
\ee
where $\hat{\CF} = \hat{B} + F$\,. In addition to the parametrizations of background fields in \eqref{eq:aexp}, we also parametrize the relativistic RR potentials $\hat{C}^{(q)}$ as \footnote{Note the pattern that a $q$-form in relativistic string theory receives a rescaling factor $\sqrt{\hat{\alpha}' / \alpha'}$ for the nonrelativistic string limit to work. However, the two-form $\ell$ is not rescaled.} 
\begin{align} \label{eq:exc}
    \hat{C}^{(q)} = 
    \lr \frac{\hat{\alpha}'}{\alpha'} \rr^{(q-2)/2} \! \lr N^{(q-2)} \wedge \ell + \frac{\hat{\alpha}'}{\alpha'} \, N^{(q)} \rr.
\end{align}
Here, $N^{(q)} = 0$ for $q < 0$ and the components of the two-form $\ell$ is defined as the pullback of
\be \label{eq:defell}
    \ell^{\phantom{(}}_{MN} = \tau^{\phantom{(}}_M{}^A \, \tau^{\phantom{(}}_N{}^B \, \epsilon^{\phantom{(}}_{AB}\,.
\ee
In terms of $\tau^A = \tau_\mu{}^A \, dY^\mu$\,, we have $\ell = \frac{1}{2} \, \tau^A \wedge \tau^B \epsilon_{AB}$\,.
%
%
Plugging the parametrizations \eqref{eq:aexp} and \eqref{eq:exc} into \eqref{eq:hs}, followed by taking the limit $\hat{\alpha}' \rightarrow 0$\,, we find that the resulting action is exactly \eqref{eq:scsm} with $\mu_p = (2\pi)^{-p} \, \alpha'{}^{-(p+1)/2}$\,.

For later references, it is useful to rescale the relativistic background fields and define the low-energy limit $\hat{\alpha}' \rightarrow 0$ in an alternative way, such that the Regge slope is not rescaled. Moreover, it is also helpful to take a generalized form of the parametrizations of the relativistic background fields such that the Stueckelberg symmetries are made manifest after the limit is taken. We now describe how such a modified limit is defined. We start in relativistic string theory with the D$p$-brane action including both the DBI and CS terms: 
\be \label{sec:hspf}
    \hat{S}_{\text{D}p} = - \int d^{p+1} Y \, e^{-\hat{\Phi}} \sqrt{-\det \lr \hat{G}_{\mu\nu} + \hat{\CF}_{\mu\nu} \rr} + \int \sum_{q} \hat{C}^{(q)} \wedge e^{\hat{\CF}} \Big|_{p+1}.
\ee
For simplicity, we assume that both the couplings in front of the DBI and CS terms are unity. We now consider the following ansatz (with $\hat{\alpha}' \rightarrow 1/c^2$):
\begin{subequations} \label{eq:cexp}
\begin{align}
    \hat{G}^{\phantom{(}}_{MN} & = c^2 \, \tau^{\phantom{(}}_{MN} + H^{\phantom{(}}_{MN}\,,     
        &%
    \hat{\CF} & = - c^2 \, \ell + \CF\,, \\[2pt]
    \hat{\Phi} & = \Phi + \ln c\,,  
        &%
    \hat{C}^{(q)} & = c^2 \, C^{(q-2)} \wedge \ell + C^{(q)}\,. \label{eq:cqexpf}
\end{align}
\end{subequations}
We have assumed that $C^{(q)} = 0$ for $q < 0$\,. Plugging \eqref{eq:cexp} into \eqref{sec:hspf} and then taking the limit $c \rightarrow \infty$\,, we find that the resulting action is
\be \label{eq:gdbbi}
    S_{\text{D}p} = - \int d^{p+1} Y \, e^{-\Phi} \sqrt{-\det
        \begin{pmatrix}
            0 &\quad \tau_\nu \\[2pt]
            \overline{\tau}_\mu &\quad H_{\mu\nu} + \CF_{\mu\nu}
        \end{pmatrix}} 
        + \int \sum_{q} C^{(q)} \wedge e^{\CF} \Big|_{p+1}\,,
\ee
which matches the DBI-like action \eqref{eq:sphb} and the CS action \eqref{eq:sphbrr}.
This $c\rightarrow\infty$ limit gives a convenient form of the nonrelativistic string limit for bosonic background fields. As we will see momentarily in \S\ref{sec:pbl}, there also exists a natural generalization of this stringy limit to the so-called $p$-brane limits. 

We also make the following remark. Note that the ansatz \eqref{eq:cexp} is a reparametrization of the relativistic background fields rather than an expansion with respect to a large $c$\,. To understand the origin of this reparametrization, we start with the Polyakov formalism of relativistic string theory, 
\be \label{eq:hsrs}
    \hat{S} = - \frac{1}{4\pi\alpha} \int d^2 \sigma \, \p X^M \, \bar{\p} X^N \lr \hat{G}_{MN} + \hat{B}_{MN} \rr,
\ee
where $\sigma^\alpha = \left( \sigma^0, \sigma^1 \right)$ are worldsheet coordinates. Here, $X^M$ are the worldsheet fields that map the worldsheet to the target space. We have taken the conformal gauge and defined the derivatives $\p = \p_{\sigma^1} + i \, \p_{\sigma^0}$ and $\bar{\p} = \p_{\sigma^1} - i \, \p_{\sigma^0}$\,. Plugging in the ansatz for $\hat{G}_{MN}$ and $\hat{B}_{MN}$ that we gave in \eqref{eq:cexp}, and introducing a pair of auxiliary fields $\lambda$ and $\bar{\lambda}$\,, we rewrite \eqref{eq:hsrs} in the following equivalent form \cite{Gomis:2000bd}:
\be
    \hat{S} = - \frac{1}{4\pi\alpha} \int d^2 \sigma \, \Bigl\{ \p X^M \, \bar{\p} X^N \lr H_{MN} + B_{MN} \rr + \lambda \, \bar{\p} X^M \tau_M + \bar{\lambda} \, \p X^N \tau_N + c^{-2} \lambda \bar{\lambda} \Bigr\}.
\ee
Now, $c^{-2}$ is associated with the functional coupling in front of the marginal operator $\lambda\bar{\lambda}$\,. Therefore, the $c \rightarrow \infty$ limit that leads to nonrelativistic string theory finds a worldsheet QFT interpretation as tuning the marginal operator $\lambda\bar{\lambda}$ to zero.~\footnote{Symmetries that protect the worldsheet QFT from being deformed by quantum corrections which generate the marginal operator $\lambda\bar{\lambda}$ have been studied in detail \cite{Gomis:2019zyu, Yan:2019xsf, Yan:2021lbe}.} On the other hand, it is also possible to consider a general Taylor series expansion of the background fields with respect to a large $c$\,. For example, one may express $\hat{G}_{MN}$ as
\be
    \hat{G}^{}_{MN} = c^2 \, \tau^{}_{MN} + H^{}_{MN} + \sum_{n=1}^\infty c^{-2n} \, H^{(n)}_{MN}\,. 
\ee
Even though this may be natural to consider from the perspective of spacetime geometry,~\footnote{See also \cite{Hartong:2021ekg} for an expansion of relativistic string theory with respect to a large $c$\,.} it appears to be an intricate choice from the worldsheet point of view, where the parameter $c$ loses the simple worldsheet QFT interpretation as a coupling constant associated with the marginal operator $\lambda\bar{\lambda}$\,. We will follow the ansatz in \eqref{eq:cexp} throughout the paper, which does not contain any subleading terms if regarded as an expansion with respect to a large $c$\,. This parametrization will be proven to be convenient for our purposes.

\subsection{Nonrelativistic Limits of Strings and \emph{p}-Branes} \label{sec:pbl}

Before we move on to building up a duality web in nonrelativistic string theory, we first review different nonrelativistic limits in string theory that have been introduced in \cite{Gomis:2000bd}.
Nonrelativistic string theory arises as a stringy limit of relativistic string theory. Such a stringy limit of relativistic strings in curved spacetime involves first parametrizing the background fields using the parameter $c$ as in \eqref{eq:cexp}, followed by taking the $c \rightarrow \infty$ limit. We have seen how such a limit of relativistic D-brane action \eqref{sec:hspf} leads to the action \eqref{eq:gdbbi} that describes the low-energy dynamics of D$p$-branes in nonrelativistic string theory \cite{Andringa:2012uz}. Moreover, the action describing nonrelativistic fundamental strings also arises from the same stringy limit of relativistic fundamental strings. To elucidate how nonrelativistic fundamental strings arise, we begin with the Nambu-Goto action for relativistic string theory
\be \label{eq:hsfs}
    \hat{S} = - T \int d^2 \sigma \lr \sqrt{- \det \hat{G}_{\alpha\beta}} + \frac{1}{2} \, \epsilon^{\alpha\beta} \, \hat{B}_{\alpha\beta} \rr.
\ee
We denote the worldsheet coordinates by $\sigma^\alpha = (\sigma^0, \sigma^1)$\,. Moreover, 
\be \label{eq:slcexp}
    \hat{G}_{\alpha\beta} = \p_\alpha X^M \, \p_\beta X^N \, \hat{G}_{MN}\,,
        \qquad%
    \hat{B}_{\alpha\beta} = \p_\alpha X^M \, \p_\beta X^N \, \hat{B}_{MN}
\ee
are the pullbacks of the background fields from the target space to the string worldsheet, with the worldsheet fields $X^M$, $M = 0, 1, \cdots, 9$ mapping the worldsheet to the ten-dimensional target space. These worldsheet fields $X^M$ play the role of spacetime coordinates.  The parametrizations in \eqref{eq:cexp} induce
\be 
    \hat{G}^{\phantom{(}}_{\alpha\beta} = c^2 \, \tau^{\phantom{(}}_{\alpha\beta} + H^{\phantom{(}}_{\alpha\beta}\,, 
        \qquad%
    \hat{B}^{\phantom{(}}_{\alpha\beta} = - c^2 \, \tau^{\phantom{(}}_\alpha{}^A \, \tau^{\phantom{(}}_\beta{}^B \, \epsilon^{\phantom{(}}_{AB} + B^{\phantom{(}}_{\alpha\beta}\,.
\ee
Here, $\tau_{\alpha\beta}$ and $\tau_\alpha{}^A$ are, respectively, pullbacks of $\tau_{MN}$ and $\tau_M{}^A$ to the worldsheet manifold.
In the $c \rightarrow \infty$ limit, we find that \eqref{eq:hsfs} becomes
\be \label{eq:ngfs}
    S = - \frac{T}{2} \int d^2 \sigma \Bigl( \sqrt{-\tau} \, \tau^{\alpha\beta} \, H_{\alpha\beta} + \epsilon^{\alpha\beta} \, B_{\alpha\beta} \Bigr)\,,
\ee
where $\tau = \det \tau_{\alpha\beta} < 0$ and $\tau^{\alpha\beta}$ is the inverse of $\tau_{\alpha\beta}$\,. Since $A = 0 ,1$\,, there is an induced codimension-two foliation structure in spacetime with $\tau^{}_M{}^A$ the vielbein field in the two-dimensional longitudinal sector and $H^{}_{MN}$ encoding the geometry on the leaves. This is the Nambu-Goto formalism of nonrelativistic string theory \cite{Andringa:2012uz}.

The nonrelativistic string limit of the relativistic string Nambu-Goto action \eqref{eq:hsfs} is a special case of the class of $p$-brane limits of a relativistic $p$-brane considered in \cite{Gomis:2000bd, Brugues:2004an, Gomis:2004pw, Gomis:2005bj, Kamimura:2005rz, Brugues:2006yd, Roychowdhury:2019qmp, Pereniguez:2019eoq, Kluson:2020rij}. Now, more generally, consider the Nambu-Goto action on a $(p+1)$-dimensional worldvolume describing a relativistic $p$-brane coupled to a $(p+1)$-form gauge field $\hat{A}^{(p+1)}$\,, 
\be \label{eq:pba}
    \hat{S}_{p\text{-brane}} = - \int d^{p+1} \sigma \sqrt{-\det \hat{G}_{\alpha\beta}} - \int \hat{A}^{(p+1)}\,,
\ee
where we have chosen the convention such that both the $p$-brane tension and the $(p+1)$-form charge are unity. The worldvolume coordinates are $\sigma^\alpha$, $\alpha = 0, \cdots, p$\,. Moreover, 
\be
    \hat{G}_{\alpha\beta} = \p_\alpha X^\CI \, \p_\beta X^\CJ \, \hat{G}_{\CI\CJ}\,,
        \qquad%
    \hat{A}_{\alpha_0 \cdots \alpha_{p}} = \p_{\alpha_0} X^{\CI_0} \, \cdots \, \p_{\alpha_p} X^{\CI_p} \, \hat{A}_{\CI_0 \cdots \CI_p}
\ee
are pullbacks of the spacetime fields from the target space to the $(p+1)$-dimensional worldvolume, with the worldvolume fields $X^\CI$, $\CI = 0, \cdots, d$ playing the role of the spacetime coordinates. Explicitly, the CS term in \eqref{eq:pba} is
\be
    \int \hat{A}^{(p+1)} = \frac{1}{(p+1)!} \int d^{p+1} \sigma \, \epsilon^{\alpha_0 \cdots \alpha_p} \, \hat{A}^{}_{\alpha_0 \cdots \alpha_p} = \int d^{p+1} \sigma \, \hat{A}^{}_{01\cdots p}\,.
\ee
Analogous to \eqref{eq:slcexp}, we now consider the following ansatz:
\be \label{eq:pbexp}
    \hat{G}^{}_{\alpha\beta} = c^2 \, \gamma^{}_{\alpha\beta} + c^{1-p} \, H^{}_{\alpha\beta}\,,
        \qquad%
    \hat{A}^{(p+1)}_{0 1 \cdots p} = - c^{p+1} \, \gamma^{}_{0}{}^{u_0} \cdots \gamma^{}_{p}{}^{u_p} \, \epsilon^{}_{u_0 \cdots u_p} + A^{(p+1)}_{0 \cdots p}\,,
\ee
where $\gamma^{}_{\alpha\beta} = \gamma^{}_\alpha{}^u \, \gamma^{}_\beta{}^v \, \eta^{}_{uv}$\,, and $u = 0, \cdots, p$\,. Plugging \eqref{eq:pbexp} into \eqref{eq:pba}, and taking the $c \rightarrow \infty$ limit leads to the non-singular action
\be \label{eq:pbaction}
    S_{p\text{-brane}} = - \frac{1}{2} \int d^{p+1} \sigma \, \sqrt{-\gamma} \, \gamma^{\alpha\beta} \, H_{\alpha\beta} - \int A^{(p+1)}\,,
\ee
where $\gamma = \det \gamma_{\alpha\beta}$ and $\gamma^{\alpha\beta}$ is the inverse of $\gamma_{\alpha\beta}$\,. These $p$-brane limits involve a cancellation of divergences between the Nambu-Goto and the CS actions. The stringy limit of relativistic fundamental strings we considered earlier is a special case of the $p$-brane limit when $p=1$\,. This is in contrast to the nonrelativistic string limit of D$p$-branes discussed in \S\ref{sec:nrsl}, where the DBI and CS term are non-singular separately.  

In \S\ref{sec:s-duality}, we will see that a membrane limit of M-theory, which is identified with the $p$-brane limit in eleven dimensions with $p=2$, arises when the S-dual of Type IIA nonrelativistic superstring theory is considered. This leads to the notion of nonrelativistic M-theory, which is related to the DLCQ of M-theory. Recall that  nonrelativistic string theory, consisting of fundamental strings together with other extended objects such as D$p$-branes, arises as a stringy limit of relativistic string theory. In contrast, nonrelativistic M-theory, consisting of M2-branes together with M5-branes as their magnetic duals, arises as a membrane limit of relativistic M-theory. 


\section{S-Duals of \gc D-Brane Actions}
\label{sec:s-duality}

In this section, we construct dual \g D-brane actions by performing a duality transformation of the $U(1)$ gauge field on the D-brane.~\footnote{See, \emph{e.g.}, \cite{Aganagic:1997zk} for similar analysis for D-brane actions in relativistic string theory.} The dual of the $U(1)$ gauge field is a $(p-2)$-form gauge field. The dual \g D1- and D3-branes give rise to nonrelativistic fundamental strings and self-dual \g D3-branes, respectively, as expected from the nonrelativistic string limit of the SL($2,\mathbb{Z}$) duality of Type IIB superstring theory.
Moreover, dualizing the $U(1)$ guage potential on \g D2- and D4-branes give rise to nonrelativistic analogs of M2- and M5-branes in the strongly coupled regime of Type IIA nonrelativistic superstring theory. This leads us to the notion of nonrelativistic M-theory that arises as a nonrelativistic membrane limit of relativistic M-theory.

\subsection{\gc D1-Brane}

The first example that we start with is the S-dual of a \g D1-brane. The effective action is obtained by setting $p=1$ in \eqref{eq:gdbbi}
\be \label{eq:gd1b}
    S_\text{D1} = - \int d^2 Y \, e^{-\Phi} \sqrt{-\CM} 
    + \int \lr C^{(2)} + C^{(0)} \, \CF \rr.
\ee
%
Here, $\CM = \det \CM_{\mu\nu}$ and
\be \label{eq:mff}
    \CM_{\mu\nu} =
        \begin{pmatrix}
            0 &\quad \tau_\nu \\[2pt]
            \overline{\tau}_\mu &\quad H_{\mu\nu} + \CF_{\mu\nu}
        \end{pmatrix},
        \qquad%
    \CF = B + F\,, 
        \qquad%
    F = d A\,.
\ee
We will show that the S-dual of the \g D1-brane action gives rise to a bound state of nonrelativistic fundamental strings and \g D-strings.

\subsubsection{Nonrelativistic fundamental strings}

To perform an S-duality transformation, we treat $F$ as an independent field and introduce the generating function
\be \label{eq:gf}
    S_\text{gen.} = \frac{1}{2} \int d^2 Y \, \tilde{\Theta}^{\mu\nu} \lr F_{\mu\nu} - 2 \, \p_\mu A_\nu \rr,
\ee
where $\tilde{\Theta}^{\mu\nu}$ is an antisymmetric field playing the role of a Lagrange multiplier. Integrating out $\tilde{\Theta}$ in $S_\text{gen.}$ leads to the constraint $F = dA$\,, and thus gives back the original D1-brane action \eqref{eq:gd1b}. To find the S-dual theory, we instead integrate out $A_\mu$\,, which leads to the constraint
$\p_\mu \tilde{\Theta}^{\mu\nu} = 0$\,.
Locally, this constraint is solved by
\be
    \tilde{\Theta}^{\mu\nu} = \epsilon^{\mu\nu} \, p\,,
\ee
where $p$ is constant.
%
%
After integrating out $A_\mu$\,, the D1-brane action $S_\text{D1} + S_\text{gen}$ now takes the following equivalent form:
\begin{align} \label{eq:sparent}
    S_\text{parent} = - \int d^2 Y \, e^{-\Phi} \sqrt{-\CM}            
    + \int T \, \bigl( \CF - A^{(2)} \bigr)\,,
\end{align}
where
\be \label{eq:ta2def}
    T = p + C^{(0)}\,,
        \qquad%
    A^{(2)} = \frac{p \, B - C^{(2)}}{T}\,.
\ee
Instead of integrating out in the path integral the non-dynamical field $F_{\mu\nu}$\,, which is now treated as an independent field, we equivalently integrate out $\CF_{\mu\nu}$\,. Varying $S_\text{D1}$ with respect to $\CF_{\mu\nu}$ yields the equation of motion,
\be \label{eq:tth}
    T = e^{-\Phi} \, \sqrt{\frac{\tau}{\CM}}\,,
        \qquad%
    \tau = \det \tau_{\mu\nu}\,,
        \qquad%
    \tau_{\mu\nu} = \tau_\mu{}^A \, \tau_\nu{}^B \, \eta_{AB}\,,
\ee
which indicates that $T > 0$\,. This equation is solved by 
\be \label{eq:fsoln3}
    \CF_{\mu\nu} = \frac{\epsilon_{\mu\nu}}{2} \ls \frac{1}{\sqrt{-\tau}} \det \! \begin{pmatrix}
            0 &\quad \tau_\sigma \\[2pt]
            \overline{\tau}_\rho &\quad H_{\rho\sigma}
        \end{pmatrix} + \frac{e^{-2 \, \Phi}\sqrt{-\tau}}{T^2} \rs.
\ee
Plugging the solution \eqref{eq:fsoln3} into \eqref{eq:sparent}, and with a constant axion field $C^{(0)}$,~\footnote{The S-duality still holds when $C^{(0)}$ is an arbitrary function. Here, we focus on the constant $C^{(0)}$ case for the clarity of this exposition.} we find that the resulting S-dual action is
\be \label{eq:tsfd1}
    S_\text{dual} = - \frac{T}{2} \int d^2 Y \lr \sqrt{-\tau} \, \tau^{\mu\nu} \tilde{H}_{\mu\nu} +  \epsilon^{\mu\nu} A^{(2)}_{\mu\nu} \rr,
\ee
where $\tau^{\mu\nu}$ is the inverse of $\tau_{\mu\nu}$ and 
\be
    T = p + C^{(0)} > 0\,,
        \qquad%
    \tilde{H}_{\mu\nu} = H_{\mu\nu} + \frac{\tau_{\mu\nu}}{2\, T^2}\, e^{-2\Phi},
        \qquad%
    A^{(2)} = \frac{p \, B - C^{(2)}}{T}\,.
\ee
This dual action takes the form of the Nambu-Goto formalism \eqref{eq:ngfs} describing fundamental strings propagating in string Newton-Cartan geometry. 

\subsubsection{Nonrelativistic string limit of (\emph{p}\hspace{0.5mm},\hspace{0.5mm}\emph{q})-string}

To fully appreciate the S-dual action \eqref{eq:tsfd1}, it is instructive to investigate how the above S-duality relation between  \g D1-branes and fundamental strings arise as a nonrelativistic string limit in relativistic string theory.
We first review the S-duality transformation of a D1-brane in relativistic string theory. Starting with the D1-brane action in Type IIB relativistic superstring theory,
\be \label{eq:reld1b}
    \hat{S}_\text{D1} = - \int d^2 Y \, e^{-\hat{\Phi}} \sqrt{-\det \lr \hat{G}_{\mu\nu} + \hat{\CF}_{\mu\nu} \rr} + \int \lr \hat{C}^{(2)} + \hat{C}^{(0)} \hat{\CF} \rr.  
\ee
Taking the nonrelativistic string limit of \eqref{eq:reld1b} as discussed in \S\ref{sec:nrsl}, by first plugging in the ansatz \eqref{eq:cexp} and then setting $c \rightarrow \infty$\,, we recover the \g D1-brane action \eqref{eq:gd1b}. Instead, we now add the generating function \eqref{eq:gf} to \eqref{eq:reld1b} and perform an S-duality transformation. This is done by first integrating out the gauge potential $A_\mu$ and then the field strength $F_{\mu\nu}$\,, which is treated as an independent field. The dual action is \cite{Tseytlin:1996it, Aganagic:1997zk}
\be \label{eq:hsddpq}
    \hat{S}_\text{dual} = - \hat{T} \int d^2 Y \lr \sqrt{-\det \hat{G}_{\mu\nu}} + \frac{1}{2} \, \epsilon^{\mu\nu} \hat{A}^{(2)}_{\mu\nu} \rr,
\ee
where
\be \label{eq:ttk}
    \hat{T} = \bigl| p + \hat{g} \bigr| = \sqrt{\bigl( p + \hat{C}^{(0)} \bigr)^2 + e^{-2\hat{\Phi}}}\,,
        \qquad%
    \hat{A}^{(2)} = \frac{p \, \hat{B} - \hat{C}^{(2)}}{\hat{T}}\,.
\ee
Here, we defined the relativistic axio-dilaton field,
\be \label{eq:hg}
    \hat{g} = \hat{C}^{(0)} + i \, e^{-\hat{\Phi}},
\ee
and also assumed that $\hat{\Phi}$ is constant. The associated Type IIB supergravity is invariant under the SL$(2,\,\mathbb{R})$ transformation of $\hat{g}$\,. In the full string theory, only the discrete subgroup SL$(2,\,\mathbb{Z})$ is preserved. The effective tension $\hat{T}$ in \eqref{eq:ttk} takes the form of the $(p, 1)$-string tension. In general, a $(p,q)$-string is a bound state of $p$ fundamental strings and $q$ D-strings, \emph{i.e.}, $q$ D1-branes. Such a bound state carries both the Kalb-Ramond and two-form RR charges. The generalized Dirac quantization condition requires that both the charges are quantized. Measured in the inverse charge carried by the five-brane that arises as a magnetic dual of strings, both $p$ and $q$ must be integers \cite{Schwarz:1995dk}.~\footnote{Moreover, $p$ and $q$ are coprimes such that the bound state cannot be decomposed into a multiple string configuration, with the number off strings given by the common divisor.} In our case, \eqref{eq:hsddpq} describes a $(p, 1)$-string in the dilaton and axion background fields, where $p = \Theta$ is required to be an integer. Here, normalized by the effective tension, $\hat{B}$ is the Kalb-Ramond field coupled to $p$ fundamental strings and $\hat{C}^{(2)}$ is the RR-potential coupled to the single D-string in the bound state. Equivalently, the dual action \eqref{eq:hsddpq} also receives an interpretation as the fundamental $(1,0)$-string with an SL$(2,\mathbb{Z})$ transformed background \cite{Aganagic:1997zk}. 

The parametrizations in \eqref{eq:cexp} imply that the ingredients in the dual relativistic $(p,1)$-string action \eqref{eq:hsddpq} now assume the following expressions in terms of $c$\,:
\be \label{eq:p1sexp}
    \hat{T} = T + O(c^{-2})\,, 
        \quad%
    \hat{G}_{\mu\nu} = c^2 \, \tau_{\mu\nu} + H_{\mu\nu}\,,
\ee
and
\be
    \hat{A}^{(2)} = - c^2 \, \ell + A^{(2)} + \frac{e^{-2\Phi}}{2 \, T^2} \, \ell + O(c^{-2})\,,
\ee
where $T$ and $A^{(2)}$ are defined in \eqref{eq:ta2def}. Also recall that $\ell$ is defined in components in \eqref{eq:defell}, with $\ell_{\mu\nu} = \tau_\mu{}^A \, \tau_\nu{}^B \, \epsilon_{AB}$\,. Plugging \eqref{eq:p1sexp} into \eqref{eq:hsddpq}, and then taking the $c \rightarrow \infty$ limit, we find that the resulting action is precisely \eqref{eq:tsfd1}. Note that this limit is reminiscent of the stringy limit discussed in \S\ref{sec:pbl}, which leads to the fundamental string action \eqref{eq:ngfs} in nonrelativistic string theory. Analogous to the $(p, 1)$-string action \eqref{eq:hsddpq} in relativistic string theory, the S-dual action \eqref{eq:tsfd1} describes a \g $(p, 1)$-string state in nonrelativistic string theory, with $p$ the number of fundamental nonrelativistic strings in the bound state. 

\subsection{\gc D2-Brane}

We now move on to \g D2-brane, whose effective action is obtained by setting $p=2$ in \eqref{eq:gdbbi}, \emph{i.e.},
\be \label{eq:gd2b}
    S_\text{D2} = - \int d^3 Y \, e^{-\Phi} \sqrt{-\CM} 
    + \int \Bigl( C^{(3)} + C^{(1)} \wedge \CF \Bigr)\,,
\ee
where
\be \label{eq:mff2}
    \CM_{\mu\nu} =
        \begin{pmatrix}
            0 &\quad \tau_\nu \\[2pt]
            \overline{\tau}_\mu &\quad H_{\mu\nu} + \CF_{\mu\nu}
        \end{pmatrix},
        \qquad%
    \CF = B + F\,, 
        \qquad%
    F = d A\,.
\ee
We have introduced the three-form and one-form RR-potentials. Explicitly, in components, the CS term can be written as
\be
    S_\text{CS} = \frac{1}{3!} \int d^3 Y \, \epsilon^{\mu\nu\rho} \Bigl( C^{(3)}_{\mu\nu\rho} + 3 \, C^{(1)}_\rho \, \CF^{}_{\mu\nu} \Bigr)\,.
\ee
As in relativistic string theory, by dualizing the $U(1)$ gauge potential $A_\mu$ in \eqref{eq:gd2b}, we are probing the strongly coupled regime of IIA nonrelativistic superstring theory, which corresponds to nonrelativistic M-theory.

\subsubsection{Dual nonrelativistic membrane} \label{sec:nmfgd2b}

To dualize the gauge potential $A_\mu$\,, we first treat $F_{\mu\nu}$ as an independent field and introduce the generating function in the same way as in \eqref{eq:gf}, but now with
\be \label{eq:gf2}
    S_\text{gen.} = \frac{1}{2} \int d^3 Y \, \tilde{\Theta}^{\mu\nu} \lr F_{\mu\nu} - 2 \, \p_\mu A_\nu \rr,
\ee
where $\tilde{\Theta}^{\mu\nu}$ is an antisymmetric field that imposes the constraint $F = dA$\,. Integrating out $\tilde{\Theta}$ in $S_\text{gen.}$ gives back the original D2-brane action \eqref{eq:gd2b}. To find the S-dual theory, we instead integrate out $A_\mu$\,, which leads to the constraint
$\p_\mu \tilde{\Theta}^{\mu\nu} = 0$\,.
Locally, on the three-dimensional worldvolume, this constraint is solved by
\be
    \tilde{\Theta}^{\mu\nu} = \epsilon^{\mu\nu\rho} \, \p_\rho \Theta\,.
\ee
The dual field $\Theta$ will play the role of the extra eleventh dimension in nonrelativistic M-theory. This extra dimension becomes decompactified in the strongly coupled regime and thus visible in the S-dual theory. Now, the ``parent" action $S_\text{D2} + S_\text{gen.}$ becomes
\be \label{eq:spd2}
    S_\text{parent} = - \int d^3 Y \, e^{-\Phi} \sqrt{-\CM} + \int \Bigl( C \wedge \CF - A^{(3)} \Bigr)\,, 
\ee
where we defined 
\be
    A^{(3)} = - C^{(3)} + B \wedge d\Theta\,,
        \qquad%
    C = C^{(1)} + d\Theta\,.
\ee
We also set the dilaton to zero in the following calculation. The dilaton can easily be  recovered at the end of this calculation by performing rescalings of various background fields,~\footnote{We chose the above rescalings such that it is easier to facilitate the later comparison with the dimensional reduction of M-theory. However, a more practical way to recover the $\Phi$ dependence is by only rescaling the longitudinal vielbein, with $\tau_\mu{}^A \rightarrow e^{-\Phi} \, \tau_\mu{}^A$.}
\begin{subequations} \label{eq:rsbf}
\begin{align}
    \tau_\mu{}^A & \rightarrow e^{-\Phi/3} \, \tau_\mu{}^A\,,
        &%
    \CF_{\mu\nu} & \rightarrow e^{-2\Phi/3} \, \CF_{\mu\nu}\,, \\[2pt]
    H_{\mu\nu} & \rightarrow e^{-2\Phi/3} \, H_{\mu\nu}\,,
        &%
    C & \rightarrow e^{2\Phi/3} \, C\,.
\end{align}
\end{subequations}

To facilitate the duality transformation, we introduce an auxiliary field $u$ and rewrite \eqref{eq:spd2} as
\begin{align} \label{eq:pacd22}
    S_\text{parent} = - \frac{1}{2} \int d^3 Y \, e^{-\Phi} \lr \frac{\CM}{u} - u \rr + \int \Bigl( C \wedge \CF - A^{(3)} \Bigr)\,,
        \qquad%
    u < 0\,.
\end{align}
Varying the parent action \eqref{eq:pacd22} with respect to $\CF$\,, we find that the resulting equations of motion constrain $u$ and two components of $\CF_{\mu\nu}$\,. Plugging the solutions for $u$ and the two constrained components of $\CF_{\mu\nu}$ back into \eqref{eq:pacd22}, 
we find that the dual action is
\be \label{eq:sduald2m2}
    S_\text{dual} = - \frac{1}{2} \int d^3 Y \sqrt{-\gamma} \, \gamma^{\mu\nu} \, H_{\mu\nu} - \int A^{(3)}\,,
        \qquad%
    \gamma_{\mu\nu} = \tau_{\mu\nu} + C_\mu \, C_\nu\,,
\ee
where $\gamma = \det \gamma_{\mu\nu}$ and $\gamma^{\mu\nu}$ is the inverse of $\gamma_{\mu\nu}$\,.
Performing the rescalings \eqref{eq:rsbf} in the dual action \eqref{eq:sduald2m2}, we find the complete dual action in an arbitrary dilaton background, 
\be \label{eq:sdualm2}
    S_\text{dual} = - \frac{1}{2} \int d^3 Y \sqrt{-\gamma} \, \gamma^{\mu\nu} \, \tilde{H}^{}_{\mu\nu} - \int A^{(3)}\,,
\ee
where
\be \label{eq:gm2def}
    \gamma^{}_{\mu\nu} = \gamma^{}_\mu{}^u \, \gamma^{}_\nu{}^v \, \eta^{}_{uv}\,,
        \qquad%
    \gamma^{}_\mu{}^v = \p^{}_\mu f^\CI \, \gamma^{\phantom{(}}_\CI{}^v,
        \qquad%
    \gamma^{}_\CI{}^v = e^{-\Phi/3}
        \begin{pmatrix}
            \tau^{}_M{}^A &\,\, 0 \\[4pt] 
            e^{\Phi} \, C_M^{(1)} &\,\, e^\Phi
        \end{pmatrix}\,,
\ee
with $\CI = 0, \cdots, 10$\,, $f^{10} = \Theta$\,, and $u = 0, 1, 10$\,. Moreover,
\begin{subequations} \label{eq:tha3def}
\begin{align}
    \tilde{H}^{}_{\mu\nu} & = \p_\mu f^\CI \, \p_\nu f^\CJ \, \tilde{H}^{}_{\CI\CJ}\,,
        &%
    \tilde{H}_{\CI\CJ} & = 
        e^{-2\Phi/3}
        \begin{pmatrix}
            H^{}_{MN} &\,\, 0 \\[4 pt]
            0 &\,\, 0
        \end{pmatrix}\,, \\[4pt]
    A^{(3)}_{\mu\nu\rho} & = \p_\mu f^\CI \, \p_\nu f^\CJ \, \p_\rho f^\CK \, A^{(3)}_{\CI\CJ\CK}\,,
        &%
    A^{(3)}_{MNL} & = - C^{(3)}_{MNL}\,, 
        \qquad%
    A^{(3)}_{MN10} = B^{}_{MN}\,.
\end{align}
\end{subequations}
%
The dual action \eqref{eq:sdualm2} defines the Nambu-Goto formalism of nonrelativistic M2-branes propagating in eleven-dimensional spacetime, with the dual field $\Theta$ playing the role of the eleventh dimension. The dual action \eqref{eq:sdualm2} coincides with \eqref{eq:pbaction} when $p=2$\,, and therefore arises as a nonrelativistic membrane limit of relativistic M2-branes. See \S\ref{sec:nmlm2b} for further details.

Recall that the ten-dimensional string Newton-Cartan geometry -- the appropriate spacetime geometry coupled to nonrelativistic superstrings -- is equipped with a codimension-two foliation structure. Here, we have a two-dimensional longitudinal sector described by the vielbein field $\tau^{}_M{}^A$\,, with $M$ the ten-dimensional curved index and $A$ the two-dimensional flat index. In contrast, the nonrelativistic M2-brane described by \eqref{eq:sdualm2}, which arises as an S-dual of the \g D2-brane \eqref{eq:gd2b}, is coupled to an eleven-dimensional spacetime geometry equipped with a codimension-three foliation structure. Now, there is a three-dimensional longitudinal sector described by the vielbein field $\gamma^{}_\CI{}^u$, with $\CI$ the eleven-dimensional curved index and $u$ the three-dimensional flat index. The quantity $\tilde{H}_{\CI\CJ}$ encodes the geometry of the eight-dimensional leaves. We refer to such a geometry with a codimension-three foliation structure as the \emph{membrane Newton-Cartan geometry}. The function $f^\CI$ describes how the M2-brane is embedded in the eleven-dimensional membrane Newton-Cartan geometry.

\subsubsection{Dimensional reductions} \label{sec:drm2}

Now that we have derived the nonrelativistic M2-brane action \eqref{eq:sdualm2}, it is interesting to consider the dimensional reductions of this action to theories in ten dimensions: (i) the double dimensional reduction leads to the nonrelativistic fundamental string action \eqref{eq:ngfs}, which has been studied in \cite{Kluson:2019uza}, and (ii) the direct dimensional reduction of the M2-brane action \eqref{eq:sdualm2} gives the \g D2-brane action \eqref{eq:gd2b}. In Appendix \ref{app:dpbpbncg}, a transverse spatial reduction of M2-brane is considered, leading to a different type of nonrelativistic D2-branes that are coupled to a ten-dimensional membrane Newton-Cartan geometry \cite{Kluson:2019uza}.~\footnote{Such a geometry is referred to as a D2 Newton-Cartan geometry in \cite{Blair:2021ycc}.} 
Also see \cite{Blair:2021ycc} for similar dimensional reductions of eleven-dimensional supergravity.~\footnote{The parametrizations of relativistic background fields in terms of $c$ in \cite{Blair:2021ycc} are different from the ones given in this paper. See, \emph{e.g.}, \eqref{eq:cqexpf} and \eqref{eq:cqdpexp} in this paper and (4.10) and (4.23) in \cite{Blair:2021ycc} for comparisons.}

We first consider the double dimensional reduction, where the dimension of the brane and ambient spacetime are reduced by one simultaneously. This procedure will lead us to the fundamental string action. In practice, we require that $\Theta = Y^2$ and all the background fields be independent of $Y^2$. We also compactify $\Theta$ over a circle of radius $R_{10}$\,. Then, the quantities in \eqref{eq:gm2def} and \eqref{eq:tha3def} become
\begin{subequations} \label{eq:ddrd}
\begin{align}
    \gamma_{\mu\nu} & = e^{-2\Phi/3}
    \begin{pmatrix}
        \tau_{\alpha\beta} + e^{2\Phi} \, C_\alpha^{(1)} \, C_\beta^{(1)} &\,\, e^{2\Phi} \, C_\beta^{(1)} \\[4pt]
        e^{2\Phi} \, C_\alpha^{(1)} &\,\, e^{2\Phi}
    \end{pmatrix}\,, \\[6pt]
    \tilde{H}_{\mu\nu} & = e^{-2\Phi/3}
    \begin{pmatrix}
        H_{\alpha\beta} &\,\, 0 \\[2pt]
        0 &\,\, 0
    \end{pmatrix}\,,
        \qquad%
    A^{(3)}_{\alpha\beta\,2} = B_{\alpha\beta}\,, 
\end{align}
\end{subequations}
where $\alpha = 0,1$ denotes the worldsheet index after the double dimensional reduction. Plugging \eqref{eq:ddrd} into the M2-brane action \eqref{eq:sdualm2}, we find
\be \label{eq:ddracfs}
    S_\text{d.d.r.} = - \pi R_{10} \int d^2 Y \lr \sqrt{-\tau} \, \tau^{\alpha\beta} \, H_{\alpha\beta} + \frac{1}{2} \, \epsilon^{\alpha\beta} \, B_{\alpha\beta} \rr,
\ee
which is the Nambu-Goto formalism \eqref{eq:ngfs} that describes nonrelativistic strings propagating in ten-dimensional string Newton-Cartan geometry and $B$-field background. 

Next, we consider a direct dimensional reduction of the nonrelativistic M2-brane action \eqref{eq:sdualm2} by requiring that the M2-brane be localized in $\Theta$\,. We continue to compactify the eleventh-dimension $\Theta$\,, which we take to be an isometry direction, over a circle of radius $R_{10}$\,. The abelian isometry is given by $\delta_\epsilon \Theta = \epsilon$\,. The shape of the M2-brane can vary in the $\Theta$-direction, and this fluctuation is captured by a Nambu-Goldstone mode. To take into account this excitation, instead of directly setting $\p_\mu \Theta = 0$\,, we need to gauge the isometry by introducing an auxiliary gauge field $v_\mu$ that transforms as $\delta_\epsilon v_\mu = - \p_\mu \epsilon$\,. The gauged version of \eqref{eq:sdualm2} is
\be \label{eq:sgm2}
    S_\text{gauged} = - \frac{1}{2} \int d^3 Y \sqrt{-{\gamma'}} \, {\gamma'}{}^{\mu\nu} \, \tilde{H}_{\mu\nu} - \int \lr {A'}{}^{(3)} + v \wedge F \rr,
\ee
where
\begin{subequations}
\begin{align}
    \gamma'_{\mu\nu} & = e^{-2\Phi/3} \Bigl[ \tau^{}_{\mu\nu} + e^{2\Phi} \lr C^{(1)}_\mu + D^{}_\mu \Theta \rr \lr C^{(1)}_\nu + D^{}_\nu \Theta \rr \Bigr]\,, \\[4pt]
    A'{}^{(3)} & = - C^{(3)} + B \wedge D \Theta\,,
        \qquad%
    D_\mu \Theta = \p_\mu \Theta + v_\mu\,.
\end{align}
\end{subequations}
Moreover, locally, $F = dA$ is an exact two-form. The one-form field $A$ will gain the interpretation as a gauge potential on the D2-brane after the dimensional reduction. The boundary term $\int v \wedge F$ is required such that, upon integrating out $A$\,, the auxiliary field $v$ is pure gauge. Instead, integrating out $v$ in the path integral will give rise to the direct dimension reduction of the M2-brane action. This procedure is essentially the inverse of the duality transformation on the \g D2-brane action that we detailed in \S\ref{sec:nmfgd2b}. 
In terms of the one-form $V = C^{(1)} + D \Theta$\,, \eqref{eq:sgm2} can be rewritten as
\be \label{eq:sg2}
    S_\text{gauged} = - \frac{1}{2} \int d^3 Y \sqrt{-\gamma'} \, {\gamma'}{}^{\mu\nu} \, H_{\mu\nu} + \int \lr {C}{}^{(3)} + C^{(1)} \wedge \CF - V \wedge \CF \rr,
\ee
where 
\be
    \gamma'_{\mu\nu} = \tau_{\mu\nu} + V_\mu \, V_\nu\,.
\ee
We have set $\Phi = 0$\,; the dependence on $\Phi$ can easily be recovered by rescaling various background fields as in \eqref{eq:rsbf} at the end of the calculation. Moreover, we neglect a global contribution $\int F \wedge d\Theta$\,.
We dualize $\Theta$ by integrating out the auxiliary field $V_\mu$\,. For simplicity, we perform the duality transfomation in the special case with $\tau_{\mu\nu} = \text{diag} (\tau^{}_{00}\,, \tau^{}_{11}\,, \tau^{}_{22})$ and $H_{\mu\nu} = \text{diag}(H_{00}\,, H_{11}\,, H_{22})$\,, and covariantize the action at the end to recover the complete dual theory. Varying \eqref{eq:sg2} with respect to $V_\mu$ gives
\begin{subequations}
\begin{align}
    \CF_{01} & = \frac{1}{2 \, \tau_{00} \, \tau_{11}} \left( \frac{\tau_{00}^2 \, \tau_{11}^2 - V_0^2 \, \tau_{11}^2 + V_1^2 \, \tau_{00}^2}{V_2^2} \, H_{22} - \tau_{00}^2 \, H_{11} + \tau_{11}^2 \, H_{00} \right), \\[2pt]
    \CF_{02} & = \frac{V_1 \, \tau_{00}}{V_2 \, \tau_{11}} \, H_{22}\,,
        \qquad%
    \CF_{12} = \frac{V_0 \, \tau_{11}}{V_2 \, \tau_{00}} \, H_{22}\,,
\end{align}
\end{subequations}
which are solved by
\be \label{eq:vsol3}
    V_0 = \frac{\tau_{00}^2}{\sqrt{-\CM}} \, \CF_{12} \,,
        \qquad%
    V_1 = \frac{\tau_{11}^2}{\sqrt{-\CM}} \, \CF_{02}\,,
        \qquad%
    V_2 = \frac{\tau_{00} \, \tau_{11} \, H_{22}}{\sqrt{-\CM}}\,,
\ee
with $\CM = \det \CM_{\mu\nu}$ and
\be \label{eq:flatm}
    \CM_{\mu\nu} =
    \begin{pmatrix}
        0 &\,\, \tau_0{}^0 &\,\, \tau_1{}^1 &\,\, 0 \\[2pt]
        \tau_0{}^0 &\,\, H_{00} &\,\, \CF_{01} &\,\, \CF_{02} \\[2pt]
        - \tau_1{}^1 &\,\, -\CF_{01} &\,\, H_{11} &\,\, \CF_{12} \\[2pt]
        0 &\,\, -\CF_{02} &\,\, -\CF_{12} &\,\, H_{22} \\[2pt]
    \end{pmatrix},
\ee 
Plugging \eqref{eq:vsol3} into \eqref{eq:sg2}, and covariantizing \eqref{eq:flatm} to be 
\be 
    \CM_{\mu\nu} =
    \begin{pmatrix}
        0 &\,\,\,\, \tau_\nu \\[2pt]
        \bar{\tau}_\mu &\,\,\,\, H_{\mu\nu} + \CF_{\mu\nu}
    \end{pmatrix},
\ee 
we find that the dualized action matches the \g D2-brane action \eqref{eq:gd2b}.

%


%
        %
%
%
%

\subsubsection{Nonrelativistic membrane limit of M2-brane} \label{sec:nmlm2b}

Finally, we discuss how the different theories discussed in this section that are related by dualizing the worldvolume $U(1)$ gauge field and dimensional reductions arise as distinct limits of relativistic string theory.  

We first recapitulate that the \g D2-brane action \eqref{eq:gd2b} arises as the nonrelativistic string limit of the relativistic D2-brane action,
\be \label{eq:relsd2}
    \hat{S}_\text{D2} = - \int d^3 Y \, e^{-\hat{\Phi}} \sqrt{-\det \lr \hat{G}_{\mu\nu} + \hat{\CF}_{\mu\nu} \rr} + \int \lr \hat{C}^{(3)} + \hat{C}^{(1)} \wedge \hat{\CF} \rr.  
\ee
Plugging in the ansatz \eqref{eq:cexp} and then taking the $c \rightarrow \infty$ limit, we recover the \g D2-brane action \eqref{eq:gd2b}. Instead, we now add the generating function \eqref{eq:gf2} to \eqref{eq:relsd2}, and perform a duality transformation by integrating out $A_\mu$ and $F_{\mu\nu}$\,. This leads to the dual M2-brane action \cite{Aganagic:1997zk},
\be \label{eq:sdualm2rel}
    \hat{S}_\text{dual} = - \int d^3 Y \sqrt{- \det \mathbb{G}_{\mu\nu}} - \int \mathbb{A}^{(3)}\,, 
\ee
where 
\begin{subequations} \label{eq:mbga}
\begin{align}
    \mathbb{G}_{\mu\nu} & = e^{-2\hat{\Phi}/3} \left[ \hat{G}_{\mu\nu} + e^{2\hat{\Phi}} \bigl( \hat{C}_\mu^{(1)} + \p_\mu \Theta \bigr) \bigl( \hat{C}_\nu^{(1)} + \p_\nu \Theta \bigr) \right], \\[4pt]
    \mathbb{A}^{(3)}_{\mu\nu\rho} & = - \hat{C}^{(3)}_{\mu\nu\rho} + \p_{\mu} \Theta \, \hat{B}_{\nu\rho} + \p_{\nu} \Theta \, \hat{B}_{\rho\mu} + \p_{\rho} \Theta \, \hat{B}_{\mu\nu}\,.
\end{align}
\end{subequations}
The dual field $\Theta$ plays the role of the eleventh dimension in M-theory.
Plugging \eqref{eq:cexp} into \eqref{eq:mbga}, we find (see, \emph{e.g.}, \cite{Blair:2021ycc})
%
        %
%
%
\begin{align} \label{eq:mbgaexp}
    \mathbb{G}_{\mu\nu} = c^{4/3} \, \gamma_{\mu\nu} + c^{-2/3} \, \tilde{H}_{\mu\nu}\,, 
        \qquad%
    \mathbb{A}^{(3)}_{\mu\nu\rho} = - c^2 \, \gamma^{}_{\mu}{}^u \, \gamma^{}_\nu{}^v \, \gamma^{}_{\rho}{}^w \, \epsilon^{}_{uvw} + A^{(3)}_{\mu\nu\rho}\,,
\end{align}
%
where $\gamma_{\mu}{}^u$\,, $\tilde{H}_{\mu\nu}$\,, and $A^{(3)}$ are defined in \eqref{eq:gm2def} and \eqref{eq:tha3def}.
After redefining $c \rightarrow c^{3/2}$\,, the parametrizations in \eqref{eq:mbgaexp} coincide with the ones in \eqref{eq:pbexp} with $p=2$\,. Taking the limit $c \rightarrow \infty$ of \eqref{eq:sdualm2rel} reproduces the nonrelativistic M2-brane action \eqref{eq:sdualm2}.

We now consider the double dimensional reduction of the relativistic M2-brane action \eqref{eq:sdualm2rel}, which requires that $\Theta = Y^2$ and all the background fields be independent of $Y^2$. Then, \eqref{eq:mbga} becomes
\be \label{eq:gaddr}
    \mathbb{G}^{}_{\mu\nu} = 
        e^{-2\hat{\Phi}/3}
        \begin{pmatrix}
            \hat{G}_{\alpha\beta} + e^{2\hat{\Phi}} \, \hat{C}^{(1)}_\alpha \, \hat{C}^{(1)}_\beta &\,\, e^{2\hat{\Phi}} \, \hat{C}_\beta^{(1)} \\[2pt]
            e^{2\hat{\Phi}} \, \hat{C}_\alpha^{(1)} &\,\, e^{2\hat{\Phi}}
        \end{pmatrix},
        \quad%
    \mathbb{A}_{\alpha\beta\gamma}^{(3)} = - \hat{C}^{(3)}_{\alpha\beta\gamma}\,,
        \quad%
    \mathbb{A}^{(3)}_{\alpha\beta\,2} = \hat{B}^{}_{\alpha\beta}\,. 
\ee
Plugging the ansatz \eqref{eq:gaddr} for double dimensional reduction into the M2-brane action \eqref{eq:sdualm2rel}, we find that the reduced action is
\be \label{eq:hsddrfs}
    \hat{S}_\text{d.d.r.} = - 2 \pi R_{10} \int d^2 Y \lr \sqrt{-\det \hat{G}_{\alpha\beta}} + \frac{1}{2} \, \epsilon^{\alpha\beta} \, \hat{B}_{\alpha\beta} \rr.
\ee
Here, $R_{10}$ is the radius of the circle along which the eleventh dimension is compactified.
The parametrizations of $\hat{G}_{MN}$ and $\hat{B}_{MN}$ are given in \eqref{eq:cexp}, \emph{i.e.},
\be
    \hat{G}_{\alpha\beta} = c^2 \, \tau_{\alpha\beta} + H_{\alpha\beta}\,,
        \qquad%
    \hat{B}_{\alpha\beta} = - c^2 \, \tau_\alpha{}^A \, \tau_\beta{}^B \, \epsilon_{AB} + B_{\alpha\beta}\,.
\ee
The $c \rightarrow \infty$ limit of \eqref{eq:hsddrfs} is therefore the nonrelativistic string limit of the fundamental relativistic string as discussed in \S\ref{sec:pbl}. The resulting action is precisely \eqref{eq:ddracfs} that describes fundamental nonrelativistic strings in string Newton-Cartan geometry. 

The direct dimensional reduction of the relativistic M2-brane action is done by gauging the isometry $\Theta$ direction and performing a duality transformation. This is essentially the inverse of the duality transformation of the relativistic D2-brane action \eqref{eq:relsd2}. As shown in \cite{Tseytlin:1996it, Aganagic:1997zk}, dualizing $\Theta$ gives back the relativistic D2-brane action, whose nonrelativistic string limit gives rise to the nonrelativistic D2-brane action \eqref{eq:gd2b}, as we have discussed in \S\ref{sec:nrsl}.

\subsection{\gc D3-Brane}

The \g D3-brane is described by the action
\be \label{eq:gd3b}
    S_{\text{D}3} = - \! \int d^4 Y \, e^{-\Phi} \sqrt{-\CM} + \int \lr C^{(4)} +  C^{(2)} \wedge \CF + \frac{1}{2} \, C^{(0)} \, \CF \wedge \CF \rr,
\ee
where $\CF = B + F$\,, and
\be 
    \CM_{\mu\nu} =
    \begin{pmatrix}
        0 &\,\,\,\, \tau_\nu \\[2pt]
        \bar{\tau}_\mu &\,\,\,\, H_{\mu\nu} + \CF_{\mu\nu}
    \end{pmatrix}.
\ee 
More explicitly,
\be \label{eq:gd3bex}
    S_{\text{D}3} = - \! \int d^4 Y \left\{ e^{-\Phi} \sqrt{-\CM} - \frac{1}{4!} \, \epsilon^{\mu\nu\rho\sigma} \Bigl( C^{(4)}_{\mu\nu\rho\sigma} + 6 \, C^{(2)}_{\mu\nu} \, \CF_{\rho\sigma} + 3 \, C^{(0)} \, \CF_{\mu\nu} \, \CF_{\rho\sigma} \Bigr) \right\}.
\ee
%
During the following calculation, it is convenient to set $\Phi = 0$ and recover the dependence on $\Phi$ at the end of the calculation by taking the rescaling 
\be \label{eq:restau}
    \tau_\mu{}^A \rightarrow e^{-\Phi} \, \tau_\mu{}^A.
\ee
The S-duality transformation of the \g D3-brane action proceeds differently depending on whether $C^{(0)}$ equals zero. We will start with analyzing the S-duality transformation in the more general case when $C^{(0)} \neq 0$ and then discuss the zero $C^{(0)}$ limit next.

\subsubsection{Self-duality transformation}

To perform an S-duality transformation in the presence of a nonzero $C^{(0)}$, we treat $F$ as an independent field and introduce the generating function,
\be \label{eq:gf3}
    S_\text{gen.} = \frac{1}{2} \int d^4 Y \, \tilde{\Theta}^{\mu\nu} \lr F_{\mu\nu} - 2 \, \p_\mu A_\nu \rr,
\ee
where $\tilde{\Theta}^{\mu\nu}$ is antisymmetric. Integrating out $\tilde{\Theta}$ in $S_\text{gen.}$ leads to the constraint $F = dA$\,, and thus gives back the original D3-brane action \eqref{eq:gd3b}. To find the S-dual theory, we instead integrate out $A_\mu$\,, which leads to the constraint
$\p_\mu \tilde{\Theta}^{\mu\nu} = 0$\,.
Locally, on the four-dimensional worldvolume, this constraint is solved by
\be
    \tilde{\Theta}^{\mu\nu} = \frac{1}{2} \, \epsilon^{\mu\nu\rho\sigma} \, \tilde{F}_{\rho\sigma}\,,
        \qquad%
    \tilde{F} = d \tilde{A}\,. 
\ee
We then write the ``parent'' action $S_{\text{D3}}+S_{\text{gen.}}$ equivalently as
\be \label{eq:sd3u}
    S_\text{parent} = - \frac{1}{2} \int d^4 Y \lr \frac{\CM}{u} - u \rr + \int \lr \Gamma^{(4)} + \frac{1}{2} \, C^{(0)} \, \CV \wedge \CV \rr,
\ee
where $u<0$ and
\be
    \CM_{\mu\nu} =
    \begin{pmatrix}
        0 &\,\,\,\, \tau_\nu \\[2pt]
        \bar{\tau}_\mu &\,\,\,\, H_{\mu\nu} - \Gamma_{\mu\nu} + \CV_{\mu\nu}
    \end{pmatrix},
\ee
and
\begin{subequations} \label{eq:g4vfg2}
\begin{align}
    \Gamma^{(4)} & = C^{(4)} - \tilde{F} \wedge B - \frac{1}{2} \, C^{(0)} \, \Gamma \wedge \Gamma\,, \\[2pt]
        \qquad%
    \CV & = \CF + \Gamma\,,
        \qquad%
    \Gamma = \frac{\tilde{F} + C^{(2)}}{C^{(0)}}\,.
\end{align}
\end{subequations}
We already set $\Phi = 0$\,, bearing in mind that the dependence on $\Phi$ will be recovered at the end of the calculation. We make the special choice $\tau_\mu{}^A = \delta_\mu^A$ and $H_{\mu\nu} = \text{diag} \bigl( 0, 0, 1, 1 \bigr)$ to facilitate the calculation, and will later covariantize the dual action with respect to the string Newton-Cartan geometry.~\footnote{Note that we also suppressed various Nambu-Goldstone modes that perturb the shape of the D-brane, but they will be recovered once we covariantize the resulting dual action.} This choice of background fields leads to a D-brane extending in the longitudinal spatial direction, and therefore the NCOS sector. However, after the covariantization at the end, the NROS sector with a Dirichlet boundary condition in the longitudinal spatial direction will also be captured.
%
%
Varying the \g D3-brane action \eqref{eq:sd3u} with respect to $\CV_\mu$\,, we find the following equations of motion:
\begin{subequations}
\begin{align}
    u \, C^{(0)} \, \CV_{01} & = \epsilon^{a'b'} \, \CF_{0a'} \, \CF_{1b'} - 2 \, \CF_{ab} \, \CF_{23}\,, 
        \qquad%
    u \, C^{(0)} \, \CV_{23} = - \lr 1 + \CF_{23}^2 \rr, \\[6pt]
    u \, C^{(0)} \, \CV_{aa'} & = \epsilon_{a}{}^b \, \epsilon_{a'}{}^{b'} \CF_{bb'} - \CF_{aa'} \, \CF_{23}\,.
\end{align}
\end{subequations}
We split $\mu = (a, a')$\,, with $a=0,1$ and $a'=2,3$\,. Recall that, from \eqref{eq:g4vfg2}, we have $\CF = \CV - \Gamma$\,. 
The Levi-Civita symbols $\epsilon_{ab}$ and $\epsilon_{a'b'}$ are defined by $\epsilon_{01} = - \epsilon_{10} = 1$ and $\epsilon_{23} = - \epsilon_{32} = 1$\,, respectively. These equations are solved by
\begin{subequations} \label{eq:vaapsol}
\begin{align}
    \CV_{01} & = \frac{\CF_{02} \, \CF_{13} - \CF_{03} \, \CF_{12} + 2 \, \Gamma_{01} \, \CF_{23}}{u \, C^{(0)} + 2 \, \CF_{23}}\,, 
        \qquad%
    \CV_{23} = - \frac{1}{2} \lr u \, C^{(0)} - 2 \, \Gamma_{23} + \CP \rr, \\[2pt]
    \CV_{aa'} & = \frac{\epsilon_{a}{}^b \, \epsilon_{a'}{}^{b'} \, \Gamma_{bb'} + \epsilon^{b'c'} \, \Gamma_{ab'} \, \Gamma_{a'c'}}{1 + \Gamma_{23}^2} \, \CV_{23}\,,
\end{align}
\end{subequations}
where
\be
    \CP = \sqrt{\lr u \, C^{(0)} - 2 \, \Gamma_{23} \rr^2 -4 \lr 1 + \Gamma_{23}^2 \rr}\,.
\ee
%
Plugging \eqref{eq:vaapsol} into the \g D3-brane action \eqref{eq:sd3u} yields
\begin{align} \label{eq:sd3u2}
    S_\text{parent} & = \frac{1}{4} \int d^4Y \lr 2 \, u - \frac{u \, C^{(0)} - 2 \, \Gamma_{23} + \CP(u)}{1 + \Gamma_{23}^2} \, C^{(0)} \, \CM' \rr + \int \Gamma^{(4)},
\end{align}
where $\CM' \equiv \det \CM'_{\mu\nu}$\,, with 
\be
    \CM'_{\mu\nu} =
        \begin{pmatrix}
            0 &\,\,\,\, 1 &\,\,\,\, 1&\,\,\,\, 0 \\[2pt]
            1 &\,\,\,\, 0 &\,\,\,\, - \Gamma_{01} &\,\,\,\, - \Gamma_{0b'} \\[2pt]
            -1 &\,\,\,\, \Gamma_{01} &\,\,\,\, 0 & \,\,\,\, - \Gamma_{1b'} \\[2pt]
            0 &\,\,\,\, \Gamma_{0a'} &\,\,\,\, \Gamma_{1b'} &\,\,\,\, \mathbb{1}_{a'b'} - \Gamma_{a'b'}
        \end{pmatrix}\,.
\ee
Varying \eqref{eq:sd3u2} with respect to $u$ gives rise to a quadratic equation, which is solved by
\be \label{eq:solud3}
    u = - \frac{1 + \ls \Gamma_{23} - C^{(0)} \bigl( -\tilde{\CM} \,\bigr)^{1/2} \rs^2}{\lr C^{(0)} \rr^2 \bigl( -\tilde{\CM} \,\bigr)^{1/2}} < 0\,,
\ee
where $\tilde{\CM} \equiv \det \tilde{\CM}_{\mu\nu}$\,, with
%
        %
        %
%
\be
    \tilde{\CM}_{\mu\nu} =
        \begin{pmatrix}
            0 &\,\,\,\, 1 &\,\,\,\, 1&\,\,\,\, 0 \\[2pt]
            1 &\,\,\,\, 0 &\,\,\,\, - \Gamma_{01} + \frac{1}{2} \lr C^{(0)} \rr^{-2} &\,\,\,\, - \Gamma_{0b'} \\[2pt]
            -1 &\,\,\,\, \Gamma_{01} - \frac{1}{2} \lr C^{(0)} \rr^{-2} &\,\,\,\, 0 & \,\,\,\, - \Gamma_{1b'} \\[2pt]
            0 &\,\,\,\, \Gamma_{0a'} &\,\,\,\, \Gamma_{1b'} &\,\,\,\, \mathbb{1}_{a'b'} - \Gamma_{a'b'}
        \end{pmatrix}\,.
\ee
Plugging \eqref{eq:solud3} back into \eqref{eq:sd3u2} gives the S-dual action,
\be
    S_\text{dual} = - \int d^4 Y \lr \sqrt{-\tilde{\CM}} - \frac{\Gamma_{23}}{C^{(0)}} \rr + \int \Gamma^{(4)}\,. 
\ee
Covariantize this action using the string Newton-Cartan data $\tau_\mu{}^A$ and $H_{\mu\nu}$\,, and then take into account \eqref{eq:restau} to recover the dependence on the dilaton, we find the S-dual action,
\be
    S_\text{dual} = - \int d^4 Y \, e^{-\Phi} \sqrt{-\tilde{\CM}} + \int \lr \Gamma^{(4)} + 2 \, C^{(0)} \, \Gamma \wedge L \rr,
\ee
where $\tilde{\CM} \equiv \det \tilde{\CM}_{\mu\nu}$ and
\begin{align} \label{eq:defL}
    \tilde{\CM}_{\mu\nu} = 
    \begin{pmatrix}
        0 &\,\,\,\, \tau_\nu \\[4pt]
        \bar{\tau}_\mu &\,\,\,\, H_{\mu\nu} - \Gamma_{\mu\nu} + L_{\mu\nu}
    \end{pmatrix},  
        \qquad%
    L_{\mu\nu} = \frac{\ell_{\mu\nu}}{2 \bigl( e^{\Phi} \, C^{(0)} \bigr)^{\!2}}\,.
\end{align}
This dual action can be brought into the form of a \g D3-brane, 
\begin{align} \label{eq:sdualgd3}
    S_\text{dual} & = \int d^4 Y \, \CL_\text{DBI}
        + \int \Omega^{(4)}\,,
\end{align}
where
$$
    \CL_\text{DBI} = - e^{-\Phi} \sqrt{-\det 
        \begin{pmatrix}
            0 &\,\, \tau_\nu \\[2pt]
            \bar{\tau}_\mu &\,\, H_{\mu\nu} + \tilde{\CF}_{\mu\nu}
        \end{pmatrix}}\,,
        \qquad%
    \Omega^{(4)} = \tilde{C}^{(4)} + \tilde{C}^{(2)} \wedge \tilde{\CF} + \frac{1}{2} \, \tilde{C}^{(0)} \, \tilde{\CF} \wedge \tilde{\CF}\,,
$$
and
\begin{subequations} \label{eq:tcdata}
\begin{align}
    \tilde{C}^{(0)} & = - C^{(0)}\,, 
        &%
    \tilde{\CF} & = - \frac{\tilde{F} + C^{(2)}}{C^{(0)}} + L\,, \label{eq:tcdataa} \\[4pt]
    \tilde{C}^{(2)} & = C^{(0)} \bigl( B - L \bigr)\,, 
        &%
    \tilde{C}^{(4)} & = C^{(4)} + \bigl( C^{(2)} - C^{(0)} \, L \bigr) \wedge B \,.
\end{align}
\end{subequations}
Note that we used the identity $\ell \wedge \ell = 0$\,. 

It is useful to consider a fixed background configuration to gain some physical intuition of the S-dual action \eqref{eq:sdualgd3}. In flat background with $\tau_\mu{}^A = \delta_\mu{}^A$, we are in the regime of NCOS, with the D-brane extending in the spacetime longitudinal directions.
For simplicity, we also set the Kalb-Ramond and all the RR potentials $C^{(2)}$ and $C^{(4)}$ to zero and consider a constant dilaton field $\Phi = \Phi_0$\,. Moreover, we take $C^{(0)}$ to be constant. Then, the Yang-Mills (YM) couplings associated with the quadratic actions of $F$ and its S-dual $\tilde{F}$ are, respectively,
\be \label{eq:dymcgd3}
    g^{}_\text{YM} = e^{\Phi_0/2}\,,
        \qquad%
    \tilde{g}^{}_\text{YM} = e^{\Phi_0/2} \, C^{(0)}\,.
\ee
The usual electric-magnetic duality (see later in \eqref{eq:emd} for the relativistic case) corresponds to the choice $C^{(0)} = 0$\,. However, the dual coupling in \eqref{eq:dymcgd3} vanishes in the zero axion limit $C^{(0)} \rightarrow 0$\,. Before concluding that this axion limit is singular, we first note that there is a loophole in \eqref{eq:dymcgd3} when we write the dual Yang-Mills coupling $\tilde{g}_\text{YM}$\,: In \eqref{eq:tcdataa}, there is an additional term $L$ in $\tilde{\CF}$ that we have defined in \eqref{eq:defL}, which acts as part of the Kalb-Ramond field. In the flat spacetime limit, we have 
\be
    L_{\mu\nu} = 
    \frac{1}{2 \bigl( e^\Phi C^{(0)} \bigr)^2}
    \begin{pmatrix}
        \epsilon_{AB} & \,\,0  \\[2pt]
        0 & \,\,0
    \end{pmatrix}.
\ee
Using the Seiberg-Witten map \eqref{eq:nstsw}, we find that the open string and noncommutative Yang-Mills (NCYM) coupling are, respectively,
\be \label{eq:ncymc}
    \tilde{G}_\text{o} = \frac{1}{C^{(0)}}\,,
        \qquad
    \tilde{g}^{}_\text{NCYM} = \tilde{g}^{1/2}_s \, C^{(0)} = \bigl( C^{(0)} \bigr)^{1/2}\,.
\ee
Unfortunately, 
both the open string and NCYM coupling are still singular in the zero axion limit, even after taking into account the Seiberg-Witten map.
Nevertheless, as we will momentarily, the zero axion limit is in fact more subtle and leads to a well-defined theory with an effective gauge coupling \eqref{eq:gncym} that depends on the transverse component of $C^{(2)}$\,.   

\subsubsection{A zero axion limit} 

We now investigate the limit where the axion field $C^{(0)}$ is set to zero in the dual action \eqref{eq:sdualgd3}. This turns out to be a highly non-trivial limit to take in practice. To derive the resulting action, we define $\omega \equiv 1 / C^{(0)}$ and expand \eqref{eq:sdualgd3} with respect to a large $\omega$\,. The DBI and CS parts of the action respectively take the following expansions with respect to a large $\omega$\,:
\begin{align}
    \CL_\text{DBI} & = \frac{\omega^2}{2 \, e^{2\Phi}} \, \tr \bigl( \tilde{\ell} \, \CC \bigr) - \frac{\omega}{4} \, \tr \bigl( \tilde{\CC} \, \CC \bigr)
    + \frac{e^{-2\Phi} \, G + \tr \bigl( \tilde{\CC} \, \tau \, \tilde{\CC} \, H \bigr) - \frac{1}{16} \, e^{2\Phi} \bigl[ \tr \bigl( \tilde{\CC} \, \CC \bigr) \bigr]^{\!2}}{\tr \bigl( \tilde{\ell} \, \CC \bigr)} + O(\omega^{-1})\,, \notag \\[2pt]
    \Omega^{(4)} & = \frac{\omega^2}{e^{2\Phi}} \, \CC \wedge \ell - \frac{\omega}{2} \, \CC \wedge \CC + C^{(4)} - \tilde{F} \wedge B \,,
\end{align}
where we chose $\tr \bigl( \tilde{\ell} \, \CC \bigr) < 0$ and defined 
\be \label{eq:tlctc}
    \tilde{\ell}^{\mu\nu} = \frac{1}{2} \, \epsilon^{\mu\nu\rho\sigma} \, \ell_{\rho\sigma}\,,
        \qquad%
    \CC = \tilde{F} + C^{(2)}\,,
        \qquad%
    \tilde{\CC}^{\mu\nu} = \frac{1}{2} \, \epsilon^{\mu\nu\rho\sigma} \, \CC_{\rho\sigma}\,,
\ee
and, with $H^{\mu\nu}$ being the inverse of $H_{\mu\nu}$\,,
\be
    G = \det \bigl( H_{\mu\nu} \bigr) \, \det \bigl(\tau_\rho{}^A \, H^{\rho\sigma} \, \tau_\sigma{}^B \bigr)\,.
\ee
Note that $G$ is non-singular even if $H_{\mu\nu}$ is degenerate \cite{Bergshoeff:2019pij}.
Also note that, 
\be
    \tr \bigl( \tilde{\ell} \, \CC \bigr) = \tilde{\ell}^{\mu\nu} \, \CC_{\nu\mu}\,,
        \qquad%
    \tr \bigl( \tilde{\CC} \, \CC \bigr) = \tilde{C}^{\mu\nu} \, \CC_{\nu\mu}\,,
        \qquad%
    \tr \bigl( \tilde{\CC} \, \tau \, \tilde{\CC} \, H \bigr) = \tilde{\CC}^{\mu\rho} \, \tau_{\rho\sigma} \, \tilde{\CC}^{\sigma\nu} \, H_{\nu\mu}\,.
\ee
In the limit $\omega \rightarrow \infty$\,, the dual action \eqref{eq:sdualgd3} gives rise to a finite action,
\begin{align} \label{eq:sdzc0}
    S'_\text{dual} = \int d^4 Y \, \frac{e^{-2\Phi} \, G + \tr \bigl( \tilde{\CC} \, \tau \, \tilde{\CC} \, H \bigr) - \frac{1}{16} \, e^{2\Phi} \bigl[ \tr \bigl( \tilde{\CC} \, \CC \bigr) \bigr]^{\!2}}{\tr \bigl( \tilde{\ell} \, \CC \bigr)} + \int \bigl( C^{(4)} - \tilde{F} \wedge B \bigr)\,.
\end{align}
This is the electric-magnetic dual of the \g D3-brane action \eqref{eq:gd3b} with $C^{(0)} = 0$\,. Since we expanded with respect to a small $C^{(0)}$ to obtain \eqref{eq:sdzc0}, and there is a ratio $\tilde{F} / C^{(0)}$ appearing in the action \eqref{eq:sdualgd3} that we started with, the $C^{(0)} \rightarrow 0$ limit and truncating the action at the quadratic order in $\tilde{F}$ do not necessarily commute.~\footnote{In contrast, the $c\rightarrow\infty$ and $C^{(0)}\rightarrow0$ limits commute, because there is no truncation involved.} For this reason, the $C^{(0)} \rightarrow 0$ limit of the dual couplings in \eqref{eq:dymcgd3} and \eqref{eq:ncymc} is not trustworthy.
In \S\ref{sec:nsldds}, we will discuss the physical meaning of this action and define the effective couplings for \eqref{eq:sdzc0}, at least for a particular class of D-brane configurations. 

The same dual action \eqref{eq:sdzc0} can be reproduced by performing an S-duality transformation on the D3-brane action \eqref{eq:gd3b} with $C^{(0)} = 0$\,. Now, the nonrelativistic D3-brane action is
\be 
    S_{\text{D}3} = - \! \int d^4 Y \, e^{-\Phi} \sqrt{-\CM} + \int \lr C^{(4)} +  C^{(2)} \wedge \CF \rr,
        \qquad%
    \CM_{\mu\nu} =
    \begin{pmatrix}
        0 &\,\,\,\, \tau_\nu \\[2pt]
        \bar{\tau}_\mu &\,\,\,\, H_{\mu\nu} + \CF_{\mu\nu}
    \end{pmatrix}.
\ee
Introducing the generating function \eqref{eq:gf3} and then integrating out the gauge potential $A_\mu$ gives the ``parent" action
\be  \label{eq:pad3zc0}
    S_\text{parent} = - \frac{1}{2} \int d^4 Y \, e^{-\Phi} \lr \frac{\CM}{u} - u \rr + \int \lr C^{(4)} - \tilde{F} \wedge B + \CC \wedge \CF \rr, 
\ee
with $u < 0$\,.
In the choice of background fields with $\tau_\mu{}^A = \delta_\mu^A$ and $H_{\mu\nu} = \text{diag} \bigl( 0, 0, 1, 1 \bigr)$\,, integrating out the non-dynamical field $\CF$ in \eqref{eq:pad3zc0} leads to the equivalent action,
\be \label{eq:spzc0}
    S_\text{parent} \! = \! \int d^4 Y \lr \frac{u}{2 \, e^\Phi} - \frac{\CC_{Ai} \, \CC^A{}_i + \frac{1}{2} \, \tilde{\CC}^{\mu\nu} \, \CC_{\mu\nu} \, \sqrt{-1 - u \, e^{\Phi} \, \CC_{23}}}{2 \, \CC_{23}} \rr + \int \lr C^{(4)} - \tilde{F} \wedge B \rr. 
\ee
The presence of $\sqrt{-1 - u \, e^\Phi \, \CC_{23}}$ implicitly requires that $\CC_{23} > - e^{-\Phi} / u > 0$\,, which is consistent with our earlier choice $\tr \bigl( \tilde{\ell} \, \CC \bigr) = - \CC_{23} < 0$\,. The equation of motion from varying the nondynamical field $u$ in \eqref{eq:spzc0} is
\be \label{eq:cstz}
    \CC_{03} \, \CC_{12} - \CC_{02} \, \CC_{13} + \CC_{01} \, \CC_{23} = - e^{-2\Phi} \, \sqrt{- 1 - u \, \CC_{23}} < 0\,.
\ee
Finally, integrating out $u$ by plugging \eqref{eq:cstz} into \eqref{eq:spzc0} gives the dual action, which matches \eqref{eq:sdzc0} after covariantizing with respect to the background string Newton-Cartan geometry. 

\subsubsection{Nonrelativistic string limit of dual D3-brane} \label{sec:nsldds}

We now examine how the self-dual of the \g D3-brane action arises from the nonrelativistic string limit. Starting with the relativistic D3-brane action
\be \label{eq:reld3}
    \hat{S}_{\text{D}3} = - \! \int \! d^4 Y \, e^{-\hat{\Phi}} \sqrt{-\det\Bigl( \hat{G}_{\mu\nu} + \hat{\CF}_{\mu\nu} \Bigr)} \, + \! \int \! \lr \hat{C}^{(4)} \! +  \hat{C}^{(2)} \! \wedge \hat{\CF} + \! \frac{1}{2} \, \hat{C}^{(0)} \, \hat{\CF} \wedge \hat{\CF} \rr\!,
\ee
as we already learned from \S\ref{sec:nrsl}, under the parametrization \eqref{eq:cexp}, the $c\rightarrow \infty$ limit of \eqref{eq:reld3} leads to the \g D3-brane worldvolume action \eqref{eq:gd3b}. The S-dual of the relativistic D3-brane action \eqref{eq:reld3} is \cite{Tseytlin:1996it, Aganagic:1997zk}
\begin{align} \label{eq:hsdd3}
    \hat{S}_\text{dual} & = - \! \int \! d^4 Y \, e^{-\hat{\Phi}} \sqrt{-\det \biggl( \hat{G}_{\mu\nu} - \frac{\hat{\CC}_{\mu\nu}}{|\hat{g}|} \biggr)} 
    + \! \int \biggl( \hat{C}^{(4)} \! - \tilde{F} \wedge \hat{B} - \frac{\hat{C}^{(0)}}{2 \, |\hat{g}|^2} \, \hat{\CC} \wedge \hat{\CC} \biggr)\,,
\end{align}
where $\hat{g} = \hat{C}^{(0)} + i \, e^{-\hat{\Phi}}$ as in \eqref{eq:hg} and $\hat{\CC} = \tilde{F} + \hat{C}^{(2)}$\,, with $\tilde{F}$ the S-dual field strengh of $F$\,. In flat spacetime with zero Kalb-Ramond and RR background fields, for the quadratic terms in $F$ and $\tilde{F}$\,, the YM coupling under S-duality transforms as 
\be \label{eq:hgym}
    \hat{g}^{}_\text{YM} = e^{\hat{\Phi}_0/2}
        \,\,\longrightarrow\,\,%
    e^{\hat{\Phi}_0/2} \, |\hat{g}| = e^{\hat{\Phi}_0/2} \, \sqrt{\bigl( \hat{C}^{(0)} \bigr)^2 + e^{-2\hat{\Phi}_0}}\,.
\ee
We have taken the dilaton field to be constant here.
In the zero axion limit, we have the usual electric-magnetic duality with
\be \label{eq:emd}
    \hat{g}^{}_\text{YM} = e^{\hat{\Phi}_0/2}
        \,\,\longrightarrow\,\,%
    e^{-\hat{\Phi}_0/2}\,.
\ee

We now consider the nonrelativistic string limit of the D3-brane action \eqref{eq:hsdd3}, which can be rewritten as
\be \label{eq:hsdd3a}
    \hat{S}_\text{dual} = - \! \int \! d^4 Y \, e^{-\hat{\Phi}} \sqrt{-\det \bigl( \hat{G}_{\mu\nu} + \hat{\CK}_{\mu\nu} \bigr)} + \! \int \biggl( \hat{A}^{(4)} + \! \hat{A}^{(2)} \wedge \hat{\CK} + \frac{1}{2} \, \hat{A}^{(0)} \, \hat{\CK} \wedge \hat{\CK} \biggr),
\ee
where
\begin{align} \label{eq:ka4a2}
    \hat{\CK} = - \frac{\hat{\CC}}{|\hat{g}|}\,, 
        \qquad%
    \hat{A}^{(0)} = \tilde{C}^{(0)}\,,
        \qquad%
    \hat{A}^{(2)} = |\hat{g}| \, \hat{B}\,,
        \qquad%
    \hat{A}^{(4)} = \hat{C}^{(4)} + \hat{C}^{(2)} \wedge \hat{B}\,.
\end{align}
Plugging in the ansatz from \eqref{eq:cexp} and expanding with respect to large $c$\,, we find, in terms of the prescriptions given in \eqref{eq:tcdata},
\begin{subequations}
\begin{align}
    \hat{G}_{\mu\nu} & = c^2 \, \tau_{\mu\nu} + H_{\mu\nu}\,,     &%
    \hat{A}^{(2)} & = c^2 \, \tilde{C}^{(0)} \, \ell + \tilde{C}^{(2)} + \frac{\CN}{c^2} + O(c^{-4})\,, \\[2pt]
    \hat{\CK} &= -c^2 \, \ell + \tilde{\CF} + \frac{\CK}{c^2} + O(c^{-4})\,, 
        &%
    \hat{A}^{(4)} & = c^2 \, \tilde{C}^{(2)} \wedge \ell + \tilde{C}^{(4)} +  \tilde{C}^{(2)} \wedge L\,.
\end{align}
\end{subequations}
and $\hat{A}^{(0)} = \tilde{C}^{(0)}$\,. Here,
\be
    \CK = - \frac{2 \, \tilde{\CF} + L}{\bigl( 2 \, e^\Phi \, \tilde{C}^{(0)} \bigr)^2}\,,
        \qquad%
    \CN = \frac{2 \, \tilde{C}^{(2)} - 3 \, \tilde{C}^{(0)} \, L}{\bigl( 2 \, e^\Phi \, \tilde{C}^{(0)} \bigr)^2}\,,
        \qquad%
    L = \frac{\ell}{2 \, \bigl( e^\Phi \, \tilde{C}^{(0)} \bigr)^2}\,.
\ee
%
In the $c \rightarrow \infty$ limit, we find that \eqref{eq:hsdd3a} gives rise to \eqref{eq:sdualgd3}. Moreover, applying the same limit to the Yang-Mills coupling \eqref{eq:hgym}, the expression in \eqref{eq:dymcgd3} is recovered. 

It is also interesting to understand how the dual \g D3-brane action \eqref{eq:sdzc0} with a zero axion arises directly as a nonrelativistic string limit of the dual relativistic D3-brane action \eqref{eq:hsdd3} with $\hat{C}^{(0)} = 0$\,, \emph{i.e.},
\begin{align} \label{eq:hsdzc0}
    \hat{S}_\text{dual} & = \int \! d^4 Y \, \hat{\CL}_\text{DBI}
    + \! \int \Omega^{(4)}\,,
\end{align}
where
\be
    \hat{\CL}_\text{DBI} = - e^{-\hat{\Phi}} \sqrt{-\det \biggl( \hat{G}_{\mu\nu} - e^{\hat{\Phi}} \, \hat{\CC}_{\mu\nu} \biggr)}\,,
        \qquad%
    \hat{\Omega}^{(4)} = \hat{C}^{(4)} \! - \tilde{F} \wedge \hat{B}\,.
\ee
Plug the ansatz \eqref{eq:cexp} with $\hat{C}^{(0)} = 0$ into the dual action \eqref{eq:hsdzc0} and then expand with respect to a large $c$ gives
\begin{subequations} \label{eq:hloexp}
\begin{align}
    \CL_\text{DBI} & = \frac{1}{2} \, c^2 \, \tr \bigl( \tilde{\ell} \, \CC \bigr)
    + \frac{e^{-2\Phi} \, G + \tr \bigl( \tilde{\CC} \, \tau \, \tilde{\CC} \, H \bigr) - \frac{1}{16} \, e^{2\Phi} \bigl[ \tr \bigl( \tilde{\CC} \, \CC \bigr) \bigr]^{\!2}}{\tr \bigl( \tilde{\ell} \, \CC \bigr)} + O(c^{-2})\,, \\[2pt]
    \Omega^{(4)} & = c^2 \, \CC \wedge \ell + C^{(4)} - \tilde{F} \wedge B \,.
\end{align}
\end{subequations}
We have chosen the $\tr \bigl( \tilde{\ell} \, \CC \bigr) < 0$ branch. Plugging \eqref{eq:hloexp} back into \eqref{eq:hsdzc0} and taking the $c \rightarrow \infty$ limit indeed recovers \eqref{eq:sdzc0}. 
However, since the limit $C^{(0)} \rightarrow 0$ does not necessarily commute with truncating at the quadratic order in the dual field strength $\tilde{F}$, the effective gauge coupling cannot be read directly by taking limits of \eqref{eq:hgym} anymore.    

To unravel the physical meaning of the novel dual action \eqref{eq:sdzc0} with a zero axion, we focus on the specific background field configuration with $\tau_\mu{}^{A} = \delta_\mu^A$ and $H_{\mu\nu} = \text{diag} \bigl(0,0,1,1)$\,. We also set $B = C^{(0)} = C^{(4)} = 0$\,. As we have discussed around \eqref{eq:ncosdbifs}, this choice of background fields in the original \g D3-brane action \eqref{eq:gd3b} describes the physics in the NCOS regime, with the D3-brane extending in the longitudinal spatial direction. In \cite{Gopakumar:2000na}, it is shown that the S-dual of NCOS on a D3-brane is spatially-noncommutative $\CN = 4$ supersymmetric Yang-Mills. 
This limit is indeed captured by the nonrelativistic string limit $c \rightarrow \infty$ of the relativistic D3-brane action \eqref{eq:hsdzc0}, which in our choice of background fields is
\begin{align} \label{eq:ncyml}
    \hat{S}_\text{dual} & = - \int d^4 Y \ls e^{-\hat{\Phi}} \sqrt{-\det \Bigl(
    \hat{G}_{\mu\nu} 
        +
    \hat{\mathscr{F}}_{\mu\nu}
    \Bigr)} + c^2 \, \bigl( \tilde{F}_{23} + C^{(2)}_{23} \bigr) \rs,
\end{align}
where $\hat{\mathscr{F}}_{\mu\nu} = - c \, e^{\Phi} \bigl( \tilde{F}_{\mu\nu} + C^{(2)}_{\mu\nu} \bigr)$\,. Since $\tilde{F}_{23} = \p_2 \tilde{A}_3 - \p_3 \tilde{A}_2$ is an exact form, the boundary term $c^2 \, \tilde{F}_{23}$ can be omitted. We also take $\Phi = \Phi_0$ to be a constant and $g_s = e^{\Phi_0}$ as the closed string coupling. We further take $C^{(2)}_{23} = - 2 \pi b / g_s$ with a constant $b$ and all the other components of $C^{(2)}$ to be zero. Then, the term $c^2 \, C^{(2)}_{23}$ in \eqref{eq:ncyml} is a constant, and thus also a boundary term that can be omitted. Finally, define $\hat{\alpha}' = 1/c$\,, 
we rewrite \eqref{eq:ncyml} as
\be \label{eq:sdzaf}
    \hat{S}_\text{dual} = - \frac{1}{\hat{g}_s \, \hat{\alpha}'{}^2} \int d^4 Y \, \sqrt{- \det \Bigl[ \mathscr{G}_{\mu\nu} + 2 \pi \hat{\alpha}' \bigl( \mathscr{B}_{\mu\nu} + \mathscr{F}_{\mu\nu} \bigr) \Bigr]},
\ee
where $\mathscr{F}_{\mu\nu} = - g_s \, \tilde{F}_{\mu\nu} / (2\pi)$ and
\be \label{eq:bgscrgb}
    \mathscr{G}_{\mu\nu} = \begin{pmatrix}
        \hat{\mathscr{G}}_{AB} &\,\, 0 \\[2pt]
        0 &\,\, \hat{\mathscr{G}}_{ij}
    \end{pmatrix},
        \qquad%
    \mathscr{B}_{\mu\nu} = \begin{pmatrix}
        0 &\,\, 0 \\[2pt]
        0 &\,\, \mathscr{B}_{ij}
    \end{pmatrix},
\ee
and
\be
    \mathscr{G}_{AB} = \eta_{AB}\,,
        \qquad%
    \mathscr{G}_{ij} = \hat{\alpha}^{\prime2} \, \delta_{ij}\,,
        \qquad%
    \mathscr{B}_{ij} = - \epsilon_{ij} \, b\,,
        \qquad%
    \hat{g}_s = \hat{\alpha}' \, g_s\,.
\ee
We performed the above redefinitions for the ease of comparison with \cite{Gopakumar:2000na}. Now, the $\hat{\alpha}' \rightarrow 0$ limit reproduces the NCYM limit in \cite{Gopakumar:2000na} up to a rescaling factor of $\hat{\mathscr{G}}_{ij}$\,. 
It is important that we kept a constant $b$ in the action \eqref{eq:sdzaf}: Applying the NCYM limit to the Seiberg-Witten map \eqref{eq:swm}, we find the following $\hat{\alpha}' \rightarrow 0$ limit of the open string background fields,
\begin{subequations} \label{eq:swncos}
\begin{align}
    \hat{\mathcal{G}}^{\mu\nu} = \lr \frac{1}{\mathscr{G} + 2 \pi \hat{\alpha}' \mathscr{B}} \, \mathscr{G} \, \frac{1}{\mathscr{G} - 2\pi\hat{\alpha}' \mathscr{B}} \rr^{\mu\nu} 
        & \longrightarrow \,\,%
    \CG^{\mu\nu} = 
    \begin{pmatrix}
        \eta_{AB} & \,\,0 \\[2pt]
        0 & \,\,\delta_{ij} / (2\pi b)^2
    \end{pmatrix}, \\[2pt]
    \hat{\Theta}^{\mu\nu} = - \bigl(2\pi\hat{\alpha}'\bigr)^2 \lr \frac{1}{\mathscr{G} + 2\pi\hat{\alpha}' \mathscr{B}} \, \mathscr{B} \, \frac{1}{\mathscr{G} - 2\pi\hat{\alpha}' \mathscr{B}} \rr^{\mu\nu} 
        & \longrightarrow \,\,%
    \Theta^{\mu\nu} = 
    \begin{pmatrix}
        0 & \,\,0 \\[2pt]
        0 & \,\,\epsilon_{ij}/b
    \end{pmatrix}, \\[2pt]
    \CG^2_\text{o} = \hat{g}_s \sqrt{\frac{\det \bigl( \mathscr{G} + 2\pi\hat{\alpha}'\mathscr{B} \bigr)}{\det \mathscr{G}}}
        & \longrightarrow \,\,%
    \hat{\CG}^2_\text{o} = 2 \pi g_s \, b\,.
\end{align}
\end{subequations}
Therefore, as expected, the action \eqref{eq:sdzaf} describes NCYM with a spatial noncommutativity $[Y^2\,, Y^3] \propto 1/b$\,. This theory has a well-defined effective NCYM coupling,
\be \label{eq:gncym}
    g^{}_\text{NCYM} = \CG_\text{o} = \sqrt{2\pi g_s \, b}\,,
\ee
as long as $b \neq 0$\,. 
%
%
In this setting, the effective action \eqref{eq:sdzc0} of the worldvolume $U(1)$ gauge potential $\tilde{A}_\mu$ can be expanded around the closed string background field configuration \eqref{eq:bgscrgb} with respect to a small field strength $\tilde{F} = d\tilde{A}$\,. In this way, the term  $\tr \bigl( \tilde{\ell} \, \CC \bigr) = \bigl( 2\pi b / g_s \bigr) - \tilde{F}_{23}$ in the denominator  of \eqref{eq:sdzc0} does not present any singular behavior in the regime where $b$ is nonzero and $|\tilde{F}_{23}| \ll |b|$\,. 

The above analysis extends to the case where the D3-brane is localized in a longitudinal direction by using the T-dual relation between NCOS and NROS (in the DLCQ), which suggests that one has to introduce a nontrivial background geometry in the NROS for the associated effective gauge theory from expanding \eqref{eq:sdzc0} to be well defined \cite{Gomis:2020izd}. Additionally, one can determine how the Seiberg-Witten map 
transforms under T-duality as in \cite{Seiberg:1999vs}.

\subsection{\gc D4-Brane}

So far, we have dualized \g D1-, D2- and D3-brane actions with respect to the worldvolume $U(1)$ gauge field. We also showed that the dual actions match the nonrelativistic string limits of the associated extended objects in relativistic string theory. Now, we apply ansatz \eqref{eq:cexp}, which is well-tested by now, to the relativistic D4-brane action and its dual. We will show that the dual of a \g D4-brane gives rise to a double dimensional reduction of a \g M5-brane, which arises as the membrane limit of a relativistic M5-brane. At the end of this section, we will construct the appropriate membrane limit by generalizing \S\ref{sec:nmlm2b} to include higher form potentials. We will argue that such a membrane limit of the covariant PST formalism of relativistic M5-branes reproduces the desired dual D4-brane after double dimensional reduction. Throughout this section, we will be content with showing that the stringy and membrane limits lead to finite results. 
However, we will not write down any explicit expressions of the D4- and M5-brane actions in nonrelativistic string/M-theory. These actions appear to be rather complicated. It deserves future studies to reveal the detailed structure of a covariant formalism of nonrelativistic M5-branes.  

\subsubsection{Dual D4-brane in relativistic string theory}

We start with the relativistic D4-brane action
\be \label{eq:reld4b}
    \hat{S}_{\text{D}4} = - \int d^5 Y \, e^{-\hat{\Phi}} \sqrt{-\det \bigl( \hat{G}_{\mu\nu} + \hat{\CF}_{\mu\nu} \bigr)} + \int \lr \hat{C}^{(5)} +  \hat{C}^{(3)} \wedge \hat{\CF} + \frac{1}{2} \, \hat{C}^{(1)} \wedge \hat{\CF} \wedge \hat{\CF} \rr.
\ee
The duality transformation is implemented by adding in the following generating function:
\be \label{eq:gf4}
    S_\text{gen.} = \frac{1}{2} \int d^{5} Y \, \tilde{\Theta}^{\mu\nu} \lr F_{\mu\nu} - 2 \, \p_\mu A_\nu \rr.
\ee
Integrating out $A_\mu$ constrains the dual field $\tilde{\Theta}$ to be 
\be
    \tilde{\Theta}^{\mu\nu} = \frac{1}{3!} \, \epsilon^{\mu\nu\rho\sigma\lambda} \, \Theta_{\rho\sigma\lambda}\,,
        \qquad%
    \Theta = d \CA\,, 
\ee
where $\CA$ is a two-form potential. The ``parent" action $\hat{S}_\text{D4} + S_\text{gen.}$ becomes
\be \label{eq:hspd4}
    \hat{S}_\text{parent} = \hat{S}_\text{DBI} + \int \lr \hat{C}^{(5)} - \Theta \wedge \hat{B} + \hat{\CH} \wedge \hat{\CF} + \frac{1}{2} \, \hat{C}^{(1)} \wedge \hat{\CF} \wedge \hat{\CF} \rr,
\ee
where
\be
    \hat{\CH} = \Theta + \hat{C}^{(3)}\,.
\ee
Integrating out $\hat{\CF}$ in \eqref{eq:hspd4} leads to the dual action \cite{Aganagic:1997zk, Tseytlin:1996it}
\begin{align} \label{eq:sdm5}
\begin{split}
    \hat{S}_\text{dual} & = - \int d^5 Y e^{-\hat{\Phi}} \, \sqrt{-\hat{G}} \, \sqrt{1 + y_1 + \tfrac{1}{2} \, y_1^2 - y_2} \\[4pt]
    & \quad - \frac{1}{8} \int d^5 Y \, e^{2\hat{\Phi}} \, \hat{G} \, \tilde{G}{}^{-1} \, \epsilon^\lambda{}_{\mu\nu\rho\sigma} \, \hat{C}^{(1)}_\lambda \, \tilde{\CH}^{\mu\nu} \, \tilde{\CH}^{\rho\sigma}
    + \int \lr \hat{C}^{(5)} - \Theta \wedge \hat{B} \rr,
\end{split}
\end{align}
where
\be
    y_1 = e^{2 \hat{\Phi}} \, \frac{\tr \bigl( \tilde{G} \, \tilde{\CH} \, \tilde{G} \, \tilde{\CH} \bigr)}{2 \, \bigl( - \tilde{G} \bigr)}\,,
        \qquad%
    y_2 = e^{4 \hat{\Phi}} \, \frac{\tr \bigl( \tilde{G} \, \tilde{\CH} \, \tilde{G} \, \tilde{\CH} \, \tilde{G} \, \tilde{\CH} \, \tilde{G} \, \tilde{\CH} \bigr)}{4 \, \bigl( - \tilde{G} \bigr)^{\!2}}\,,
\ee
and
\be
   \tilde{G}_{\mu\nu} = \hat{G}_{\mu\nu} + e^{2\,\hat{\Phi}} \, \hat{C}^{(1)}_\mu \, \hat{C}^{(1)}_\nu\,,
        \qquad%
    \tilde{\CH}^{\mu\nu} = \frac{1}{3!} \, \epsilon^{\mu\nu\rho\sigma\lambda} \, \hat{\CH}_{\rho\sigma\lambda}\,.
\ee
We have defined $\hat{G} = \det \hat{G}_{\mu\nu}$ and $\tilde{G} = \det \tilde{G}_{\mu\nu}$\,, which are related to each other by
\be
    \tilde{G} = \Bigl( 1 + e^{2\hat{\Phi}} \, \hat{G}^{\mu\nu} \, \hat{C}_\mu^{(1)} \, \hat{C}_\nu^{(1)} \Bigr) \, \hat{G}\,.
\ee
The indices are lowered (and raised) by the metric $\hat{G}_{\mu\nu}$ (and its inverse).

The dual action \eqref{eq:sdm5} arises as a double dimensional reduction of the M5-brane action \cite{Aganagic:1997zq}. To understand how this works, we first present the covariant Pasti-Sorokin-Tonin (PST) formalism \cite{Pasti:1997gx, Pasti:1995tn, Pasti:1996vs} of the M5-brane action in eleven-dimensional spacetime with a six-dimensional worldvolume,
\begin{align} \label{eq:pst}
\begin{split}
    \hat{S}_\text{dual} & = - \int d^6 Y \, \sqrt{-\det \lr \mathbb{G}_{\bar{\mu}\bar{\nu}} + i \, \tilde{\mathbb{H}}_{\bar{\mu}\bar{\nu}} \rr} \\[2pt] 
    & \quad + \frac{1}{4} \int d^6 Y \sqrt{-\mathbb{G}} \, \tilde{\mathbb{H}}^{\bar{\mu}\bar{\nu}} \, \mathbb{H}_{\bar{\mu}\bar{\nu}\bar{\rho}} \, n^{\bar{\rho}} + \frac{1}{2} \int \lr \mathbb{A}^{(6)} - \Theta \wedge \mathbb{A}^{(3)} \rr.
\end{split}
\end{align}
Here, we have introduced the six-dimensional index $\bar{\mu} = (\mu\,, 5)$\,, with $\mu = 0, 1, \cdots, 4$\,, together with the metric $\mathbb{G}_{\bar{\mu}\bar{\nu}}$ as well as the three-form potential $\mathbb{A}^{(3)}$\,. The $\bar{\mu}$ index is lowered (and raised) by $\mathbb{G}_{\bar{\mu}\bar{\nu}}$ (and its inverse).
We also defined $\mathbb{G} = \det \mathbb{G}_{\bar{\mu}\bar{\nu}}$ and
\be
     \tilde{\mathbb{H}}^{\bar{\mu}\bar{\nu}} = \frac{1}{3!} \, \frac{\epsilon^{\bar{\mu} \bar{\nu} \bar{\rho} \bar{\sigma} \bar{\lambda} \bar{\kappa}} \, \mathbb{H}_{\bar{\rho}\bar{\sigma}\bar{\lambda}} \, n_{\bar{\kappa}}}{\sqrt{-\mathbb{G}}}\,,
        \qquad%
    \mathbb{H}_{\bar{\mu}\bar{\nu}\bar{\rho}} = \Theta^{\phantom{(}}_{\bar{\mu}\bar{\nu}\bar{\rho}} - \mathbb{A}^{(3)}_{\bar{\mu}\bar{\nu}\bar{\rho}}\,,
        \qquad%
    n_{\bar{\mu}} = \frac{\p_{\bar{\mu}} a}{\sqrt{\p_{\bar{\mu}} a \, \p^{\bar{\mu}} a}}\,.
\ee
The scalar field $a$ is introduced to ensure the covariance of the six-dimensional worldvolume. Note that $a$ is purely auxiliary and imposes the self-dual condition on the three-form field strength $\Theta$\,. 

We now consider the double dimensional reduction of the PST action by wrapping the M5-brane around a compactified spatial circle. This amounts to take the tenth spatial coordinate to be $X^{10} = Y^5$ and fixing the auxiliary field $a = Y^5$. The reduction map include the ones for the metric $\mathbb{G}_{MN}$ and three-form $\mathbb{A}^{(3)}$ as in \eqref{eq:gaddr} in M-theory, which we transcribe below:
\be \label{eq:rmm5}
    \mathbb{G}^{}_{\bar{\mu}\bar{\nu}} = 
        e^{-\frac{2}{3} \hat{\Phi}} \!
        \begin{pmatrix}
            \hat{G}_{\mu\nu} + e^{2\hat{\Phi}} \, \hat{C}^{(1)}_\mu \, \hat{C}^{(1)}_\nu &\,\, e^{2\hat{\Phi}} \, \hat{C}_\nu^{(1)} \\[4pt]
            e^{2\hat{\Phi}} \, \hat{C}_\mu^{(1)} &\,\, e^{2\hat{\Phi}}
        \end{pmatrix},
        \quad\,%
    \mathbb{A}^{(3)}_{\mu\nu\rho} = - C^{(3)}_{\mu\nu\rho}\,,
        \quad\,%
    \mathbb{A}^{(3)}_{\mu\nu5} = \hat{B}^{\phantom{(}}_{\mu\nu}. 
\ee
Moreover,
\be \label{eq:rmm52}
    \hat{\Theta}_{\mu\nu5} = \tilde{\mathbb{H}}^{\mu5} = n_\mu = 0\,, 
        \qquad%
    \tilde{\mathbb{H}}^{\mu\nu} = \frac{1}{3!} \frac{\epsilon^{\mu\nu\rho\sigma\lambda} \, \CH_{\rho\sigma\lambda}}{\sqrt{-\mathbb{G} \, \hat{G}^{55}}}\,,
        \qquad%
    n_5 = \frac{1}{\sqrt{\hat{G}^{55}}}\,.
\ee
Consequently, $\mathbb{H}_{\mu\nu\rho} = \hat{\CH}_{\mu\nu\rho}$\,, $\mathbb{H}_{\mu\nu5} = -\hat{B}_{\mu\nu}$\,, and
\be
    \mathbb{G}^{\bar{\mu}\bar{\nu}} = e^{2\hat{\Phi}/3} 
    \begin{pmatrix}
        \hat{G}^{\mu\nu} &\,\,\,\, - \hat{G}^{\nu\sigma} \, \hat{C}^{(1)}_\sigma \\[4pt]
        - \hat{G}^{\mu\rho} \, \hat{C}^{(1)}_{\rho} &\,\,\,\, e^{-2\hat{\Phi}} + \hat{C}^{(1)}_\rho \, \hat{G}^{\rho\sigma} \, \hat{C}^{(1)}_\sigma
    \end{pmatrix},
        \qquad%
    \mathbb{G} = e^{-2\hat{\Phi}} \, \hat{G}\,.
\ee
Finally, we take the reduction prescription for the six-form potential $\mathbb{A}^{(5)}$ as
\be \label{eq:rmm53}
    \mathbb{A}^{(6)}_{\mu\nu\rho\sigma\lambda5} = 2 \, \hat{C}^{(5)}_{\mu\nu\rho\sigma\lambda} + 10 \, \hat{C}^{(3)}_{[\mu\nu\rho} \, \hat{B}^{\phantom{(}}_{\sigma\lambda]}
        \,\,\longrightarrow\,\,%
    2 \, \hat{C}^{(5)} + \hat{C}^{(3)} \wedge \hat{B}\,.
\ee
Plugging \eqref{eq:rmm5} $\sim$ \eqref{eq:rmm53} back into \eqref{eq:pst} reproduces the dual D4-brane action \eqref{eq:sdm5}. 

\subsubsection{Nonrelativistic membrane limit of M5-brane}

From \S\ref{sec:nrsl}, we already learned that applying the parametrization \eqref{eq:cexp} to the D4-brane action \eqref{eq:reld4b} and taking the nonrelativistic string limit by sending $c\rightarrow \infty$ leads to the \g D4-brane worldvolume action,
\begin{align} \label{eq:gd4b}
\begin{split}
    S_\text{D4} & = - \int d^5 Y \, e^{-\Phi} \sqrt{- \det 
    \begin{pmatrix}
        0 &\,\,\,\, \tau_\nu \\[2pt]
        \bar{\tau}_\mu &\,\,\,\, H_{\mu\nu} + \CF_{\mu\nu}
    \end{pmatrix}} \\[2pt]
    & \quad + \int \lr {C}^{(5)} +  {C}^{(3)} \wedge {\CF} + \frac{1}{2} \, C^{(1)} \wedge {\CF} \wedge {\CF} \rr.
\end{split}
\end{align}
The dual of this action with respect to the worldvolume $U(1)$ gauge potential is given by the same nonrelativistic string limit of \eqref{eq:sdm5}. This limit gives a rather lengthy result, which requires future studies to reveal its detailed structure, with the hope that a compact and understandable form can be acquired. Instead of presenting the detailed expression of the dual \g D4-brane action, which is not very illuminating at this stage, we will focus on showing that this limit is indeed well defined and gives a finite action. 

We start with the first line in \eqref{eq:sdm5}.
In terms of the parametrizations in \eqref{eq:cexp}, we find
\be \label{eq:sqrtterm}
    e^{-\hat{\Phi}} \sqrt{-\hat{G}} = c \, e^{-\Phi} \sqrt{\det \bigl( H_{\mu\nu} \bigr) \, \det \bigl( \tau_\rho{}^A \, H^{\rho\sigma} \, \tau_\sigma{}^B \bigr)} + O(c^{-1})\,.
\ee
Moreover, by power counting, we have
\be
    y_1 = O(c^0)\,,
        \qquad%
    y_2 = O(c^0)\,.
\ee
However, intriguingly, there is a non-trivial cancellation among the terms under the square root in \eqref{eq:sdm5}, such that
\be \label{eq:yex}
    1 + y_1 + \tfrac{1}{2} \, y_1^2 - y_2 = O (c^{-2})\,.
\ee
We explicitly checked this cancellation when $\tau_\mu{}^A = \delta_\mu^A$ and $H_{\mu\nu} = \text{diag} \bigl( 0, 0, 1, 1, 1 \bigr)$\,. Instead of presenting this lengthy calculation, in the following, we demonstrate that this cancellation works in the simple case where $\Theta + C^{(3)} = 0$\,. We then have
\begin{align}
    \tilde{G} & = - c^6 \, e^{2\Phi} \lr C_2^2 + C_3^2 + C_4^2 \rr + O(c^4)\,, 
\end{align}
and
\begin{align}
    \tr \! \ls \bigl( \tilde{G} \cdot \tilde{\CH} \bigr)^{2n} \rs = 2 \, \bigl( -c^4 \bigr)^n \lr C_2^2 + C_3^2 + C_4^2 \rr^n\,.
\end{align}
It then follows that
\be
    y_1  = - 1 + O(c^{-2})\,,
        \qquad%
    y_2 = \tfrac{1}{2} + O(c^{-2})\,.
\ee
As expected in \eqref{eq:yex}, the zeroth order terms in $c$ exactly cancel. Combining \eqref{eq:sqrtterm} and \eqref{eq:yex}, we find
\be \label{eq:dbidivm5}
    - \int d^5 Y e^{-\hat{\Phi}} \, \sqrt{-\hat{G}} \, \sqrt{1 + y_1 + \tfrac{1}{2} \, y_1^2 - y_2} = O(c^0)\,.
\ee
Next, we turn to the second line in \eqref{eq:sdm5}. Applying the parametrizations in \eqref{eq:cexp}, we find
\begin{subequations}
\begin{align}
    \frac{1}{8} \, e^{2\hat{\Phi}} \, \hat{G} \, \tilde{G}{}^{-1} \, \epsilon^\mu{}_{\nu\rho\sigma\lambda} \, \hat{C}^{(1)}_\mu \, \tilde{\CH}^{\nu\rho} \, \tilde{\CH}^{\sigma\lambda} & = \frac{1}{12} \, c^2 \, \epsilon^{\mu\nu\rho\sigma\lambda} \, \ell_{\mu\nu} \, \CH_{\rho\sigma\lambda} + O(c^0) \,, \\[4pt]
    \hat{C}^{(5)} - \Theta \wedge \hat{B} & = c^2 \, \CH \wedge \ell + O(c^0)\,,
\end{align}
\end{subequations}
where $\CH = \Theta + C^{(3)}$\,.
These two divergences in $O(c^2)$ are canceled in \eqref{eq:sdm5}. Together with \eqref{eq:dbidivm5}, we find that the nonrelativistic string limit of the dual action \eqref{eq:sdm5} gives a finite S-dual of the \g D4-brane action \eqref{eq:gd4b}. 

The ansatz in \eqref{eq:cexp} can be lifted to be a membrane limit in eleven dimensions that generalizes the one discussed in \S\ref{sec:nmlm2b}, now applied to the relativistic M5-brane action with a six-dimensional worldvolume. The appropriate parametrizations for the background fields in \eqref{eq:pst} are given by
\begin{subequations} \label{eq:gha36}
\begin{align}
    \mathbb{G}^{}_{\bar{\mu}\bar{\nu}} & = c^{4/3} \, \gamma^{}_{\bar{\mu}\bar{\nu}} + c^{-2/3} \, \tilde{H}^{}_{\bar{\mu}\bar{\nu}}\,, \\[4pt]
    \mathbb{H} & = \tfrac{1}{3!} \, c^2 \, \gamma^u \wedge \gamma^v \wedge \gamma^w \, \epsilon^{}_{uvw} + \CH\,, \\[4pt]
    \mathbb{A}^{(3)} & = - \tfrac{1}{3!} \, c^2 \, \gamma^u \wedge \gamma^v \wedge \gamma^w \, \epsilon^{}_{uvw} + A^{(3)}\,, \\[4pt]
    \mathbb{A}^{(6)} & = - \tfrac{1}{3!} \, c^2 \, A^{(3)} \wedge \gamma^u \wedge \gamma^v \wedge \gamma^w \, \epsilon^{}_{uvw} + A^{(6)}\,.
\end{align}
\end{subequations}
Here, $\gamma_\mu{}^v$, $\tilde{H}_{\mu\nu}$\,, and $A^{(3)}$ are defined in the same way as in \eqref{eq:gm2def} and \eqref{eq:tha3def}. Note that the parametrizations of $\mathbb{G}_{\bar{\mu}\bar{\nu}}$ and $\mathbb{A}^{(3)}$ match the ones in \eqref{eq:mbgaexp} for M2-branes.
We also require that, under the double dimensional reduction, 
\be
    A^{(6)} \rightarrow 2 \, C^{(5)} + C^{(3)} \wedge B\,.
\ee
These parametrizations are constructed such that they reproduce the ones in \eqref{eq:cexp} after plugging in the double dimensional reduction prescriptions in \eqref{eq:rmm5} $\sim$ \eqref{eq:rmm53}. 

\section{T-Duals of \gc D-Brane Actions} \label{sec:T-duality}

In \cite{Bergshoeff:2018yvt}, T-duality transformations in the path integral of the sigma model that describes nonrelativistic string theory have been studied in detail, where nonrelativistic strings are coupled to an arbitrary string Newton-Cartan geometry background and a Kalb-Ramond and dilaton field. Due to its codimension-two foliation structure, the string Newton-Cartan geometry of nonrelativistic string theory admits two distinct classes of T-duality transformations, depending on whether the isometry lies in the  longitudinal or transverse sector. These T-duality transformations are then classified in \cite{Bergshoeff:2018yvt} according to the nature of the associated isometry directions. These analyses are then generalized in \cite{Gomis:2020izd} to include open string background fields and also applied to the DBI actions for various D-branes. T-duality transformations in nonrelativistic string theory are summarized as follows:
\begin{enumerate}
    \item \emph{Longitudinal spatial T-duality.} The T-duality transformation is performed along a compact longitudinal spacelike isometry in nonrelativistic string theory coupled to a string Newton-Cartan background. The resulting theory is relativistic string theory coupled to a Lorentzian background geometry with a compact lightlike isometry, \emph{i.e.}, the DLCQ of relativistic string theory. 
    
    \item \emph{Longitudinal lightlike T-duality.} The T-duality transformation is performed along a compact longitudinal lightlike isometry in nonrelativistic string theory coupled to a string Newton-Cartan background. The resulting theory is nonrelativistic string theory coupled to a T-dual string Newton-Cartan background with a longitudinal lightlike isometry of an inverse radius. Applied to nonrelativistic open strings, this T-duality relates nonrelativistic and noncommutative open strings to each other. 
    
    \item \emph{Transverse T-duality.} The T-duality transformation is performed along a compact transverse (spacelike) isometry in nonrelativistic string theory on a string Newton-Cartan background. The resulting theory is nonrelativistic string theory on a T-dual string Newton-Cartan background with a transverse isometry.
\end{enumerate}
In this section, we include RR potentials to this discussion and derive the associated Buscher rules. Historically, the relativistic Buscher rules for the RR-potentials were derived from analyzing type II supergravity \cite{Meessen:1998qm}. However, the same results can also be derived by using probe D-branes (see, \emph{e.g.}, \cite{Simon:2011rw} and references therein). In \S\ref{sec:tdrdba}, we will first review how the generalized Buscher rules that incorporate the RR potentials are derived by analyzing relativistic D-brane actions. Then, in \S\ref{sec:tdgdba}, we apply the same analysis to \g D-branes in the presence of RR potentials and different isometries. We will generalized Buscher rules that also act on the RR potentials in nonrelativistic string theory. Finally, in \S\ref{eq:tdsi}, we show how these generalized Buscher rules are reproduced by taking the nonrelativistic limit of relativistic Buscher rules, which serves as an extra sanity check of the generalize nonrelativistic string limit that we proposed in \S\ref{sec:nrsl}.  

\subsection{T-duals of Relativistic D-Brane Actions} \label{sec:tdrdba}

We start with a brief review of T-duality transformations of the Buscher rules for the RR potentials in relativistic string theory. 
Such Buscher rules can be derived by using the relativistic D$p$-brane action that consists of both the DBI and CS parts is given in \eqref{sec:hspf},
which we repeat below for convenience:
\be \label{eq:relsm}
    \hat{S}_{\text{D}p} = - \int d^{p+1} Y \, e^{-\hat{\Phi}} \sqrt{-\det \lr \hat{G}_{\mu\nu} + \hat{B}_{\mu\nu} \rr} + \int \sum_{q} \hat{C}^{(q)} \wedge e^{\hat{B}} \Big|_{p+1}.
\ee 
Recall that $\hat{G}_{\mu\nu} = \p_\mu X^M \, \p_\nu X^N \, \hat{G}_{MN}$ and $\hat{B}_{\mu\nu} = \p_\mu X^M \, \p_\nu X^N \, \hat{B}_{MN}$ are respectively the pullbacks of the background metric $\hat{G}_{MN}$ and Kalb-Ramond field $\hat{B}_{MN}$ to the D$p$-brane's worldvolume. For simplicity, we set the gauge field strength $F_{\mu\nu}$ on the D-brane to zero and focus on the closed string background fields. The generalization that incorporates the worldvolume $U(1)$ gauge field can be obtained straightforwardly by following \cite{Alvarez:1996up}. 
 
Assume that there is a Killing vector $k^M$ in the target space. We defined the target-space coordinates $X^M =(y\,, X^m)$ that are adapted to $k^M$\,, with $k^M \p_M = \p_y$\,. Therefore, the translation in $y$ represents an abelian isometry. The standard Buscher rules for the metric field $\hat{G}_{MN}$\,, Kalb-Ramond field $\hat{B}_{MN}$ and dilaton field $\hat{\Phi}$ are derived from performing a T-duality transformation in the path integral of the worldsheet sigma model \cite{Buscher:1987sk, Buscher:1987qj}, with
\begin{subequations} \label{eq:relbuscher}
\begin{align}
    \hat{G}'_{yy} & = \frac{1}{\hat{G}_{yy}}\,,
        &
    \hat{G}'_{my} & = \frac{\hat{B}_{my}}{\hat{G}_{yy}}\,, 
        &
    \hat{G}'_{mn} & = \hat{G}_{mn} + \frac{\hat{B}_{my} \, \hat{B}_{ny} - \hat{G}_{my} \, \hat{G}_{ny}}{\hat{G}_{yy}}\,, \\[2pt]
    \hat{\Phi}' & = \hat{\Phi} - \frac{1}{2} \ln \hat{G}_{yy}\,,
        &
    \hat{B}'_{my} & = \frac{\hat{G}_{my}}{\hat{G}_{yy}}\,,
        &
    \hat{B}'_{mn} & = \hat{B}_{mn} - \frac{\hat{B}_{my} \, \hat{G}_{ny} - \hat{B}_{ny} \, \hat{G}_{my}}{\hat{G}_{yy}}\,. 
\end{align}
\end{subequations}
%
%
        %
%
In the following, we perform the T-duality transformation of the D-brane action \eqref{eq:relsm} along the isometry direction in two different cases, depending on what boundary conditions in the isometry direction that the open strings satisfy. 

\vspace{3mm}

\noindent $\bullet$ \textbf{Neumann boundary condition.}
We first consider a D-brane that is extending in the isometry $y$ direction, \emph{i.e.}, the open strings satisfy the Neumann boundary condition in $y$\,. The D-brane action is already given in \eqref{eq:relsm}.
We choose the adapted coordinates $Y^\mu = \lr Y^\alpha, y \rr$ with $\alpha = 0, 1, \cdots, p-1$ on the worldvolume of the D$p$-brane. Under the Buscher rules \eqref{eq:relbuscher}, there holds the following identity between different DBI Lagrangians:
\be \label{eq:dbiid}
    e^{-\hat{\Phi}} \, \sqrt{ - \det 
    \lr \hat{G}_{\mu\nu} + \hat{B}_{\mu\nu} \rr} = e^{-\hat{\Phi}'} \, \sqrt{ - \det 
    \lr \hat{G}'_{\alpha\beta} + \hat{B}'_{\alpha\beta} \rr}\,,
\ee
where the LHS describes a D$p$-brane extending along $y$ and the RHS describes the dual D$(p-1)$-brane localized at a point in $y$\,. For consistency, the following condition must hold for CS Lagrangians: 
\be \label{eq:relcs}
    \sum_q \hat{C}{}^{(q)} \wedge e^{\hat{B}} \bigg|_{p+1} = \lr \sum_q \hat{C}'{}^{(q)} \wedge e^{\hat{B}'} \bigg|_{p} \rr \wedge dy\,,
\ee
such that the T-dual D-brane action is
\be 
    \hat{S}'_{\text{D}(p-1)} = - \int d^{p} Y \, e^{-\hat{\Phi}'} \sqrt{-\det \lr \hat{G}'_{\mu\nu} + \hat{B}'_{\mu\nu} \rr} + \int \sum_{q} \hat{C}'{}^{(q)} \wedge e^{\hat{B}'} \Big|_{p}\,,
\ee 
which is in the same form as the original action \eqref{eq:relsm} but in terms of the T-dual fields and describes a D($p-1$)-brane transverse to the dual isometry $y$ direction.
%
%
%
%
%
It is convenient to define the worldvolume differential forms
\begin{align} \label{eq:defcy}
    \hat{C}^{(q+1)}_y & = \frac{1}{q!} \, \hat{C}^{{(q+1)}}_{\alpha_1 \cdots \alpha_q \, y} \, dY^{\alpha_1} \wedge \cdots \wedge dY^{\alpha_q}\,, 
        \quad%
    \hat{G}_y = \hat{G}_{\alpha y} \, dY^\alpha\,,
        \quad%
    \hat{B}_y = \hat{B}_{\alpha y} \, dY^\alpha\,,
\end{align}
and similarly for the primed fields.
We will use $\hat{B}$ (also for other worldvolume forms) to denote both $\frac{1}{2} \, \hat{B}_{\alpha\beta} \, dY^\alpha \! \wedge dY^\beta$ and $\frac{1}{2} \, \hat{B}_{\mu\nu} \, dY^\mu \! \wedge dY^\nu$\,, with $\mu = (\alpha, y)$ to avoid a cluster of notation.
%
%
The difference should be clear from the context. 
Then, \eqref{eq:relbuscher} becomes
\begin{subequations} \label{eq:bdf}
\begin{align}
    \hat{G}'_{yy} & = \frac{1}{\hat{G}_{yy}}\,,
        &
    \hat{G}'_{y} & = \frac{\hat{B}_{y}}{\hat{G}_{yy}}\,, 
        &
    \hat{G}'_{mn} & = \hat{G}_{mn} + \frac{\hat{B}_{my} \, \hat{B}_{ny} - \hat{G}_{my} \, \hat{G}_{ny}}{\hat{G}_{yy}}\,, \\[2pt]
    \hat{\Phi}' & = \hat{\Phi} - \frac{1}{2} \ln \hat{G}_{yy}\,,
        &
    \hat{B}'_{y} & = \frac{\hat{G}_{y}}{\hat{G}_{yy}}\,,
        &
    \hat{B}' & = \hat{B} - \frac{\hat{B}_{y} \wedge \hat{G}_{y}}{\hat{G}_{yy}}\,. 
\end{align}
\end{subequations}
Note that we always choose to place the index $y$ at the end of the subscript in the form's components. 
Now, the LHS of \eqref{eq:relcs} gives
\be
    \sum_q \hat{C}^{{(q)}} \wedge e^{\hat{B}} \bigg|_{p+1} = \CL'_\text{CS} \wedge dy\,,
        \qquad%
    \CL'_\text{CS} = \sum_q \lr \hat{C}^{{(q)}}_{y} + \hat{C}^{{(q)}} \wedge \hat{B}_y \rr \wedge e^{\hat{B}} \bigg|_{p}\,.
\ee
Using \eqref{eq:bdf}, we find
\begin{align} \label{eq:lpcs}
    \CL'_\text{CS} 
        %
        %
    = \sum_q \lr \hat{C}^{(q+1)}_y + \frac{\hat{C}^{(q-1)}_y \wedge \hat{B}_y \wedge \hat{G}_y}{\hat{G}_{yy}} + \hat{C}^{{(q-1)}} \wedge \hat{B}_y \rr \wedge e^{\hat{B}'} \bigg|_{p}\,. 
\end{align}
%
Here, we rewrote $\hat{B}$ as 
$\hat{B} = \hat{B}' + ( \hat{B}_y \wedge \hat{G}_y / \hat{G}_{yy} )$.
Without repeating further, this is also the trick that we use for other similar derivations in this section.
Finally, \eqref{eq:relcs} gives
\be \label{eq:RRtd1}
    \hat{C}^{\prime(q)} = \hat{C}^{(q+1)}_y + \hat{C}^{{(q-1)}} \wedge \hat{B}_y + \frac{\hat{C}^{(q-1)}_y \wedge \hat{B}_y \wedge \hat{G}_y}{\hat{G}_{yy}}\,.
\ee

\vspace{3mm}

\noindent $\bullet$ \textbf{Dirichlet boundary condition.} Now, we consider a D$p$-brane that is transverse to the isometry $y$ direction, \emph{i.e.}, the open strings satisfy the Dirichlet boundary condition in $y$\,. It then follows that $\p_\mu y = 0$\,. Note that this D$p$-brane is still described by the action \eqref{eq:relsm}. Under the Buscher rules in \eqref{eq:relbuscher}, there holds an identity between different DBI Lagrangians that is in form the same as \eqref{eq:dbiid},
but now with ${Y'}^\alpha = (Y^\mu, y')$\,. Here, $y'$ is dual to $y$ in the target space. This implies that \eqref{eq:dbiid} receives a different interpretation with the LHS describing a D$p$-brane transverse to $y$ and the RHS describing the dual D$(p+1)$-brane extending in $y'$\,. For consistency, we require
\be \label{eq:csdndual}
    \sum_q \hat{C}{}^{(q)} \wedge e^{\hat{B}} \bigg|_{p+1} = \CL'_\text{CS}\,,
        \qquad%
    \CL'_\text{CS} \equiv \sum_q \lr \hat{C}^{\prime(q+1)}_y + \hat{C}^{\prime(q)} \wedge \hat{B}'_y \rr \wedge e^{\hat{B}'} \bigg|_{p+1}\,,
\ee
with
\be \label{eq:csdndual3}
    \CL'_\text{CS} \wedge dy' = \sum_q \hat{C}^{\prime(q+1)} \wedge e^{\hat{B}'} \bigg|_{p+2}
\ee
The LHS of the first equation in \eqref{eq:csdndual} is the CS term for the D$p$-brane transverse to $y$\,. Moreover, \eqref{eq:csdndual3}
denotes the dual CS term for the D$(p+1)$-brane extending in $y'$.
Using \eqref{eq:bdf}, we find that \eqref{eq:csdndual} implies
\be \label{eq:RRtd2}
    \hat{C}^{\prime(q+1)}_y
    = \hat{C}^{(q)} + \frac{\hat{C}^{(q-2)} \wedge \hat{B}_y \wedge \hat{G}_y}{\hat{G}_{yy}} - \frac{\hat{C}^{\prime(q-1)} \wedge \hat{G}_y}{\hat{G}_{yy}}\,.
\ee

\vspace{3mm}

\noindent $\bullet$ \textbf{Buscher rules for RR potential.}
Finally, combining \eqref{eq:RRtd1} and \eqref{eq:RRtd2} that relate the RR potentials to their T-duals, we find:
\begin{subequations} \label{eq:relbuscherrr}
\begin{align} 
    \hat{C}^{\prime(q+1)}_y & = \hat{C}^{(q)} - \frac{\hat{C}^{(q)}_y \wedge \hat{G}_y}{\hat{G}_{yy}}\,, \\[4pt]
    \hat{C}^{\prime(q)} & = \hat{C}^{(q+1)}_y + \hat{C}^{(q-1)} \wedge \hat{B}_y + \frac{\hat{C}^{(q-1)}_y \wedge \hat{B}_y \wedge \hat{G}_y}{\hat{G}_{yy}}\,.
\end{align}
\end{subequations}
In the following subsections, we will derive the generalized Buscher rules for RR-potentials in nonrelativistic string theory, analogous to the ones in \eqref{eq:relbuscherrr}. Note that the differential forms here do not contain $dy$ or $dy'$.

\subsection{T-duals of \gc D-Brane Actions} \label{sec:tdgdba}

Now, we return to the \g D$p$-brane action \eqref{eq:gdbbi} and study its T-dual along a compact isometry direction in the target space. We address all the three cases with a longitudinal spacelike, longitudinal lightlike and transverse isometry, respectively. 

\subsubsection{Longitudinal spatial T-duality} \label{sec:lstd}

We start with the case where there is a longitudinal spacelike Killing vector $k^M$ satisfying
\be \label{eq:LongSpatialTdual}
    k^M \, \tau^{}_M{}^0 = 0\,,
        \qquad
    k^M \, \tau^{}_M{}^1 \neq 0\,,
        \qquad
    k^M E_M{}^{A'} = 0\,.
\ee
Define the coordinates $X^M = (X^m,\, y)$ adapted to $k^M$, with $k^M \, \partial_M = \partial_y$\,. In terms of these adapted coordinates, \eqref{eq:LongSpatialTdual} implies $\tau_y{}^0 =  E_{y}{}^{A'} = 0$ and $\tau_y{}^1 \neq 0$\,. Then, $y$ is a longitudinal spacelike isometry direction. In the absence of RR potentials,  the Buscher rules for nonrelativistic string theory with a longitudinal spatial isometry in spacetime are given in \cite{Bergshoeff:2018yvt}. These Buscher rules are derived by using the worldsheet formalism. The dual worldsheet sigma model describes relativistic string theory in the DLCQ, where a dual Lorentzian metric field $\tilde{G}_{MN}$ exists, in addition to the dual Kalb-Ramond field $\tilde{B}_{MN}$ and dilaton field $\tilde{\Phi}$\,. The Buscher rules that relate these dual fields to the nonrelativistic closed string background fields $\tau_\mu{}^A$\,, $H_{\mu\nu}$\,, $B_{\mu\nu}$ and $\Phi$ are
\begin{subequations} \label{eq:relsncbuscher}
\begin{align}
    \tilde{G}_{yy} & = 0\,, 
        \qquad%
     \tilde{G}_{my} = - \frac{\ell_{my}}{\tau_{yy}}\,,
        \qquad%
    \tilde{B}_{my} = \frac{\tau_{my}}{\tau_{yy}}\,,
        \qquad%
    \tilde{\Phi} = \Phi - \frac{1}{2} \, \ln \tau_{yy}\,, \\[2pt] 
    \tilde{G}_{mn} &= H_{mn} - \frac{\left( B_{my} \, \ell_{ny} + B_{ny} \, \ell_{my} \right) + \left( H_{my} \, \tau_{ny} + H_{ny} \, \tau_{my} - H_{yy} \, \tau_{mn} \right)}{\tau_{yy}}\,, \\[2pt] 
    \tilde{B}_{mn} &= B_{mn} - \frac{\left( B_{my} \, \tau_{ny} - B_{ny} \, \tau_{my} \right) + \left(H_{my} \, \ell_{ny} - H_{ny} \, \ell_{my} + H_{yy} \, \ell_{mn} \right)}{\tau_{yy}}\,. 
\end{align}
\end{subequations}
See \eqref{eq:defell} for the definition of $\ell^{}_{MN}$\,. Using differential forms, we rewrite the Buscher rules associated with $\tilde{G}_{my}$\,, $\tilde{B}_{my}$\,, and $\tilde{B}_{mn}$ as
\be
\label{lstdcomponents}
    \tilde{G}_y = - \frac{\ell_y}{\tau_{yy}}\,,
        \qquad%
    \tilde{B}_y = \frac{\tau^{(1)}_y}{\tau_{yy}}\,,
        \qquad%
    \tilde{B} = B - \frac{B_y \wedge \tau^{(1)}_y + H_y \wedge \ell_y + H_{yy} \, \ell}{\tau_{yy}}\,.
\ee
Here, $\tau^{(1)}_y = \tau_{\alpha y} \, dY^\alpha$\,, where $\alpha$ is a worldvolume index excluding $y$\,.
The fact that $\tilde{G}_{yy} = 0$ implies that there is a lightlike isometry in the dual relativistic target-space geometry. We also defined
\be \label{eq:deflly}
    \ell = \frac{1}{2} \, \tau^{}_\alpha{}^A \, \tau^{}_\beta{}^B \, \epsilon_{AB} \, dY^\alpha \wedge dY^\beta\,,
        \qquad%
    \ell_y = \tau_\alpha{}^A \, \tau_y{}^B \, \epsilon_{AB} \, dY^\alpha\,.
\ee
Note that the two-form $\ell$ here is defined differently from \eqref{eq:defell}. We consider how the following \g D$p$-brane action transforms under the above Buscher rules: 
\begin{equation} \label{eq:nrdba4}
    S_{\text{D}p} = - \int d^{p+1} Y \, e^{-\Phi} \sqrt{-\operatorname{det}\left(\begin{array}{cc}
0 & \tau_\nu \\
\bar{\tau}_{\mu} & \quad H_{\mu\nu} + B_{\mu\nu}
\end{array}\right)}
    + \int \sum_q C^{(q)} \wedge e^{B} \bigg|_{p+1}\,.
\end{equation}
The same \g D$p$-brane action has been given in \eqref{eq:gdbbi}, but now we set the gauge field strength $F = 0$ on the D-brane. For the inclusion of the worldvolume $U(1)$ gauge field in the Buscher rules, see \cite{Gomis:2020izd}.



We first consider the case where the \g D$p$-brane described by the action \eqref{eq:nrdba4} extends in the longitudinal spacelike isometry $y$ direction.
The T-dual of this \g D$p$-brane is a relativistic D$(p-1)$-brane transverse to the isometry direction $\tilde{y}$\,. Note that $\tilde{y}$ is dual to $y$ and is compactified over a lightlike circle. As we already explained earlier, this is because the dual metric component $\tilde{G}_{yy}$ vanishes in \eqref{eq:relsncbuscher}. We choose the adapted coordinates $Y^\mu = (Y^\alpha, y)$\,, with $\alpha = 0, 1, \cdots,\, p-1$ on and $Y^\mu$ the coordinates on the ($p+1$)-dimensional worldvolume of the \g D$p$-brane. Under the Buscher rules \eqref{eq:relsncbuscher},the following identity between different DBI Lagrangians holds \cite{Gomis:2020izd}:
\be \label{eq:dbilongtd}
    e^{-\Phi} \sqrt{- \det 
    \begin{pmatrix}
        0 &\quad \tau_\nu \\[2pt]
        \bar{\tau}_\mu &\quad H_{\mu\nu} + B_{\mu\nu}
    \end{pmatrix}}
    =
    e^{-\tilde{\Phi}} \sqrt{-\det
    \begin{pmatrix}
        \tilde{G}_{\alpha\beta} + \tilde{B}_{\alpha\beta}
    \end{pmatrix}}\,.
\ee
The LHS of \eqref{eq:dbilongtd} describes a \g D$(p-1)$-brane extending in $y$ and the RHS describes the dual relativistic D$p$-brane transverse to $\tilde{y}$\,. For consistency, we require for the CS Lagrangians that
\be \label{eq:csrells}
	\sum_q C^{(q)} \wedge e^B \bigg|_{p+1} = \lr \sum_q \tilde{C}^{(q)} \wedge e^{\tilde{B}} \bigg|_{p} \rr \wedge dy\,,
\ee	
in analog with the relativistic case in \eqref{eq:relcs}. Applying \eqref{lstdcomponents} to \eqref{eq:csrells}, we find
\begin{align} \label{eq:ltdc}
\begin{split}
	\tilde{C}^{(q)} & = C^{(q+1)}_y + C^{(q-1)} \wedge B_y + \frac{C^{(q-1)}_y \wedge \lr B_y \wedge \tau_y^{(1)} + H_y \wedge \ell_y + H_{yy} \, \ell \rr}{\tau_{yy}} \\[2pt]
	& \hspace{1.65cm} + \frac{C^{(q-3)} \wedge B_y \wedge \lr H_y \wedge \ell_y + H_{yy} \, \ell \rr}{\tau_{yy}} + \frac{C^{(q-3)}_y \wedge B_y \wedge H_y \wedge \ell}{\tau_{yy}}\,.
\end{split}
\end{align}


        %
        %

Next, we consider a \g D$p$-brane described by the action \eqref{eq:nrdba4} but now transverse to the longitudinal spacelike isometry $y$\,. The T-dual of the \g D$p$-brane is a relativistic D$(p+1)$-brane extending in the lightlike isometry direction $\tilde{y}$ that is dual to $y$\,.  
Under the Buscher rules in \eqref{eq:relsncbuscher}, there holds an identity between different DBI Lagrangians that is in form the same as \eqref{eq:dbilongtd}, but now with $Y'{}^\alpha = (Y^\mu, \tilde{y})$\,. Then, \eqref{eq:dbilongtd} receives the interpretation that the LHS describes a \g D$(p-1)$-brane transverse to $y$ and the RHS describes a relativistic D$p$-brane extending in $\tilde{y}$\,. 
For consistency, like \eqref{eq:csdndual} and \eqref{eq:csdndual3} in relativistic string theory, the following condition for CS Lagrangians must hold:
\be \label{eq:relcs1}
    \lr \sum_q C^{(q)} \wedge e^{B} \bigg|_{p+1} \rr \wedge d\tilde{y}
    =
    \sum_q \tilde{C}^{(q)} \wedge e^{\tilde{B}} \bigg|_{p+2}\,.
\ee
Using \eqref{lstdcomponents}, we find in analog with \eqref{eq:RRtd1} that
%
        %
%
\begin{align} \label{eq:ltdcy}
\begin{split}
    & \quad \tilde{C}^{(q)}_y + \tilde{C}^{(q-2)} \wedge \frac{\tau_y^{(1)}}{\tau_{yy}} \\[2pt]
    & = C^{(q-1)} + C^{(q-3)} \wedge \frac{B_y \wedge \tau^{(1)}_y + H_y \wedge \ell_y + H_{yy} \, \ell}{\tau_{yy}} + \frac{C^{(q-5)} \wedge B_y \wedge H_y \wedge \ell}{\tau_{yy}}\,.
\end{split}
\end{align}

Finally, combining \eqref{eq:ltdcy} and \eqref{eq:ltdc}, we find the following map between the RR potentials in nonrelativistic string theory and their T-dual RR potentials in relativistic string theory:
\begin{subequations} \label{eq:lscb}
\begin{align} 
	\tilde{C}^{(q)}_y & = C^{(q-1)} - \frac{C^{(q-1)}_y \wedge \tau_y^{(1)}}{\tau_{yy}} + \frac{C^{(q-3)} \wedge \lr H_y \wedge \ell_y + H_{yy} \, \ell \rr}{\tau_{yy}} - \frac{C^{(q-3)}_y \wedge H_y \wedge \ell}{\tau_{yy}}\,, \label{eq:tcqy} \\[10pt]
	\tilde{C}^{(q)} & = C^{(q+1)}_y + C^{(q-1)} \wedge B_y + \frac{C^{(q-1)}_y \wedge \lr B_y \wedge \tau_y^{(1)} + H_y \wedge \ell_y + H_{yy} \, \ell \rr}{\tau_{yy}} \notag \\[2pt]
	& \hspace{1.65cm} + \frac{C^{(q-3)} \wedge B_y \wedge \lr H_y \wedge \ell_y + H_{yy} \, \ell \rr}{\tau_{yy}} + \frac{C^{(q-3)}_y \wedge B_y \wedge H_y \wedge \ell}{\tau_{yy}}\,. \label{eq:tcq}
\end{align}
\end{subequations}
The same Buscher rules \eqref{eq:lscb} for RR potentials can be reproduced by plugging the ansatz \eqref{eq:cexp} into the relativistic Buscher rules \eqref{eq:relbuscherrr} and then taking the $c \rightarrow \infty$ limit. Note that $\hat{C}'{}^{(q)}$ in \eqref{eq:relbuscherrr} will be identified with $\tilde{C}^{(q)}$ in \eqref{eq:lscb} after the limit is taken. We will study this nonrelativistic string limit of relativistic Buscher rules in \S\ref{eq:tdsi}.
Moreover, as a nontrivial check, we will further show in Appendix \ref{app:stbr} that the Buscher rules \eqref{eq:lscb} are invariant under the infinitesimal version of the Stueckelberg transformations in \eqref{eq:ss} and \eqref{eq:ssC}. 

\subsubsection{Longitudinal lightlike T-duality} \label{sec:lltd}

Previously, we considered the T-duality transformation of D-branes in nonrelativistic string theory along a longitudinal spacelike isometry, and the dual D-branes are coupled to a Lorentzian background geometry with a compact lightlike isometry. For completeness, we now perform a T-duality transformation along a lightlike isometry for a \g D-brane coupled to string Newton-Cartan geometry. This leads to a dual \g D-brane that is coupled to a dual string Newton-Cartan geometry. This lightlike T-duality transformation maps between two lightlike circles with dual radii \cite{Bergshoeff:2018yvt}. This lightlike T-duality provides a formal relation between nonrelativistic and noncommutative open strings \cite{Gomis:2020izd}.

We start with a longitudinal lightlike Killing vector $k^M$ satisfying
\be \label{eq:kkll}
	k^M \, \tau^{}_M \neq 0\,,
		\qquad%
	k^M \, \bar{\tau}^{}_M = 0\,,
		\qquad%
	k^M \, E_M{}^{A'} = 0\,.
\ee
We then have $\bar{\tau}_y = E_y{}^{A'} = 0$ in the coordinates $X^M = (X^m, \,y)$ adapted to $k^M$, satisfying $k^M \, \p_M = \p_y$\,. These prescriptions require that $y$ be a lightlike isometry. In the absence of RR potentials, the Buscher rules for nonrelativistic string theory with a longitudinal lightlike isometry are derived in \cite{Bergshoeff:2018yvt} using the worldsheet formalism, with the dual fields $T_M{}^A$, $\tilde{H}_{MN}$\,, $\tilde{B}_{MN}$ and $\tilde{\Phi}$ given by
\begin{subequations} \label{eq:relsncbuscherll}
\begin{align}
    T_y & = \frac{1}{\tau_y}\,,
    	\qquad\,\,\,%
    \tilde{\Phi} = \Phi - \ln \, \tau_y\,, \\[2pt]
    \bar{T}_y & = 0\,, 
    	\qquad\,\,\,%
    \bar{T}_m = \bar{\tau}_m\,, 
        \qquad%
    T_m = \frac{\bigl( B_{my} - H_{my} \bigr) \, \tau_y + H_{yy} \, \tau_m}{\tau^2_y}\,, \\[2pt]
    \tilde{H}_{My} & = 0\,, 
    	\qquad%
    \tilde{B}_{my} = \frac{\tau_m}{\tau_y}\,, 
    	\qquad\,%
    \tilde{B}_{mn} = B_{mn} - \frac{B_{my} \, \tau_n - B_{ny} \, \tau_m}{\tau_y}\,, \\[2pt]
    \tilde{H}_{mn} & = H_{mn} + \frac{H_{yy} \, \tau_m \, \tau_n - \bigl( H_{my} \, \tau_n + H_{ny} \, \tau_m \bigr) \, \tau_y}{\tau_y^2}\,.
\end{align}
\end{subequations}
Here, we used $T_M{}^A$ to denote the T-dual of the longitudinal vielbein field $\tau^{}_M{}^A$ in string Newton-Cartan geometry.

We first consider a \g D$p$-brane that is described by the action \eqref{eq:nrdba4} and extends in the longitudinal lightlike isometry $y$\,. This theory is in the sector of NCOS. We choose the adapted coordinates $Y^\mu = (Y^\alpha, y)$ with $\alpha = 0, 1, \cdots, p-1$\,. Using \eqref{eq:relsncbuscherll}, we obtain the following identity between different DBI Lagrangians \cite{Gomis:2020izd}:

\be \label{eq:dbilongtd1}
    e^{-\Phi} \sqrt{- \det 
    \begin{pmatrix}
        0 &\quad \tau_\nu \\[2pt]
        \bar{\tau}_\mu &\quad H_{\mu\nu} + B_{\mu\nu}
    \end{pmatrix}}
    =
    e^{-\tilde{\Phi}} \sqrt{-\det
    \begin{pmatrix}
        0 &\quad T_\beta \\[2pt]
        \bar{T}_\alpha &\quad \tilde{H}_{\alpha\beta} + \tilde{B}_{\alpha\beta}
    \end{pmatrix}}\,,
\ee
where the RHS describes the dual \g D$(p-1)$-brane transverse to the dual lightlike isometry $\tilde{y}$\,. This dual theory is in the sector of NROS. Requiring that the analog of the identity between the CS Lagrangians in \eqref{eq:csrells} hold, but now with $\tilde{B}_{MN}$ related to $B_{MN}$ as in \eqref{eq:relsncbuscherll}, we find
\be \label{eq:cll}
    \tilde{C}^{(q)} = C^{(q+1)}_y + C^{(q-1)} \wedge B_y + \frac{C_y^{(q-1)} \wedge B_y \wedge \tau}{\tau_y}\,,
        \qquad%
    \tau = \tau^{}_\alpha \, dY^\alpha\,.
\ee
%

%

Next, we consider NROS and a D$p$-brane that is described by the action \eqref{eq:nrdba4} and transverse to the longitudinal lightlike isometry $y$\,. The T-dual action describes a D$(p+1)$-brane in NCOS that extends in the dual lightlike isometry $\tilde{y}$\,. In this case, we continue to have the identity \eqref{eq:dbilongtd1} between different DBI actions, but with $Y'{}^{\alpha} = (Y^\mu, \tilde{y})$\,. For consistency, the same relation \eqref{eq:relcs1} between different CS Lagrangians has to hold, but now with the background fields satisfying the Buscher rules in \eqref{eq:relsncbuscherll}. Then, the relation \eqref{eq:relcs1} implies 
\be \label{eq:cyll}
	\tilde{C}^{(q)}_y + \frac{\tilde{C}^{(q-2)} \wedge \tau}{\tau_y} = C^{(q-1)} + \frac{C^{(q-3)} \wedge B_y \wedge \tau}{\tau_y}\,.
\ee

Finally, combining \eqref{eq:cyll} and \eqref{eq:cll}, we find the Buscher rules for the RR potentials,
\begin{subequations} \label{eq:llrrbr}
\begin{align} 
    \tilde{C}^{(q)}_y & = C^{(q-1)} - \frac{C^{(q-1)}_y \wedge \tau}{\tau_y}\,, \\[4pt]
    \tilde{C}^{(q)} & = C^{(q+1)}_y + C^{(q-1)} \wedge B_y + \frac{C^{(q-1)}_y \wedge B_y \wedge \tau}{\tau_y}\,.
\end{align}
\end{subequations}
Later in \S\ref{eq:tdsi}, we will discuss how the generalized Buscher rules \eqref{eq:relsncbuscherll} and \eqref{eq:llrrbr} arise as a nonrelativistic string limit of the relativistic Buscher rules. In Appendix \ref{app:stbr}, we will analyze how these Buscher rules transform under the infinitesimal Stueckelberg symmetry, whose finite form is given in \eqref{eq:ss} and \eqref{eq:ssC}.

\subsubsection{Transverse T-duality} \label{sec:ttd}

It is also possible to perform a T-duality transformation along a transverse isometry direction that is compactified over a circle.~\footnote{Note that all transverse directions are spacelike.} 
Consider a transverse Killing vector $k^M$ that satisfies
\be \label{eq:tkv}
    k^M \tau^{}_M{}^A = 0\,,
        \qquad%
    k^M E^{}_{MN} \neq 0\,.
\ee
Recall that $E_{MN} = E_M{}^{A'} E_N{}^{A'}$\,. The derivation of the Buscher rules for transverse T-duality is in form the same as the relativistic case in \S\ref{sec:tdrdba}. The resulting Buscher rules are 
\begin{subequations} \label{eq:ttdbr}
\begin{align}
    \tilde{H}_{yy} & = \frac{1}{H_{yy}}\,,
        &
    \tilde{H}_{my} & = \frac{B_{my}}{H_{yy}}\,, 
        &
    \tilde{H}_{mn} & = {H}_{mn} + \frac{{B}_{my} \, {B}_{ny} - {H}_{my} \, H_{ny}}{\hat{G}_{yy}}\,, \\[2pt]
    \tilde{\Phi} & = {\Phi} - \frac{1}{2} \ln H_{yy}\,,
        &
    \tilde{B}_{my} & = \frac{H_{my}}{H_{yy}}\,,
        &
    \tilde{B}_{mn} & = B_{mn} - \frac{B_{my} \, H_{ny} - B_{ny} \, H_{my}}{H_{yy}}\,, 
\end{align}
\end{subequations}
and
\begin{subequations} \label{eq:tbrrrp}
\begin{align} 
    \tilde{C}^{(q)}_y & = C^{(q-1)} - \frac{C^{(q-1)}_y \wedge H_y}{H_{yy}}\,, \\[4pt]
    \tilde{C}^{(q)} & = C^{(q+1)}_y + C^{(q-1)} \wedge B_y + \frac{C^{(q-1)}_y \wedge B_y \wedge H_y}{H_{yy}}\,.
\end{align}
\end{subequations}
Here, $H_y = H_{\alpha y} \, dY^\alpha$\,, where the worldvolume index $\alpha$ does not include $y$\,.
The background field $\tau_M{}^A$ remains unchanged under the transverse T-duality transformation, while the background fields $H_{MN}\,, B_{MN}\,, \Phi$ and $C^{(q)}$ are mapped to their T-duals, $\tilde{H}_{MN}\,, \tilde{B}_{MN}\,, \tilde{\Phi}$ and $\tilde{C}^{(q)}$\,.
Both the original and T-dual D-brane actions are in the form of \eqref{eq:gdbbi}. Also see Appendix \ref{app:stbr} for how these Buscher rules transform under the infinitesimal Stuckelberg symmetry.

\subsection{Buscher Rules from Nonrelativistic String Limit} \label{eq:tdsi}

Finally, we discuss how the Buscher rules derived in \S\ref{sec:tdgdba} can be reproduced by taking appropriate nonrelativisitc string limits of the relativistic Buscher rules in \S\ref{sec:tdrdba}. 
Our starting point are the background fields' parametrizations given in \eqref{eq:cexp}, which we rewrite below as:
\begin{subequations} \label{eq:expbfc}
\begin{align}
    \hat{G}^{\phantom{(}}_{MN} & = c^2 \, \tau^{\phantom{(}}_{MN} + H^{\phantom{(}}_{MN}\,, 
        &%
    \hat{\Phi} & = \Phi + \ln c\,, \\[2pt]  
    \hat{B} & = - c^2 \, \ell + B\,, 
        &%
    \hat{C}^{(q)} & = c^2 \, C^{(q-2)} \wedge \ell + C^{(q)}\,,
    \\[2pt]
    \hat{B}_y & = - c^2 \, \ell_y + B_y\,,
        &%
    \hat{C}^{(q)}_y & = c^2 \bigl( C^{(q-2)}_y \wedge \ell + C^{(q-2)} \wedge \ell_y \bigr) + C^{(q)}_y\,.
\end{align}
\end{subequations}
For $q<0$\,, we set $\hat{C}^{(q)} = \hat{C}^{(q+1)}_y = 0$\,. We have set the worldvolume field strength $F_{\mu\nu}$ to zero. Also note that $\ell$ and $\ell_y$ are defined in \eqref{eq:deflly}.

\vspace{3mm}

\noindent $\bullet$ \textbf{Longitudinal spatial T-duality.} In the presence of the longitudinal spatial Killing vector as specified in \eqref{eq:LongSpatialTdual}, we learned from \S\ref{sec:lstd} that the T-dual of nonrelativistic string theory along the longitudinal spacelike isometry direction describes the DLCQ of relativistic string theory \cite{Bergshoeff:2018yvt}. It has been shown in \cite{Bergshoeff:2019pij} that the Buscher rules \eqref{eq:relsncbuscher} for longitudinal spatial T-duality arise as the nonrelativistic string limit of the relativistic Buscher rules in \eqref{eq:relbuscher}. This is done by plugging the ansatz \eqref{eq:expbfc} of relativistic background fields into the relativistic Buscher rules in \eqref{eq:relbuscher}, and then taking the $c \rightarrow \infty$ limit. Note that we also need to identify $\hat{G}'_{MN} \rightarrow \tilde{G}_{MN}$\,, $\hat{B}'_{MN} \rightarrow \tilde{B}_{MN}$\,, $\hat{\Phi}' \rightarrow \tilde{\Phi}$ after applying the $c\rightarrow\infty$ limit. This procedure also applies to the RR potentials. 
Starting with the relativistic Buscher rules for RR potentials in \eqref{eq:relbuscherrr} and plugging in \eqref{eq:expbfc}, we find,
\begin{subequations} \label{eq:cpexp}
\begin{align}
\begin{split}
    C^{\prime(q+1)}_y 
    & = c^2 \! \ls C^{(q-2)} \wedge \ell - \lr 1 - \frac{H_{yy}}{c^2 \, \tau_{yy}} \rr \frac{C^{(q-2)} \wedge \ell_y \wedge \tau^{(1)}_y}{\tau_{yy}} \rs \\[2pt]
    & \quad + C^{(q)} - \frac{C^{(q-2)}_y\wedge \ell \wedge H_y + C^{(q-2)} \wedge \ell_y \wedge H_y + C^{(q)}_y \wedge \tau_y^{(1)}}{\tau_{yy}} + O(c^{-2})\,, 
\end{split}
\end{align}
and
\begin{align}
\begin{split}
    C^{\prime(q)} 
        %
        %
    & = c^2 \lr C^{(q-1)}_y \wedge \ell + C^{(q-3)} \wedge \ell \wedge B_y \rr + C^{(q+1)}_y + C^{(q-1)} \wedge B_y \\[2pt]
    & \quad + \lr c^2 - \frac{H_{yy}}{\tau_{yy}} \rr \frac{C^{(q-3)} \wedge \ell_y \wedge B_y \wedge \tau_y^{(1)} - C^{(q-1)}_y \wedge \ell_y \wedge\tau_y^{(1)}}{\tau_{yy}} \\[2pt]
        %
        %
    & \quad + \tau^{-1}_{yy} \Bigl( C^{(q-3)}_y \wedge \ell \wedge B_y \wedge H_y + C^{(q-3)} \wedge \ell_y \wedge B_y \wedge H_y \\[2pt]
    & \hspace{3.2cm} - C^{(q-1)}_y \wedge \ell_y \wedge H_y + C^{(q-1)}_y \wedge B_y \wedge \tau_y^{(1)} \Bigr) + O(c^{-2})\,. 
\end{split}
\end{align}
\end{subequations}
In the $c \rightarrow \infty$\, limit, $\hat{C}'{}^{(q)} \rightarrow \tilde{C}^{(q)}$ and $\hat{C}_y^{\prime(q)} \rightarrow \tilde{C}^{(q)}_y$. Using the identity $\ell_y \wedge\tau_y^{(1)} = \ell \, \tau_{yy}$\,, we find that \eqref{eq:cpexp} becomes \eqref{eq:lscb} at $c\rightarrow \infty$\,.


\vspace{3mm}

\noindent $\bullet$ \textbf{Longitudinal lightlike T-duality.} The construction of the nonrelativistic string limit that reproduces the Buscher rules associated with the longitudinal lightlike T-duality transformation appears to be rather delicate and requires a careful treatment. We start with revisiting the ansatz \eqref{eq:expbfc} of background fields in relativistic string theory, but now in the presence of the longitudinal lightlike Killing vector defined in \eqref{eq:kkll}. 
To facilitate the following discussion,
we fix the Stueckelberg symmetry \eqref{eq:ss} and \eqref{eq:ssC} by setting $\Xi^{}_M{}^A = m^{}_M{}^A$ as in \S\ref{sec:nrsl}. Accordingly, we modify the reparametrizations of background fields in \eqref{eq:expbfc} as
\begin{subequations} \label{eq:exfs}
\begin{align}
    \hat{G}^{\phantom{(}}_{MN} & = c^2 \, \tau^{\phantom{(}}_{MN} + E^{\phantom{(}}_{MN}\,, 
        &%
    \hat{\Phi} & = \Phi + \ln c\,, \\[2pt]  
    \hat{B} & = - c^2 \, \ell + M\,, 
        &%
    \hat{C}^{(q)} & = c^2 \, N^{(q-2)} \wedge \ell + N^{(q)}\,,
    \\[2pt]
    \hat{B}_y & = - c^2 \, \ell_y + M_y\,,
        &%
    \hat{C}^{(q)}_y & = c^2 \bigl( N^{(q-2)}_y \wedge \ell + N^{(q-2)} \wedge \ell_y \bigr) + N^{(q)}_y\,.
\end{align}
\end{subequations}
We recall that $E_{MN} = E_M{}^{A'} \, E_N{}^{A'}$\,, $M_{MN}$ and $N^{(q+1)}_{M_0 \cdots M_q}$ are defined in \eqref{eq:mbrelation} and \eqref{eq:mq}, respectively. Additionally, 
\be
    \ell = \tfrac{1}{2} \, \bar{\tau} \wedge \tau\,,
        \qquad%
    \ell_y = \tfrac{1}{2} \bigl( \bar{\tau} \, \tau_y - \bar{\tau}_y \, \tau \bigr)\,,
        \qquad%
    \tau = \tau_\alpha \, dY^\alpha\,,
        \qquad
    \bar{\tau} = \bar{\tau}_\alpha \, dY^\alpha\,.
\ee
Recall that the worldvolume $\alpha$ index excludes $y$\,.
Moreover, as detailed in \S\ref{sec:lltd}, the existence of the longitudinal lightlike isometry in $y$ implies that $\bar{\tau}_y = E_y{}^{A'} = 0$ in the adapted coordinates $X^M = (X^m, \,y)$\,. Using \eqref{eq:exfs}, we find
$\hat{G}_{yy} = 0$\,.
Upon initial inspection, 
it seems impossible to take the $c \rightarrow \infty$ limit of the relativistic Buscher rules because of $\hat{G}_{yy}$ appearing in several denominators in \eqref{eq:relbuscher}. 

Fortunately, the above difficulty is avoidable by considering a double scaling limit instead of the original nonrelativistic string limit that only sets $c \rightarrow \infty$\,. We first construct such a double scaling limit without any RR potential and show how the Buscher rules \eqref{eq:relsncbuscherll} for longitudinal lightlike T-duality can be reproduced. In addition to the parameter $c$ that controls the nonrelativistic string limit in the original theory, 
we now introduce a second parameter $\tilde{c}$ that controls the nonrelativistic string limit on the T-dual side, with
\begin{align} \label{eq:lltd}
    \hat{G}'_{MN} & = \tilde{c}^{\,2} \, T^{\phantom{(}}_{MN} + \tilde{E}^{\phantom{(}}_{MN}\,,     
        &%
    \hat{B}'_{MN} & = - \tilde{c}^{\,2} \, \tilde{\ell}_{MN} + \tilde{M}_{MN}\,, 
        &%
    \hat{\Phi}' & = \tilde{\Phi} + \ln \tilde{c}\,.  
\end{align}
Here, $T_{MN} = T_M{}^A \, T_N{}^B \, \eta_{AB}$ and $\tilde{\ell}_{MN} = T_M{}^A \,
T_N{}^B \, \epsilon_{AB}$\,. In addition, we define
\be \label{eq:btaubT}
    \bar{\tau}_y = - \frac{\tau_y}{\tilde{c}^{\,2}}\,,
        \qquad%
    \bar{T}_y = - \frac{T_y}{c^2}\,,
        \qquad%
    \bar{T}_m = \bar{\tau}_m - \frac{M_{my}}{c^2 \, \tau_y} + \frac{\tau_m}{\tilde{c}^{\, 2}}\,,
\ee
where $T_m = T_m{}^0 + T_m{}^1$ and $\bar{T}_m = T_m{}^0 - T_m{}^1$\,.
In the limit $\tilde{c} \rightarrow \infty$\,, $\bar{\tau}_y$ becomes zero; this is required such that $y$ is a longitudinal lightlike isometry. Further taking $c \rightarrow \infty$\,, we find $\bar{T}_m \rightarrow - \bar{\tau}_m$\,. Moreover, we also require that $c^2 / \tilde{c}^{\,2} \rightarrow 0$\,, such that the ansatz \eqref{eq:exfs} is consistent with the one from setting $\bar{\tau}_m$ to zero identically.~\footnote{Note that the definition of the longitudinal lightlike Killing vector originally presented in \cite{Bergshoeff:2018yvt} has $\tilde{c}^{\,-1} = 0$ identically. This does not cause any problem there, since \eqref{eq:exfs} uses the path integral of the sigma model that describes nonrelativistic string theory and does not rely on any limits of relativistic string theory.} This double scaling limit is reminiscent of defining the DLCQ of string/M-theory as a subtle infinite boost limit \cite{Seiberg:1997ad}.
Plugging \eqref{eq:exfs}, \eqref{eq:btaubT} and \eqref{eq:lltd} into the relativistic Buscher rules \eqref{eq:relbuscher}, we find
\begin{subequations}  \label{eq:fsllbrrp}
\begin{align}
    T_y & = \frac{1}{\tau_y}\,,
        \qquad\qquad\!%
    \bar{T}_y = - \frac{T_y}{c^2}\,, 
    	\qquad%
    \bar{T}_m = \bar{\tau}_m - \frac{M_{my}}{c^2 \tau_y} + \frac{\tau_m}{\tilde{c}^{\,2}}\,,  \\[2pt]
    T_m & = \frac{M_{my}}{\tau_y}\,,
        \qquad%
    \tilde{E}_{mn} = E_{mn}\,,
        \qquad%
    \tilde{E}_{My} = 0\,, 
    	\qquad\,\,\,%
    \tilde{\Phi} = \Phi - \ln \tau_y\,, \\[4pt]
    \tilde{M}_{my} & = \frac{\tau_m}{\tau_y}\,, 
    	\qquad\,\,\,%
    \tilde{M}_{mn} = M_{mn} - \frac{M_{my} \, \tau_n - M_{ny} \, \tau_m}{\tau_y}\,,
\end{align}
\end{subequations}
Furthermore, taking the double scaling limit $c\,, \tilde{c} \rightarrow \infty$ of \eqref{eq:fsllbrrp} reproduces the expressions in \eqref{eq:relsncbuscherll} with a fixed Stueckelberg symmetry (see \S\ref{sec:nrsl}).

Similarly, we parametrize the dual RR potential in relativistic string theory as 
\be \label{eq:llr1}
    \hat{C}'{}^{(q)} = \tilde{c}^{\,2} \, \tilde{N}^{(q-2)} \wedge \tilde{\ell} + \tilde{N}^{(q)}\,,
        \qquad%
    \hat{C}^{\prime(q)}_y = \tilde{c}^{\,2} \, \bigl( \tilde{N}^{(q-2)}_y \wedge \tilde{\ell} + \tilde{N}^{(q-2)} \wedge \tilde{\ell}_y \bigr) + \tilde{N}^{(q)}_y\,,
\ee
where $\tilde{N}^{(q)}$ and $\tilde{N}^{(q)}_y$ are dual RR potentials in \g string theory. Note that 
\be
    \tilde{\ell} = \tfrac{1}{2} \, \bar{T} \wedge T\,,
        \qquad%
    \tilde{\ell}_y = \tfrac{1}{2} \bigl( \bar{T} \, T_y - \bar{T}_y \, T \bigr),
        \qquad%
    T = T_\alpha \, dY^\alpha\,,
        \qquad
    \bar{T} = \bar{T}_\alpha \, dY^\alpha\,.
\ee
To facilitate the calculation, we first rewrite $\bar{T}_m$\,, $T_m$\,, $\tilde{M}_{my}$ and $\tilde{M}_{mn}$ from \eqref{eq:fsllbrrp} in terms of differential forms, with 
\be \label{eq:llr2}
     T = \frac{M_y}{\tau_y}\,,
        \qquad%
    \bar{T} = \bar{\tau} - \frac{M_{y}}{c^2 \, \tau_y} + \frac{\tau}{\tilde{c}^{\, 2}}\,,
        \qquad%
    \tilde{M_y} = \frac{\tau}{\tau_y}\,,
        \qquad%
    \tilde{M} = M - \frac{M_y \wedge \tau}{\tau_y}\,.
\ee
%
Plugging the above ingredients into the relativistic RR Buscher rules \eqref{eq:relbuscherrr} gives
\begin{subequations}
\begin{align}
\begin{split}
    & \quad \frac{\tilde{c}^{\,2}}{2 \, \tau_y} \lr \tilde{N}_y^{(q-2)} \wedge \bar{\tau} \wedge M_y + \tilde{N}^{(q-2)} \wedge \bar{\tau} \rr + \tilde{N}_y^{(q)} + \frac{\tilde{N}_y^{(q-2)} \wedge \tau \wedge M_y + \tilde{N}^{(q-2)} \wedge \tau}{2 \, \tau_y} \\[2pt]
    & = \frac{\tilde{c}^{\,2} \, N_y^{(q-1)} \wedge \bar{\tau}}{2 \, \tau_y} + N^{(q-1)} - \frac{N_y^{(q-1)} \wedge \tau}{2 \, \tau_y}\,, 
\end{split}
\end{align}
and
\begin{align}
\begin{split}
    & \quad \frac{\tilde{c}^{\,2} \, \tilde{N}^{(q-2)} \wedge \bar{\tau} \wedge M_y}{2 \,\tau_y} + \tilde{N}^{(q)} + \frac{\tilde{N}^{(q-2)} \wedge \tau \wedge M_y}{2 \, \tau_y} \\[2pt]
    & = - \frac{\tilde{c}^{\,2} \, N_y^{(q-1)} \wedge M_y \wedge \bar{\tau}}{2 \tau_y} + N_y^{(q+1)} + N^{(q-1)} \wedge M_y + \frac{N_y^{(q-1)} \wedge M_y \wedge \tau}{2 \, \tau_y}\,.
\end{split}
\end{align}
\end{subequations}
Solving for $\tilde{N}^{(q)}_y$ and $\tilde{N}^{(q)}$, we find
\begin{subequations}
\begin{align}
    \tilde{N}^{(q)}_y & = N^{(q-1)} \!- \frac{N^{(q-1)}_y \wedge \tau}{\tau_y}\,, \\[4pt]
    \tilde{N}^{(q)} & = N^{(q+1)}_y \!+ N^{(q-1)} \wedge M_y + \frac{N^{(q-1)}_y \wedge M_y \wedge \tau}{\tau_y}\,,
\end{align}
\end{subequations}
reproducing the Buscher rules in \eqref{eq:llrrbr} with a fixed Stueckelberg symmetry.

\vspace{3mm}

\noindent $\bullet$ \textbf{Transverse T-duality.}
In the presence of the transverse Killing vector $k^M$ as specified in \eqref{eq:tkv}, we learned from \S\ref{sec:ttd} that the T-duality transformation along the associated spacelike isometry in the transverse sector relates nonrelativistic string theories compactified over dual spacelike circles. The Buscher rules for transverse T-duality have been shown in \cite{Bergshoeff:2018yvt} to arise from taking the nonrelativistic string limit of the relativistic Buscher rules in \eqref{eq:relbuscher}. This is done by first rewriting the background fields as in \eqref{eq:expbfc} and similarly for the dual fields as
\begin{align} \label{eq:bdpexp}
    \hat{G}'^{\phantom{(}}_{MN} & = c^2 \, \tau^{\phantom{(}}_{MN} + \tilde{H}^{\phantom{(}}_{MN}\,,     
        &%
    \hat{B}'_{MN} & = - c^2 \, \ell_{MN} + \tilde{B}_{MN}\,, 
        &%
    \hat{\Phi}' & = \tilde{\Phi} + \ln c\,,  
        &%
\end{align}
followed by plugging \eqref{eq:expbfc} and \eqref{eq:bdpexp} into the relativistic Buscher rules in \eqref{eq:relbuscher}. Note that the longitudinal vielbein field $\tau_\mu{}^A$ does not transform under the transverse T-duality. The same procedure can also be applied to the RR potentials. In analog to \eqref{eq:expbfc}, we parametrize the dual relativistic RR potentials as
\be \label{eq:bdpexpc}
    \hat{C}'{}^{(q)} = c^2 \, \tilde{C}^{(q-2)} \wedge \ell + \tilde{C}^{(q)}\,,
        \qquad%
    \hat{C}{}^{\prime(q)}_y = c^2 \, \tilde{C}^{(q-2)}_y \wedge \ell + \tilde{C}^{(q)}_y\,.
\ee
While $\hat{C}^{(q)}_{} = \tilde{C}^{(q+1)}_y = 0$ for $q < 0$\,. Plugging \eqref{eq:expbfc}, \eqref{eq:bdpexp} and \eqref{eq:bdpexpc} into the Buscher rules \eqref{eq:relbuscherrr} for RR potentials in relativistic string theory, and noting that $\tau_y{}^A = 0$ in the presence of the transverse Killing vector $k^M$\,, we find 
\begin{subequations}
\begin{align}
\begin{split}
    c^2 \, \tilde{C}^{(q-1)}_y \wedge \ell + \tilde{C}^{(q+1)}_y 
        %
        %
    & = c^2 \lr C^{(q-2)} - \frac{C^{(q-2)}_y \wedge H_y}{H_{yy}} \rr \wedge \ell + C^{(q)} - \frac{C^{(q)}_y \wedge H_y}{H_{yy}}\,, 
\end{split}
\end{align}
and
\begin{align}
\begin{split}
    c^2 \, \tilde{C}^{(q-2)} \wedge \ell + \tilde{C}^{(q)} & = c^2 \lr C^{(q-1)}_y + \frac{C^{(q-3)} \wedge H_y + C^{(q-2)}_y \wedge H_y \wedge B_y}{H_{yy}} \rr \wedge \ell \\[2pt]
    & \quad + C^{(q+1)}_y + \frac{ C^{(q-1)} \wedge H_y}{H_{yy}} + \frac{ C^{(q-1)}_y  \wedge H_y \wedge B_y}{H_{yy}}\,.
\end{split}
\end{align}
\end{subequations}
Solving for $\tilde{C}^{(q)}$ reproduces the RR Buscher rules \eqref{eq:tbrrrp} under the transverse T-duality. 
%

\section{Conclusions}
\label{sec:conclusions}

In this paper, we generalized the worldvolume actions that describe D-branes in nonrelativistic string theory by including RR potentials. Using nonrelativistic D-branes as probes, we initiated a systematic classification of duality transformations for D-brane actions with diverse worldvolume dimensions in nonrelativistic string theory. This study uncovers a class of nonrelativistic duality transformations that are distinct in nature from the ones in relativistic string theory, and lead to novel dual D-brane actions. These results are further corroborated by carefully performing the stringy and membrane limits of relativistic string and M-theory, respectively. Such limits involve nontrivial cancellations among the background metric, Kalb-Ramond and RR fields and elegantly reproduce the finite dual D-brane actions that we found from the first principles method. A general $p$-brane limit of the associated relativistic D$p$-brane action has been considered in Appendix \ref{app:dpbpbncg}, leading to other corners of relativistic string/M-theory that exhibit nonrelativistic behaviors. 

While our analyses focused on the bosonic sector, we have not addressed the structure of the fermionic sector for the full supersymmetric \g brane action propagating on a supergravity background that generalizes the string Newton-Cartan geometry.~\footnote{See \cite{Bergshoeff:2021tfn} for a supersymmetrization of string Newton-Cartan geometry.} For example, it is important to understand how to incorporate local kappa symmetry for \g D-actions, which arise in the nonrelativistic string limit of the kappa-symmetric relativistic D-brane constructed in \cite{Aganagic:1996pe, Aganagic:1996nn}. Also see \cite{Gomis:2004pw, Gomis:2005bj, Kamimura:2005rz, Gomis:2005pg} for previous works on the kappa-symmetric nonrelativistic $p$-brane action that arises as the $p$-brane limit of the relativistic $p$-brane. The supersymmetric generalization of the $p$-brane limit in flat spacetime is studied in \cite{Gomis:2004pw}, and its curved-spacetime generalization is recently introduced in \cite{Bergshoeff:2021tfn}.







We have studied the duality transformation of the nonrelativistic D4-brane by dualizing the worldvolume $U(1)$ gauge field. This leads to the nonrelativistic M5-brane in ten-dimensional membrane Newton-Cartan geometry. Nevertheless, future work is still required for attaining a closed form of the nonrelativistic M5-brane action. Moreover, it would be highly interesting to apply the techniques developed in this paper to dualize the worldvolume $U(1)$ gauge field for D$p$-branes with $p>4$\,. For example, the nonrelativistic NS5-brane action can found from S-dualizing the nonrelativistic D5-brane action (or, from a direct dimensional reduction of the nonrelativistic M5-brane). 
Along these lines, a hierarchy of nonrelativistic D$p$-brane actions and their duals can be built. This would also generalize the atlas of relativistic exotic branes to their nonrelativistic counterparts.  

There are also numerous other future directions for which the concepts and techniques derived in this paper can be useful. First, the stringy limit has been applied to ten-dimensional heterotic superstring theory in \cite{Bergshoeff:2021tfn}, where there are no RR potentials. It would be fascinating to apply the stringy limit proposed in this paper to Type I and II supergravity in ten-dimensions, where RR potentials are present, and compare with the membrane limit of eleven-dimensional supergravity that has been studied in \cite{Blair:2021ycc}. This would also make it possible to construct an S-dual invariant nonrelativistic Type IIB supergravity action. 
Moreover, the inclusion of RR potentials forms an essential step for looking for possible black hole-like solutions to the nonrelativistic supergravity equations of motion, as well as a top-down construction of nonrelativistic holography. 
Secondly, one may also add a cosmological constant term (or a total derivative term) as \cite{Green:1996bh, Hassan:1999bv} did for massive IIA supergravity theory. It would be interesting to examine how this modifies \g brane actions and T-duality transformations. Finally, we only considered the low-energy effective action of \g D-branes. The full D-brane dynamics are characterized by Witten's cubic bosonic open string field theory \cite{Witten:1985cc}, while the (non-abelian) Born-Infeld action arises from integrating out all the massive modes in the string field theory. It would be interesting to understand how the analog of open string field theory can be formulated in nonrelativistic string theory. 


\acknowledgments

We would like to thank Eric Bergshoeff, Jaume Gomis, Niels Obers, Gerben Oling, Peter Schupp and Matthew Yu for useful discussions.
We also thank the organizers and participants at the \emph{First School on Non-relativistic Quantum Field Theory, Gravity, and Geometry} (August 23--27, 2021) for stimulating discussions, where part of this paper's results were first presented. S.E. is supported from the Bhaumik Institute. H.-Y.S. is supported in part by the Simons Collaborations on Ultra-Quantum Matter, a grant No.~651440 (A.K.) from the Simons Foundation. Nordita is supported in part by NordForsk.


\newpage

\appendix

\section{D\texorpdfstring{$p$}{p}-Branes in a \texorpdfstring{$p$}{p}-Brane Newton-Cartan Geometry} \label{app:dpbpbncg}  

In \S\ref{sec:drm2}, we discussed both double and direct dimensional reductions of \g M2-branes along a longitudinal spacelike isometry. These \g M2-branes are coupled to an eleven-dimensional membrane Newton-Cartan geometry equipped with a co-dimensional three foliation. These reductions lead to the actions that respectively describe fundamental strings and D2-branes in nonrelativistic string theory, where the spacetime geometry is string Newton-Cartan geometry equipped with a codimension-two foliation. In this appendix, we consider a direct dimensional reduction of the same nonrelativistic M2-brane action but now along a transverse isometry \cite{Kluson:2019uza}. We first review how this procedure gives rise to a \g D2-brane action that is coupled to a ten-dimensional membrane Newton-Cartan geometry, which inherits the three-dimensional foliation structure from nonrelativistic M-theory. We then propose a membrane limit of relativistic string theory that leads to the same nonrelativistic D2-brane action. One may then use such D2-branes as probes for understanding the membrane limit of relativistic string theory. The corners of the string and membrane limits of relativistic string theory are in this sense unified under the notion of nonrelativistic M-theory. We then discuss general $p$-brane limits of relativistic string theory \cite{Gomis:2000bd}.

We begin with a review of the transverse dimensional reduction of \g M2-brane \cite{Kluson:2019uza}. 
Consider the M2-brane action \eqref{eq:sdualm2} in the most general form,
\be \label{eq:sm2gf}
    S_\text{M2} = - \frac{1}{2} \int d^3 Y \sqrt{-\gamma'} \, {\gamma'}^{\mu\nu} \, H'_{\mu\nu} - \int A^{\prime(3)}\,,
\ee
where 
\begin{subequations}
\begin{align}
    H'_{\mu\nu} & = \p_\mu X^\CI \, \p_\nu X^\CJ \, H'_{\CI\CJ}\,,
        &%
    A^{\prime(3)} & = \p_\mu X^\CI \, \p_\nu X^\CJ \, \p_\rho X^\CK \, A^{\prime(3)}_{\CI\CJ\CK}\,,
        &%
    \CI & = 0, 1, \cdots, 10\,, \\[4pt]
    \gamma'_{\mu\nu} & = \gamma'_\mu{}^u \, \gamma'_\nu{}^v \, \eta_{uv}\,,
        &%
    \gamma'_\mu{}^u & = \p_\mu X^\CI \, \gamma'_\CI{}^u\,, 
        &%
    u & = 0,1,2\,.
\end{align}
\end{subequations}
Here, $X^\CI$ are coordinates of the eleven-dimensional target space.
Unlike $\tilde{H}_{\CI\CJ}$ in \eqref{eq:tha3def}, where the components that contain the tenth spatial index are set to zero, now, all entries in $H'_{\mu\nu}$ can be nonzero. We require that the M2-brane is localized in the transverse isometry direction $\Theta = X^{10}$, which we compactify over a circle of radius $R_{10}$\,. Gauging the isometry by introducing a pure gauge field $v_\mu$\,, we write the gauged form of \eqref{eq:sm2gf} as
\be \label{eq:sm2gf2}
    S_\text{gauged} = - \frac{1}{2} \int d^3 Y \sqrt{-\gamma'} \, \gamma'{}^{\mu\nu} \, H'_{\mu\nu} - \int \lr A^{\prime(3)} + v \wedge F \rr.
\ee
We take the following Kaluza-Klein reduction ansatz:
\begin{subequations}
\begin{align}
        %
    H'_{\mu\nu} & = e^{-2\Phi/3} \ls H_{\mu\nu} + e^{2\Phi} \lr C_\mu^{(1)} + D_\mu \Theta \rr \lr C_\nu^{(1)} + D_\nu \Theta \rr \rs\,, \\[2pt]
    \gamma'_{\mu\nu} & = e^{-2\Phi/3} \, \gamma_{\mu\nu}\,, 
        \qquad%
    A^{\prime(3)} = - C^{(3)} + B \wedge D\Theta\,, 
        \qquad%
    D_\mu \Theta = \p_\mu \Theta + v_\mu\,,
\end{align}
\end{subequations}
where $\gamma_{\mu\nu} = \gamma_\mu{}^u \, \gamma_\nu{}^v \, \eta^{}_{uv}$ is a rank-three matrix. In terms of the one-form $V = C^{(1)} + D\Theta$\,, we write \eqref{eq:sm2gf2} as
\begin{align}
    S_\text{gauged} & = - \frac{1}{2} \int d^3Y \, e^{-\Phi} \sqrt{-\gamma} \, \gamma^{\mu\nu} \lr H_{\mu\nu} + e^{2\Phi} \, V_\mu \, V_\nu \rr
    + \! \int \! \lr C^{(3)} + \CF \wedge C^{(1)} - \CF \wedge V \rr\!.
\end{align}
Integrating out $V_\mu$ in the path integral leads to the dimensionally reduced action
\begin{align} \label{eq:gd2p}
    S'_\text{D2} = - \frac{1}{2} \int d^3 Y \, e^{-\Phi} \sqrt{-\gamma} \, \Bigl( \gamma^{\mu\nu} \, H_{\mu\nu} + \tfrac{1}{2} \, \gamma^{\mu\nu} \, \gamma^{\rho\sigma} \, \CF_{\mu\rho} \, \CF_{\nu\sigma} \Bigr) + \int \lr C^{(3)} + \CF  \wedge C^{(1)} \rr.
\end{align}
This action reproduces the one in \cite{Kluson:2019uza} and describes Galilean D2-branes~\footnote{See \cite{Gomis:2000bd} for the origin of the terminology ``Galilean D$p$-brane."} coupled to ten-dimensional membrane Newton-Cartan geometry, which has a codimension-three foliation. This theory is different in nature from \eqref{eq:gd2b} that describes \g D2-branes coupled to string Newton-Cartan geometry. Upon performing an S-duality transformation on the D2-brane action \eqref{eq:gd2p} by following the same procedure detailed in \S\ref{sec:nmfgd2b}, the original M2-brane action \eqref{eq:sm2gf} is recovered.

In a flat limit with $\gamma_{\mu\nu} \!=\!\text{diag} (-1, 1, 1)$, $H_{\mu\nu} \!=\! \p_\mu \pi^{u'} \, \p_\nu \pi^{u'}$ and $B \!=\! C^{(1)} \!\!=\! C^{(3)}\! \!=\! 0$\,, the D2-brane is orthogonal to the transverse directions and extends in the longitudinal directions. Here, $\pi^{u'}$ are Nambu-Goldstone bosons that perturb the shape of the D2-brane in the transverse directions, with $u' = 3, \cdots, 9$\,. We also assume that $\Phi = \Phi_0$ is a constant, which determines the string coupling $g_s \equiv e^{\Phi_0} = g^2_\text{YM}$\,, with $g^{}_\text{YM}$ being the Yang-Mills coupling. At the quadratic order in field perturbations, the DBI action \eqref{eq:gd2p} gives
\be
    S^{\prime(2)}_\text{D2} = - \frac{1}{g^2_\text{YM}} \int d^3 Y \, \lr \tfrac{1}{4} \, F_{\mu\nu} \, F^{\mu\nu} + \tfrac{1}{2} \, \p_\mu \pi^{u'} \, \p^\mu \pi^{u'} \rr. 
\ee
This quadratic action is relativistic. It is also possible to consider Galilean D2-branes transverse to one or both of the longitudinal spatial directions, in which case a nontrivial geometry background is required for the effective gauge theory to be well defined.
%
%
        %
            %
            %
        %
        %



It is also possible to derive the same D2-brane action \eqref{eq:gd2p} as a limit of the action \eqref{eq:relsd2} describing D2-branes in relativistic string theory. We start with the following ansatz:
\begin{subequations} \label{eq:gd2mngl}
\begin{align}
    \hat{G}_{MN} & = c^{4/3} \, \gamma_{MN} + c^{-2/3} \, H_{MN}\,,
        &%
    \hat{C}^{(1)} & = c^{-1/3} \, C^{(1)}\,, 
        \qquad%
    \hat{\Phi} = \Phi\,, \\[2pt]
    \hat{\CF}_{MN} & = c^{1/3} \, \CF_{MN}\,,     
        &%
    \hat{C}^{(3)}_{MNL} & = c^2 \, e^{-\Phi} \, \gamma^{}_M{}^u \, \gamma^{}_N{}^v \, \gamma^{}_L{}^w \, \epsilon^{}_{uvw} + C^{(3)}_{MNL}\,.
\end{align}
\end{subequations}
Plugging the above ansatz into \eqref{eq:relsd2}, which we transcribe as
\be 
    \hat{S}_\text{D2} = - \int d^3 Y \, e^{-\hat{\Phi}} \sqrt{-\det \lr \hat{G}_{\mu\nu} + \hat{\CF}_{\mu\nu} \rr} + \int \lr \hat{C}^{(3)} + \hat{C}^{(1)} \wedge \hat{\CF} \rr\,,  
\ee
we find that
\begin{subequations}
\begin{align}
    e^{-\hat{\Phi}} \sqrt{-\det \lr \hat{G}_{\mu\nu} + \hat{\CF}_{\mu\nu} \rr} & = e^{-\Phi} \sqrt{-\gamma} \, \Bigl( c^2 + \tfrac{1}{2} \, \gamma^{\mu\nu} \, H_{\mu\nu} + \tfrac{1}{4} \, \CF^{\mu\nu} \CF_{\mu\nu} \Bigr) + O(c^{-2})\,, \\[2pt]
    \frac{1}{3!} \, \epsilon^{\mu\nu\rho} \lr \hat{C}^{(3)}_{\mu\nu\rho} + 3 \, \hat{C}^{(1)}_\mu \, \hat{\CF}_{\nu\rho} \rr & = c^2 \, e^{-\Phi} \, \sqrt{-\gamma} + \frac{1}{3!} \, \epsilon^{\mu\nu\rho} \lr {C}^{(3)}_{\mu\nu\rho} + 3 \, {C}^{(1)}_\mu \, {\CF}_{\nu\rho} \rr,
\end{align}
\end{subequations}
and thus \eqref{eq:gd2p} is recovered in the $c \rightarrow \infty$ limit. This membrane limit of relativistic string theory generalizes the one initially considered in \cite{Gomis:2000bd}.

To further continue the study of this sector that is defined from the $c \rightarrow \infty$ limit with the prescriptions given in \eqref{eq:gd2mngl}, it would be useful to understand whether fundamental strings can be defined. 
It would also be intriguing to consider T-duality transformations of the action \eqref{eq:gd2p} and look for a notion of general D$p$-brane actions in ten-dimensional membrane Newton-Cartan geometry. In particular, if one could make sense of a D1-string action in ten-dimensional membrane Newton-Cartan geometry, it would be possible to study the associated fundamental strings by performing an S-duality transformation of D1-strings. We will leave the studies of extended objects other than Galilean D2-branes in membrane Newton-Cartan geometry to the future. 

Finally, it is also possible to generalize the two-brane limit of the relativistic D2-brane action, defined by the prescriptions in \eqref{eq:gd2mngl}, to other $p$-brane limits of the associated relativistic D$p$-brane action. We start with the D$p$-brane action \eqref{sec:hspf} in relativistic string theory, 
which we transcribe below:
\be \label{eq:relsdpz}
    \hat{S}_{\text{D}p} = - \int d^{p+1} Y \, e^{-\hat{\Phi}} \sqrt{-\det \lr \hat{G}_{\mu\nu} + \hat{\CF}_{\mu\nu} \rr} + \int \sum_q \hat{C}^{(q)} \wedge e^{\hat{\CF}} \bigg|_{p+1}.
\ee
Consider the following ansatz that generalizes \eqref{eq:gd2mngl}:
\begin{align}
    \hat{G}_{MN} & = c^{2} \, \gamma_{MN} + c^{1-p} \, H_{MN}\,,
        &%
    \hat{\CF}_{MN} & = c^{(3-p)/2} \, \CF_{MN}\,, 
        &%
    \hat{\Phi} & = \Phi\,,
\end{align}
and
\begin{subequations} \label{eq:cqdpexp}
\begin{align}
    \hat{C}^{(p+1)}_{M_0 \cdots M_{p}} & = c^{p+1} \, e^{-\Phi} \, \gamma^{}_{M_0}{}^{u_0} \cdots \gamma^{}_{M_p}{}^{u_p} \, \epsilon_{u_0 \cdots u_p} + C^{(p+1)}_{M_0\cdots M_p}\,, \\[4pt]
    \hat{C}^{(q)} & = c^{\frac{1}{4}(p-3)(p-q+1)} \, C^{(q)}\,,
        \qquad%
    q < p+1\,.
\end{align}
\end{subequations}
When $p = 2$\,, upon redefining $c \rightarrow c^{2/3}$\,, the ansatz \eqref{eq:gd2mngl} is recovered. Note that the parametrizations for $\hat{G}_{MN}$ and $\hat{C}^{(p+1)}$ match \eqref{eq:pbexp} for the $p$-brane limit (up to a rescaling of $\hat{C}^{(p+1)}$ before identifying it with $\hat{A}^{(p+1)}$ in \eqref{eq:pbexp}). The $c \rightarrow \infty$ limit of \eqref{eq:relsdpz} gives
\begin{align} \label{eq:gdpp}
    S'_{\text{D}p} = - \frac{1}{2} \int d^{p+1} Y \, e^{-\Phi} \sqrt{-\gamma} \, \Bigl( \gamma^{\mu\nu} \, H_{\mu\nu} + \tfrac{1}{2} \, \gamma^{\mu\nu} \, \gamma^{\rho\sigma} \, \CF_{\mu\rho} \, \CF_{\nu\sigma} \Bigr) + \int \sum_q C^{(q)} \wedge e^{\CF} \bigg|_{p+1}\,,
\end{align}
which describes the so-called Galilean D$p$-brane first proposed in \cite{Gomis:2000bd}. A Galilean D$p$-brane is coupled to a ten-dimensional $p$-brane Newton-Cartan geometry. 
It is then a straightforward exercise to dualize the $U(1)$ gauge field in \eqref{eq:gdpp} and derive various actions that describe S-dual objects in Type IIB or membrane configurations in M-theory. This may improve our understanding of $p$-brane limits of relativistic string/M-theory. Moreover, it would be intriguing to derive the Buscher rules associated with T-duality transformations under the $p$-brane limits. This will help us understand how to define extended objects other than the D$p$-branes in \eqref{eq:gdpp} within these corners.

\section{Stueckelberg Transformations of Buscher Rules for RR Potentials}
\label{app:stbr}

In this appendix, we apply the infinitesimal version of the Stueckelberg transformations \eqref{eq:ss} and \eqref{eq:ssC} to the Buscher rules containing RR potentials. The infinitesimal Stueckelberg transformations of $B_{\mu\nu}$\,, $H_{\mu\nu}$ and $C^{(q)}$ are
\begin{subequations}  \label{eq:strep}
\begin{align} 
    \delta_\xi H^{}_{MN} & = - \lr \tau^{}_M{}^A \, \xi^{}_N{}^B + \tau^{}_N{}^A \, \xi^{}_M{}^B \rr \eta^{}_{AB}\,, 
        &%
    \delta_\xi C^{(q)} & = - C^{(q-2)} \wedge \delta_\xi B\,,
        \quad%
    q \geq 2\,, \\[2pt]
    \delta_\xi B^{}_{MN} & = \lr \tau^{}_M{}^A \, \xi^{}_N{}^B - \tau^{}_N{}^A \, \xi^{}_M{}^B \rr \epsilon^{}_{AB}\,,
        &%
    \delta_\xi C^{(0)} & = \delta_\xi C^{(1)} = 0\,.
\end{align}
\end{subequations}
In the following, we analyze how the Buscher rules \eqref{eq:lscb}, \eqref{eq:lltdrep} and \eqref{eq:ttdrep} for RR-potentials transform under these Stueckelberg symmetries. This serves as an extra check of the Buscher rules derived in \S\ref{sec:T-duality}. 

\vspace{3mm}

\noindent $\bullet$ \textbf{Longitudinal spacelike T-duality.} For the T-duality transformation along a longitudinal spatial isometry that we denote by $y$\,, 
the associated Buscher rules are given in \eqref{eq:lscb}, with 
\begin{subequations} \label{eq:cqtdual}
\begin{align} 
	\tilde{C}^{(q)}_y & = C^{(q-1)} - \frac{C^{(q-1)}_y \wedge \tau_y^{(1)}}{\tau_{yy}} + \frac{C^{(q-3)} \wedge \lr H_y \wedge \ell_y + H_{yy} \, \ell \rr}{\tau_{yy}} - \frac{C^{(q-3)}_y \wedge H_y \wedge \ell}{\tau_{yy}}\,, \label{eq:cqyrep} \\[4pt]
	\tilde{C}^{(q)} & = C^{(q+1)}_y + C^{(q-1)} \wedge B_y + \frac{C^{(q-1)}_y \wedge \lr B_y \wedge \tau_y^{(1)} + H_y \wedge \ell_y + H_{yy} \, \ell \rr}{\tau_{yy}} \notag \\[2pt]
	& \hspace{1.65cm} + \frac{C^{(q-3)} \wedge B_y \wedge \lr H_y \wedge \ell_y + H_{yy} \, \ell \rr}{\tau_{yy}} + \frac{C^{(q-3)}_y \wedge B_y \wedge H_y \wedge \ell}{\tau_{yy}}\,. \label{eq:cqrep}
\end{align}
\end{subequations}
%
Note that $\delta_\xi \tilde{C}^{(q)} = \delta_\xi \tilde{C}^{(q)}_y = 0$ because $\tilde{C}^{(q)}$ and $\tilde{C}^{(q)}_y$ are RR potentials in the DLCQ of relativistic string theory. Therefore, the RHS of both equations in \eqref{eq:cqtdual} must vanish under \eqref{eq:strep}. 
We start with analyzing \eqref{eq:cqyrep} and varying with respect to \eqref{eq:strep} gives 
\begin{align} \label{eq:LSsimplified1}
\begin{split}
    \delta_\xi \tilde{C}^{(q)}_y & = - C^{(q-3)} \wedge \delta_\xi \biggl( B - \frac{B_y \wedge \tau_y^{(1)} + H_y \wedge \ell_y + H_{yy} \, \ell}{\tau_{yy}} \biggr) \\[2pt]
    & \quad + \frac{1}{\tau_{yy}} \, C^{(q-3)}_y \wedge \Bigl( \delta_\xi B \wedge \tau_y^{(1)} + \delta_\xi H_y \Bigr) + \frac{1}{\tau_{yy}} \, C^{(q-5)}_y \wedge \delta_\xi B \wedge H_y \wedge \ell \\[2pt]
    & \quad - \frac{1}{\tau_{yy}} \, C^{(q-5)} \wedge \Bigl[ \delta_\xi B \wedge \bigl( H_y \wedge \ell_y + H_{yy} \, \ell \bigr) - \delta_\xi B_y \wedge H_y \wedge \ell \Bigr]\,.
\end{split}
\end{align}
The terms proportional to $C^{(q-3)}$ and $C^{(q-5)}_y$ vanish due to
\begin{equation} \label{eq:dxitb}
    \delta_\xi \tilde B = B - \frac{B_y \wedge \tau^{(1)}_y + H_y \wedge \ell_y + H_{yy} \, \ell}{\tau_{yy}} = 0, 
        \qquad%
    \delta_\xi B \wedge \ell = 0\,,
\end{equation}
where $\tilde{B}$ is the T-dual Kalb-Ramond field in \eqref{lstdcomponents}.
%
%
%
        %
%
Furthermore, we write the components of \eqref{eq:strep} that are relevant to \eqref{eq:LSsimplified1} in terms of differential forms as 
\begin{subequations} \label{eq:DifferentialFormOfSection2}
\begin{align}
    \delta_\xi H_y = - \left( \tau^A \, \xi_y{}^B + \tau_y{}^A \, \xi^B \right) \eta_{AB} \,, 
        \qquad%
    & \delta_\xi B = \tau^A \wedge \xi^B \epsilon_{AB}\,, \\[2pt] 
    & \delta_\xi B_y = \left( \tau^A \, \xi^B_y - \tau^A_y \, \xi^B \right) \epsilon_{AB}\,, 
\end{align}
\end{subequations}
where $\xi^A = \xi_m{}^A \, dY^m$\,.
Substituting \eqref{eq:DifferentialFormOfSection2} into \eqref{eq:LSsimplified1}, and using the identities, 
%
\begin{align} \label{eq:epsilionidentity}
    \epsilon^{}_{AB} \, \eta^{}_{CD} =  - \epsilon^{}_{CA} \, \eta^{}_{BD} - \epsilon^{}_{BC} \, \eta_{AD}\,,
        \qquad%
    \epsilon^{}_{AB} \, \epsilon^{}_{CD} = \eta^{}_{AD} \, \eta^{}_{BC} - \eta^{}_{AC} \, \eta^{}_{BD}\,,
\end{align}
%
%
%
%
we find that the $C^{(q-3)}_y$ and $C^{(q-5)}$ terms also vanish which is consistent with $\delta_\xi \tilde{C}^{(q)}_y = 0$\,.
%
%

Next, we prove the invariance of $\tilde{C}^{(q)}_y$ under \eqref{eq:strep}. Varying the RHS of \eqref{eq:cqrep} with respect to \eqref{eq:strep}, and applying \eqref{eq:dxitb}, we find
%
        %
%
%
%
\begin{equation}
    \begin{aligned} \label{LSContributions}
\delta_{\xi} \tilde{C}^{(q)} & = \frac{1}{\tau_{yy}} \, C_{y}^{(q-3)} \wedge \ls \delta_{\xi} B \wedge \Bigl( \ell_y \wedge H_y - B_y \wedge \tau_y^{(1)} \Bigr) 
+ \Bigl( \delta_{\xi} B_{y} \wedge H_{y} + B_{y} \wedge \delta_{\xi} H_{y} \Bigr) \wedge \ell \rs \\[2pt]
    & \quad - \frac{1}{\tau_{yy}} \, C^{(q-5)} \wedge B_y \wedge H_y \wedge \lr \delta_\xi B \wedge \ell_y + \delta_\xi B_y \wedge \ell \rr. 
\end{aligned}
\end{equation}
These contributions vanish upon substituting \eqref{eq:DifferentialFormOfSection2} directly into \eqref{LSContributions} and using 
\eqref{eq:epsilionidentity}. 
%
%
%
%

\vspace{3mm}

\noindent $\bullet$ \textbf{Longitudinal lightlike T-duality.} The Buscher rules of the RR potentials along a longitudinal lightlike isometry $y$ are given in \eqref{eq:llrrbr}, with
\begin{subequations}  \label{eq:lltdrep}
\begin{align}
    \tilde{C}^{(q)}_y & = C^{(q-1)} - \frac{C^{(q-1)}_y \wedge \tau}{\tau_y}\,, \\[2pt]
    \tilde{C}^{(q)} & = C^{(q+1)}_y + C^{(q-1)} \wedge B_y + \frac{C^{(q-1)}_y \wedge B_y \wedge \tau}{\tau_y}\,.
\end{align}
\end{subequations}
We use \eqref{eq:ssC} to find that
\begin{subequations} \label{eq:Stueckelbergll} 
\begin{align} 
    \delta_\xi \tilde{C}^{(q)}_y = -C^{(q-3)} \wedge\left(\delta_{\xi} B-\frac{\delta_{\xi} B_{y} \wedge \tau}{\tau_{y}}\right) & +C_{y}^{(q-3)} \wedge \frac{\delta_{\xi} B \wedge \tau}{\tau_{y}}\,, \\[2pt]
    \delta_\xi \tilde{C}^{(q)} =-C_{y}^{(q-1)} \wedge\left(\delta_{\xi} B-\frac{\delta_{\xi} B_{y} \wedge \tau}{\tau_{y}}\right) & -C^{(q-3)} \wedge B_{y} \wedge\left(\delta_{\xi} B -\frac{\delta_{\xi} B_{y} \wedge \tau}{\tau_{y}}\right) \notag \\[2pt]
    & -C_{y}^{(q-3)} \wedge \frac{\delta_{\xi} B \wedge B_{y} \wedge \tau}{\tau_{y}}\,.
\end{align}
\end{subequations}
%
Applying \eqref{eq:strep} and the Buscher rules \eqref{eq:relsncbuscherll}, \eqref{eq:Stueckelbergll} can be rewritten in terms of the T-dual fields, with
\begin{subequations} \label{LHSofStuckSymmetries}
\begin{align}
   \delta_\xi \tilde{C}^{(q)} &= - \tilde{C}^{(q-2)} \wedge \delta_\xi \tilde{B}\,, \\
      \delta_\xi \tilde{C}^{(q)}_y &= - \tilde{C}^{(q-2)} \wedge \delta_\xi \tilde{B}_y - \tilde{C}^{(q-2)}_y \wedge \delta_\xi \tilde{B}\,,
\end{align}
\end{subequations}
reproducing the Stueckelberg transformations of $\tilde{C}^{(q)}$ and $\tilde{C}^{(q)}_y$ in nonrelativistic string theory. 

\vspace{3mm}

\noindent $\bullet$ \textbf{ Transverse T-duality.} In the presence of a transverse isometry $y$\,, we record the RR potentials' Buscher rules \eqref{eq:tbrrrp} for convenience 
\begin{subequations} \label{eq:ttdrep} 
\begin{align}
    \tilde{C}^{(q)}_y & = C^{(q-1)} - \frac{C^{(q-1)}_y \wedge H_y}{H_{yy}}\,, \\[2pt]
    \tilde{C}^{(q)} &= C^{(q+1)}_y + C^{(q-1)} \wedge B_y + \frac{C^{(q-1)}_y \wedge B_y \wedge H_y}{H_{yy}}\,.
\end{align}
\end{subequations}
Using \eqref{eq:ssC}, we find
\begin{subequations} \label{eq:TransverseBuscherStueckelberg}
\begin{align} 
    \delta_{\xi} \tilde{C}_{y}^{(q)} & = -C^{(q-3)} \wedge \delta_{\xi} B+\frac{\lr C^{(q-3)} \wedge \delta_{\xi} B_{y} + C_{y}^{(q-3)} \wedge \delta_{\xi} B \rr \wedge H_{y} - C_{y}^{(q-1)} \wedge \delta_{\xi} H_{y}}{H_{y y}}\,, \\[4pt] 
    \delta_\xi \tilde{C}^{(q)} & = -C_{y}^{(q-1)} \wedge\left(\delta_{\xi} B-\frac{\delta_{\xi} B_{y} \wedge H_{y}+B_{y} \wedge \delta_{\xi} H_{y}}{H_{y y}}\right) \notag \\[2pt]
    & \quad -C^{(q-3)} \wedge\left(\delta_{\xi} B \wedge B_{y}+\frac{\delta_{\xi} B_{y} \wedge B_{y} \wedge H_{y}}{H_{y y}}\right) 
    +C_{y}^{(q-3)} \wedge \frac{\delta_{\xi} B \wedge B_{y} \wedge H_{y}}{H_{y y}}\,.
\end{align}
\end{subequations}
Applying \eqref{eq:strep} and the transverse Buscher rules \eqref{eq:ttdbr}, \eqref{eq:TransverseBuscherStueckelberg} can be rewritten in terms of the T-dual fields as in \eqref{LHSofStuckSymmetries}. This reproduces the Stueckelberg transformations of the T-dual RR potentials in nonrelativistic string theory. 

\newpage


\bibliographystyle{JHEP}
\bibliography{ddbanst}

\providecommand{\href}[2]{#2}\begingroup\raggedright\begin{thebibliography}{10}

\bibitem{Banks:1996vh}
T.~Banks, W.~Fischler, S.~H. Shenker and L.~Susskind, \emph{{M theory as a
  matrix model: A Conjecture}},
  \href{http://dx.doi.org/10.1103/PhysRevD.55.5112}{\emph{Phys. Rev. D} {\bf
  55} (1997) 5112--5128}, [\href{https://arxiv.org/abs/hep-th/9610043}{{\tt
  hep-th/9610043}}].

\bibitem{Susskind:1997cw}
L.~Susskind, \emph{{Another conjecture about M(atrix) theory}},
  \href{https://arxiv.org/abs/hep-th/9704080}{{\tt hep-th/9704080}}.

\bibitem{Seiberg:1997ad}
N.~Seiberg, \emph{{Why is the matrix model correct?}},
  \href{http://dx.doi.org/10.1103/PhysRevLett.79.3577}{\emph{Phys. Rev. Lett.}
  {\bf 79} (1997) 3577--3580},
  [\href{https://arxiv.org/abs/hep-th/9710009}{{\tt hep-th/9710009}}].

\bibitem{Sen:1997we}
A.~Sen, \emph{{D0-branes on T${}^{ n}$ and matrix theory}},
  \href{http://dx.doi.org/10.4310/ATMP.1998.v2.n1.a2}{\emph{Adv. Theor. Math.
  Phys.} {\bf 2} (1998) 51--59},
  [\href{https://arxiv.org/abs/hep-th/9709220}{{\tt hep-th/9709220}}].

\bibitem{Motl:1997th}
L.~Motl, \emph{{Proposals on nonperturbative superstring interactions}},
  \href{https://arxiv.org/abs/hep-th/9701025}{{\tt hep-th/9701025}}.

\bibitem{Banks:1996my}
T.~Banks and N.~Seiberg, \emph{{Strings from matrices}},
  \href{http://dx.doi.org/10.1016/S0550-3213(97)00278-2}{\emph{Nucl. Phys. B}
  {\bf 497} (1997) 41--55}, [\href{https://arxiv.org/abs/hep-th/9702187}{{\tt
  hep-th/9702187}}].

\bibitem{Dijkgraaf:1997vv}
R.~Dijkgraaf, E.~P. Verlinde and H.~L. Verlinde, \emph{{Matrix string theory}},
  \href{http://dx.doi.org/10.1016/S0550-3213(97)00326-X}{\emph{Nucl. Phys. B}
  {\bf 500} (1997) 43--61}, [\href{https://arxiv.org/abs/hep-th/9703030}{{\tt
  hep-th/9703030}}].

\bibitem{Gomis:2000bd}
J.~Gomis and H.~Ooguri, \emph{{Nonrelativistic closed string theory}},
  \href{http://dx.doi.org/10.1063/1.1372697}{\emph{J. Math. Phys.} {\bf 42}
  (2001) 3127--3151}, [\href{https://arxiv.org/abs/hep-th/0009181}{{\tt
  hep-th/0009181}}].

\bibitem{Klebanov:2000pp}
I.~R. Klebanov and J.~M. Maldacena, \emph{{(1+1)-dimensional NCOS and its U(N)
  gauge theory dual}},
  \href{http://dx.doi.org/10.1142/S0217751X01004001}{\emph{Adv. Theor. Math.
  Phys.} {\bf 4} (2000) 283--302},
  [\href{https://arxiv.org/abs/hep-th/0006085}{{\tt hep-th/0006085}}].

\bibitem{Danielsson:2000gi}
U.~H. Danielsson, A.~Guijosa and M.~Kruczenski, \emph{{IIA/B, wound and
  wrapped}}, \href{http://dx.doi.org/10.1088/1126-6708/2000/10/020}{\emph{JHEP}
  {\bf 10} (2000) 020}, [\href{https://arxiv.org/abs/hep-th/0009182}{{\tt
  hep-th/0009182}}].

\bibitem{Danielsson:2000mu}
U.~H. Danielsson, A.~Guijosa and M.~Kruczenski, \emph{{Newtonian gravitons and
  D-brane collective coordinates in wound string theory}},
  \href{http://dx.doi.org/10.1088/1126-6708/2001/03/041}{\emph{JHEP} {\bf 03}
  (2001) 041}, [\href{https://arxiv.org/abs/hep-th/0012183}{{\tt
  hep-th/0012183}}].

\bibitem{Andringa:2012uz}
R.~Andringa, E.~Bergshoeff, J.~Gomis and M.~de~Roo, \emph{{`Stringy'
  Newton-Cartan gravity}},
  \href{http://dx.doi.org/10.1088/0264-9381/29/23/235020}{\emph{Class. Quant.
  Grav.} {\bf 29} (2012) 235020},
  [\href{https://arxiv.org/abs/arXiv:1206.5176}{{\tt arXiv:1206.5176}}].

\bibitem{Bergshoeff:2018yvt}
E.~Bergshoeff, J.~Gomis and Z.~Yan, \emph{{Nonrelativistic String Theory and
  T-Duality}}, \href{http://dx.doi.org/10.1007/JHEP11(2018)133}{\emph{JHEP}
  {\bf 11} (2018) 133}, [\href{https://arxiv.org/abs/arXiv:1806.06071}{{\tt
  arXiv:1806.06071}}].

\bibitem{Harmark:2018cdl}
T.~Harmark, J.~Hartong, L.~Menculini, N.~A. Obers and Z.~Yan, \emph{{Strings
  with Non-Relativistic Conformal Symmetry and Limits of the AdS/CFT
  Correspondence}},
  \href{http://dx.doi.org/10.1007/JHEP11(2018)190}{\emph{JHEP} {\bf 11} (2018)
  190}, [\href{https://arxiv.org/abs/arXiv:1810.05560}{{\tt
  arXiv:1810.05560}}].

\bibitem{Bergshoeff:2019pij}
E.~A. Bergshoeff, J.~Gomis, J.~Rosseel, C.~\c{S}im\c{s}ek and Z.~Yan,
  \emph{{String theory and string Newton-Cartan geometry}},
  \href{http://dx.doi.org/10.1088/1751-8121/ab56e9}{\emph{J. Phys. A} {\bf 53}
  (2020) 014001}, [\href{https://arxiv.org/abs/arXiv:1907.10668}{{\tt
  arXiv:1907.10668}}].

\bibitem{Harmark:2019upf}
T.~Harmark, J.~Hartong, L.~Menculini, N.~A. Obers and G.~Oling, \emph{{Relating
  non-relativistic string theories}},
  \href{http://dx.doi.org/10.1007/JHEP11(2019)071}{\emph{JHEP} {\bf 11} (2019)
  071}, [\href{https://arxiv.org/abs/arXiv:1907.01663}{{\tt
  arXiv:1907.01663}}].

\bibitem{Bergshoeff:2021bmc}
E.~A. Bergshoeff, J.~Lahnsteiner, L.~Romano, J.~Rosseel and C.~\c{S}im\c{s}ek,
  \emph{{A non-relativistic limit of NS-NS gravity}},
  \href{http://dx.doi.org/10.1007/JHEP06(2021)021}{\emph{JHEP} {\bf 06} (2021)
  021}, [\href{https://arxiv.org/abs/arXiv:2102.06974}{{\tt
  arXiv:2102.06974}}].

\bibitem{Yan:2021lbe}
Z.~Yan, \emph{{Torsional Deformation of Nonrelativistic String Theory}},
  \href{https://arxiv.org/abs/arXiv:2106.10021}{{\tt arXiv:2106.10021}}.

\bibitem{Bidussi:2021ujm}
L.~Bidussi, T.~Harmark, J.~Hartong, N.~A. Obers and G.~Oling, \emph{{Torsional
  string Newton-Cartan geometry for non-relativistic strings}},
  \href{https://arxiv.org/abs/arXiv:2107.00642}{{\tt arXiv:2107.00642}}.

\bibitem{Bergshoeff:2021tfn}
E.~A. Bergshoeff, J.~Lahnsteiner, L.~Romano, J.~Rosseel and C.~Simsek,
  \emph{{Non-Relativistic Ten-Dimensional Minimal Supergravity}},
  \href{https://arxiv.org/abs/arXiv:2107.14636}{{\tt arXiv:2107.14636}}.

\bibitem{Oling:2022fft}
G.~Oling and Z.~Yan, \emph{{Aspects of Nonrelativistic Strings}},
  \href{https://arxiv.org/abs/2202.12698}{{\tt 2202.12698}}.

\bibitem{Gomis:2005pg}
J.~Gomis, J.~Gomis and K.~Kamimura, \emph{{Non-relativistic superstrings: A New
  soluble sector of AdS$_5\times S^5$}},
  \href{http://dx.doi.org/10.1088/1126-6708/2005/12/024}{\emph{JHEP} {\bf 12}
  (2005) 024}, [\href{https://arxiv.org/abs/hep-th/0507036}{{\tt
  hep-th/0507036}}].

\bibitem{Ko:2015rha}
S.~M. Ko, C.~Melby-Thompson, R.~Meyer and J.-H. Park, \emph{{Dynamics of
  Perturbations in Double Field Theory \& Non-Relativistic String Theory}},
  \href{http://dx.doi.org/10.1007/JHEP12(2015)144}{\emph{JHEP} {\bf 12} (2015)
  144}, [\href{https://arxiv.org/abs/arXiv:1508.01121}{{\tt
  arXiv:1508.01121}}].

\bibitem{Morand:2017fnv}
K.~Morand and J.-H. Park, \emph{{Classification of non-Riemannian
  doubled-yet-gauged spacetime}},
  \href{http://dx.doi.org/10.1140/epjc/s10052-017-5257-z}{\emph{Eur. Phys. J.
  C} {\bf 77} (2017) 685}, [\href{https://arxiv.org/abs/arXiv:1707.03713}{{\tt
  arXiv:1707.03713}}].

\bibitem{Harmark:2017rpg}
T.~Harmark, J.~Hartong and N.~A. Obers, \emph{{Nonrelativistic strings and
  limits of the AdS/CFT correspondence}},
  \href{http://dx.doi.org/10.1103/PhysRevD.96.086019}{\emph{Phys. Rev. D} {\bf
  96} (2017) 086019}, [\href{https://arxiv.org/abs/arXiv:1705.03535}{{\tt
  arXiv:1705.03535}}].

\bibitem{Gomis:2019zyu}
J.~Gomis, J.~Oh and Z.~Yan, \emph{{Nonrelativistic string theory in background
  fields}}, \href{http://dx.doi.org/10.1007/JHEP10(2019)101}{\emph{JHEP} {\bf
  10} (2019) 101}, [\href{https://arxiv.org/abs/arXiv:1905.07315}{{\tt
  arXiv:1905.07315}}].

\bibitem{Gallegos:2019icg}
A.~D. Gallegos, U.~G\"ursoy and N.~Zinnato, \emph{{Torsional Newton Cartan
  gravity from non-relativistic strings}},
  \href{http://dx.doi.org/10.1007/JHEP09(2020)172}{\emph{JHEP} {\bf 09} (2020)
  172}, [\href{https://arxiv.org/abs/arXiv:1906.01607}{{\tt
  arXiv:1906.01607}}].

\bibitem{Blair:2019qwi}
C.~D.~A. Blair, \emph{{A worldsheet supersymmetric Newton-Cartan string}},
  \href{http://dx.doi.org/10.1007/JHEP10(2019)266}{\emph{JHEP} {\bf 10} (2019)
  266}, [\href{https://arxiv.org/abs/arXiv:1908.00074}{{\tt
  arXiv:1908.00074}}].

\bibitem{Yan:2019xsf}
Z.~Yan and M.~Yu, \emph{{Background field method for nonlinear sigma models in
  nonrelativistic string theory}},
  \href{http://dx.doi.org/10.1007/JHEP03(2020)181}{\emph{JHEP} {\bf 03} (2020)
  181}, [\href{https://arxiv.org/abs/arXiv:1912.03181}{{\tt
  arXiv:1912.03181}}].

\bibitem{Harmark:2020vll}
T.~Harmark, J.~Hartong, N.~A. Obers and G.~Oling, \emph{{Spin Matrix Theory
  String Backgrounds and Penrose Limits of AdS/CFT}},
  \href{http://dx.doi.org/10.1007/JHEP03(2021)129}{\emph{JHEP} {\bf 03} (2021)
  129}, [\href{https://arxiv.org/abs/arXiv:2011.02539}{{\tt
  arXiv:2011.02539}}].

\bibitem{Gallegos:2020egk}
A.~D. Gallegos, U.~G\"ursoy, S.~Verma and N.~Zinnato, \emph{{Non-Riemannian
  gravity actions from Double Field Theory}},
  \href{http://dx.doi.org/10.1007/JHEP06(2021)173}{\emph{JHEP} {\bf 06} (2021)
  173}, [\href{https://arxiv.org/abs/arXiv:2012.07765}{{\tt
  arXiv:2012.07765}}].

\bibitem{Blair:2020gng}
C.~D.~A. Blair, G.~Oling and J.-H. Park, \emph{{Non-Riemannian isometries from
  Double Field Theory}},
  \href{http://dx.doi.org/10.1007/JHEP04(2021)072}{\emph{JHEP} {\bf 04} (2021)
  072}, [\href{https://arxiv.org/abs/arXiv:2012.07766}{{\tt
  arXiv:2012.07766}}].

\bibitem{Blair:2021ycc}
C.~D.~A. Blair, D.~Gallegos and N.~Zinnato, \emph{{A non-relativistic limit of
  M-theory and 11-dimensional membrane Newton-Cartan}},
  \href{https://arxiv.org/abs/arXiv:2104.07579}{{\tt arXiv:2104.07579}}.

\bibitem{Fontanella:2021hcb}
A.~Fontanella, J.~M. Nieto~Garc\'\i{}a and A.~Torrielli, \emph{{Light-Cone
  Gauge in Non-Relativistic AdS$_5\times S^5$ String Theory}},
  \href{https://arxiv.org/abs/arXiv:2102.00008}{{\tt arXiv:2102.00008}}.

\bibitem{Fontanella:2021btt}
A.~Fontanella and J.~M.~N. Garc\'\i{}a, \emph{{Classical string solutions in
  Non-Relativistic AdS$_5\times S^5$: Closed and Twisted sectors}},
  \href{https://arxiv.org/abs/arXiv:2109.13240}{{\tt arXiv:2109.13240}}.

\bibitem{Kluson:2021tub}
J.~Kluso\v{n}, \emph{{New Non-Relativistic String in AdS$_5\times S^5$}},
  \href{https://arxiv.org/abs/arXiv:2111.12075}{{\tt arXiv:2111.12075}}.

\bibitem{Yan:2021hte}
Z.~Yan and M.~Yu, \emph{{KLT Factorization of Nonrelativistic String
  Amplitudes}},  \href{https://arxiv.org/abs/2112.00025}{{\tt 2112.00025}}.

\bibitem{Gomis:2020fui}
J.~Gomis, Z.~Yan and M.~Yu, \emph{{Nonrelativistic Open String and Yang-Mills
  Theory}}, \href{http://dx.doi.org/10.1007/JHEP03(2021)269}{\emph{JHEP} {\bf
  03} (2021) 269}, [\href{https://arxiv.org/abs/arXiv:2007.01886}{{\tt
  arXiv:2007.01886}}].

\bibitem{Tseytlin:1996it}
A.~A. Tseytlin, \emph{{Self-duality of Born-Infeld action and Dirichlet
  three-brane of type IIB superstring theory}},
  \href{http://dx.doi.org/10.1016/0550-3213(96)00173-3}{\emph{Nucl. Phys. B}
  {\bf 469} (1996) 51--67}, [\href{https://arxiv.org/abs/hep-th/9602064}{{\tt
  hep-th/9602064}}].

\bibitem{Aganagic:1997zk}
M.~Aganagic, J.~Park, C.~Popescu and J.~H. Schwarz, \emph{{Dual D-brane
  actions}}, \href{http://dx.doi.org/10.1016/S0550-3213(97)00257-5}{\emph{Nucl.
  Phys. B} {\bf 496} (1997) 215--230},
  [\href{https://arxiv.org/abs/hep-th/9702133}{{\tt hep-th/9702133}}].

\bibitem{Berman:2000jw}
D.~S. Berman, V.~L. Campos, M.~Cederwall, U.~Gran, H.~Larsson, M.~Nielsen
  et~al., \emph{{Holographic noncommutativity}},
  \href{http://dx.doi.org/10.1088/1126-6708/2001/05/002}{\emph{JHEP} {\bf 05}
  (2001) 002}, [\href{https://arxiv.org/abs/hep-th/0011282}{{\tt
  hep-th/0011282}}].

\bibitem{Berman:2001rka}
D.~S. Berman, M.~Cederwall, U.~Gran, H.~Larsson, M.~Nielsen, B.~E.~W. Nilsson
  et~al., \emph{{Deformation independent open brane metrics and generalized
  theta parameters}},
  \href{http://dx.doi.org/10.1088/1126-6708/2002/02/012}{\emph{JHEP} {\bf 02}
  (2002) 012}, [\href{https://arxiv.org/abs/hep-th/0109107}{{\tt
  hep-th/0109107}}].

\bibitem{Gopakumar:2000na}
R.~Gopakumar, J.~M. Maldacena, S.~Minwalla and A.~Strominger, \emph{{S duality
  and noncommutative gauge theory}},
  \href{http://dx.doi.org/10.1088/1126-6708/2000/06/036}{\emph{JHEP} {\bf 06}
  (2000) 036}, [\href{https://arxiv.org/abs/hep-th/0005048}{{\tt
  hep-th/0005048}}].

\bibitem{Gopakumar:2000ep}
R.~Gopakumar, S.~Minwalla, N.~Seiberg and A.~Strominger, \emph{{OM theory in
  diverse dimensions}},
  \href{http://dx.doi.org/10.1088/1126-6708/2000/08/008}{\emph{JHEP} {\bf 08}
  (2000) 008}, [\href{https://arxiv.org/abs/hep-th/0006062}{{\tt
  hep-th/0006062}}].

\bibitem{Bergshoeff:2000ai}
E.~Bergshoeff, D.~S. Berman, J.~P. van~der Schaar and P.~Sundell,
  \emph{{Critical fields on the M5-brane and noncommutative open strings}},
  \href{http://dx.doi.org/10.1016/S0370-2693(00)01081-9}{\emph{Phys. Lett. B}
  {\bf 492} (2000) 193--200}, [\href{https://arxiv.org/abs/hep-th/0006112}{{\tt
  hep-th/0006112}}].

\bibitem{Kamimura:2005rz}
K.~Kamimura and T.~Ramirez, \emph{{Brane dualities in non-relativistic limit}},
  \href{http://dx.doi.org/10.1088/1126-6708/2006/03/058}{\emph{JHEP} {\bf 03}
  (2006) 058}, [\href{https://arxiv.org/abs/hep-th/0512146}{{\tt
  hep-th/0512146}}].

\bibitem{Kluson:2019uza}
J.~Kluso\v{n} and P.~Novosad, \emph{{Non-Relativistic M2-Brane}},
  \href{http://dx.doi.org/10.1007/JHEP06(2019)072}{\emph{JHEP} {\bf 06} (2019)
  072}, [\href{https://arxiv.org/abs/arXiv:1903.12450}{{\tt
  arXiv:1903.12450}}].

\bibitem{Kluson:2018vfd}
J.~Kluso\v{n}, \emph{{Note about T-duality of non-relativistic string}},
  \href{http://dx.doi.org/10.1007/JHEP08(2019)074}{\emph{JHEP} {\bf 08} (2019)
  074}, [\href{https://arxiv.org/abs/arXiv:1811.12658}{{\tt
  arXiv:1811.12658}}].

\bibitem{Kluson:2018egd}
J.~Kluso\v{n}, \emph{{Remark about non-relativistic string in Newton-Cartan
  background and null reduction}},
  \href{http://dx.doi.org/10.1007/JHEP05(2018)041}{\emph{JHEP} {\bf 05} (2018)
  041}, [\href{https://arxiv.org/abs/arXiv:1803.07336}{{\tt
  arXiv:1803.07336}}].

\bibitem{Gomis:2020izd}
J.~Gomis, Z.~Yan and M.~Yu, \emph{{T-Duality in Nonrelativistic Open String
  Theory}}, \href{http://dx.doi.org/10.1007/JHEP02(2021)087}{\emph{JHEP} {\bf
  02} (2021) 087}, [\href{https://arxiv.org/abs/arXiv:2008.05493}{{\tt
  arXiv:2008.05493}}].

\bibitem{Bershadsky:1995qy}
M.~Bershadsky, C.~Vafa and V.~Sadov, \emph{{D-branes and topological field
  theories}}, \href{http://dx.doi.org/10.1016/0550-3213(96)00026-0}{\emph{Nucl.
  Phys. B} {\bf 463} (1996) 420--434},
  [\href{https://arxiv.org/abs/hep-th/9511222}{{\tt hep-th/9511222}}].

\bibitem{Green:1996dd}
M.~B. Green, J.~A. Harvey and G.~W. Moore, \emph{{I-Brane Inflow and Anomalous
  Couplings on D-Branes}},
  \href{http://dx.doi.org/10.1088/0264-9381/14/1/008}{\emph{Class. Quant.
  Grav.} {\bf 14} (1997) 47--52},
  [\href{https://arxiv.org/abs/hep-th/9605033}{{\tt hep-th/9605033}}].

\bibitem{Morales:1998ux}
J.~F. Morales, C.~A. Scrucca and M.~Serone, \emph{{Anomalous couplings for
  D-branes and O-planes}},
  \href{http://dx.doi.org/10.1016/S0550-3213(99)00217-5}{\emph{Nucl. Phys. B}
  {\bf 552} (1999) 291--315}, [\href{https://arxiv.org/abs/hep-th/9812071}{{\tt
  hep-th/9812071}}].

\bibitem{Bachas:1999um}
C.~P. Bachas, P.~Bain and M.~B. Green, \emph{{Curvature terms in D-brane
  actions and their M theory origin}},
  \href{http://dx.doi.org/10.1088/1126-6708/1999/05/011}{\emph{JHEP} {\bf 05}
  (1999) 011}, [\href{https://arxiv.org/abs/hep-th/9903210}{{\tt
  hep-th/9903210}}].

\bibitem{Andreev:1988cb}
O.~D. Andreev and A.~A. Tseytlin, \emph{{Partition Function Representation for
  the Open Superstring Effective Action: Cancellation of Mobius Infinities and
  Derivative Corrections to Born-Infeld Lagrangian}},
  \href{http://dx.doi.org/10.1016/0550-3213(88)90148-4}{\emph{Nucl. Phys. B}
  {\bf 311} (1988) 205--252}.

\bibitem{Wyllard:2000qe}
N.~Wyllard, \emph{{Derivative corrections to D-brane actions with constant
  background fields}},
  \href{http://dx.doi.org/10.1016/S0550-3213(00)00780-X}{\emph{Nucl. Phys. B}
  {\bf 598} (2001) 247--275}, [\href{https://arxiv.org/abs/hep-th/0008125}{{\tt
  hep-th/0008125}}].

\bibitem{Polchinski:1995mt}
J.~Polchinski, \emph{{Dirichlet Branes and Ramond-Ramond charges}},
  \href{http://dx.doi.org/10.1103/PhysRevLett.75.4724}{\emph{Phys. Rev. Lett.}
  {\bf 75} (1995) 4724--4727},
  [\href{https://arxiv.org/abs/hep-th/9510017}{{\tt hep-th/9510017}}].

\bibitem{Seiberg:2000ms}
N.~Seiberg, L.~Susskind and N.~Toumbas, \emph{{Strings in background electric
  field, space / time noncommutativity and a new noncritical string theory}},
  \href{http://dx.doi.org/10.1088/1126-6708/2000/06/021}{\emph{JHEP} {\bf 06}
  (2000) 021}, [\href{https://arxiv.org/abs/hep-th/0005040}{{\tt
  hep-th/0005040}}].

\bibitem{Festuccia:2016caf}
G.~Festuccia, D.~Hansen, J.~Hartong and N.~A. Obers, \emph{{Symmetries and
  Couplings of Non-Relativistic Electrodynamics}},
  \href{http://dx.doi.org/10.1007/JHEP11(2016)037}{\emph{JHEP} {\bf 11} (2016)
  037}, [\href{https://arxiv.org/abs/arXiv:1607.01753}{{\tt
  arXiv:1607.01753}}].

\bibitem{Santos:2004pq}
E.~S. Santos, M.~de~Montigny, F.~C. Khanna and A.~E. Santana, \emph{{Galilean
  covariant Lagrangian models}},
  \href{http://dx.doi.org/10.1088/0305-4470/37/41/011}{\emph{J. Phys. A} {\bf
  37} (2004) 9771--9789}.

\bibitem{Bergshoeff:2015sic}
E.~Bergshoeff, J.~Rosseel and T.~Zojer, \emph{{Non-relativistic fields from
  arbitrary contracting backgrounds}},
  \href{http://dx.doi.org/10.1088/0264-9381/33/17/175010}{\emph{Class. Quant.
  Grav.} {\bf 33} (2016) 175010},
  [\href{https://arxiv.org/abs/arXiv:1512.06064}{{\tt arXiv:1512.06064}}].

\bibitem{Chapman:2020vtn}
S.~Chapman, L.~Di~Pietro, K.~T. Grosvenor and Z.~Yan, \emph{{Renormalization of
  Galilean Electrodynamics}},
  \href{http://dx.doi.org/10.1007/JHEP10(2020)195}{\emph{JHEP} {\bf 10} (2020)
  195}, [\href{https://arxiv.org/abs/arXiv:2007.03033}{{\tt
  arXiv:2007.03033}}].

\bibitem{Seiberg:1999vs}
N.~Seiberg and E.~Witten, \emph{{String theory and noncommutative geometry}},
  \href{http://dx.doi.org/10.1088/1126-6708/1999/09/032}{\emph{JHEP} {\bf 09}
  (1999) 032}, [\href{https://arxiv.org/abs/hep-th/9908142}{{\tt
  hep-th/9908142}}].

\bibitem{Kluson:2019avy}
J.~Kluso\v{n}, \emph{{Non-Relativistic D-brane from T-duality Along Null
  Direction}}, \href{http://dx.doi.org/10.1007/JHEP10(2019)153}{\emph{JHEP}
  {\bf 10} (2019) 153}, [\href{https://arxiv.org/abs/arXiv:1907.05662}{{\tt
  arXiv:1907.05662}}].

\bibitem{Kluson:2020aoq}
J.~Kluso\v{n}, \emph{{Unstable D-brane in Torsional Newton-Cartan Background}},
  \href{http://dx.doi.org/10.1007/JHEP09(2020)191}{\emph{JHEP} {\bf 09} (2020)
  191}, [\href{https://arxiv.org/abs/arXiv:2001.11543}{{\tt
  arXiv:2001.11543}}].

\bibitem{Brugues:2004an}
J.~Brugues, T.~Curtright, J.~Gomis and L.~Mezincescu, \emph{{Non-relativistic
  strings and branes as non-linear realizations of Galilei groups}},
  \href{http://dx.doi.org/10.1016/j.physletb.2004.05.024}{\emph{Phys. Lett. B}
  {\bf 594} (2004) 227--233}, [\href{https://arxiv.org/abs/hep-th/0404175}{{\tt
  hep-th/0404175}}].

\bibitem{Brugues:2006yd}
J.~Brugues, J.~Gomis and K.~Kamimura, \emph{{Newton-Hooke algebras,
  non-relativistic branes and generalized pp-wave metrics}},
  \href{http://dx.doi.org/10.1103/PhysRevD.73.085011}{\emph{Phys. Rev. D} {\bf
  73} (2006) 085011}, [\href{https://arxiv.org/abs/hep-th/0603023}{{\tt
  hep-th/0603023}}].

\bibitem{Kim:2007pc}
B.~S. Kim, \emph{{Non-relativistic superstring theories}},
  \href{http://dx.doi.org/10.1103/PhysRevD.76.126013}{\emph{Phys. Rev. D} {\bf
  76} (2007) 126013}, [\href{https://arxiv.org/abs/arXiv:0710.3203}{{\tt
  arXiv:0710.3203}}].

\bibitem{Li:1995pq}
M.~Li, \emph{{Boundary states of D-branes and Dy strings}},
  \href{http://dx.doi.org/10.1016/0550-3213(95)00630-3}{\emph{Nucl. Phys. B}
  {\bf 460} (1996) 351--361}, [\href{https://arxiv.org/abs/hep-th/9510161}{{\tt
  hep-th/9510161}}].

\bibitem{Douglas:1995bn}
M.~R. Douglas, \emph{{Branes within branes}}, {\emph{NATO Sci. Ser. C} {\bf
  520} (1999) 267--275}, [\href{https://arxiv.org/abs/hep-th/9512077}{{\tt
  hep-th/9512077}}].

\bibitem{Hartong:2021ekg}
J.~Hartong and E.~Have, \emph{{On the Non-Relativistic Expansion of Closed
  Bosonic Strings}},  \href{https://arxiv.org/abs/arXiv:2107.00023}{{\tt
  arXiv:2107.00023}}.

\bibitem{Gomis:2004pw}
J.~Gomis, K.~Kamimura and P.~K. Townsend, \emph{{Non-relativistic
  superbranes}},
  \href{http://dx.doi.org/10.1088/1126-6708/2004/11/051}{\emph{JHEP} {\bf 11}
  (2004) 051}, [\href{https://arxiv.org/abs/hep-th/0409219}{{\tt
  hep-th/0409219}}].

\bibitem{Gomis:2005bj}
J.~Gomis, F.~Passerini, T.~Ramirez and A.~Van~Proeyen, \emph{{Non relativistic
  Dp branes}},
  \href{http://dx.doi.org/10.1088/1126-6708/2005/10/007}{\emph{JHEP} {\bf 10}
  (2005) 007}, [\href{https://arxiv.org/abs/hep-th/0507135}{{\tt
  hep-th/0507135}}].

\bibitem{Roychowdhury:2019qmp}
D.~Roychowdhury, \emph{{Probing tachyon kinks in Newton-Cartan background}},
  \href{http://dx.doi.org/10.1016/j.physletb.2019.06.031}{\emph{Phys. Lett. B}
  {\bf 795} (2019) 225--229},
  [\href{https://arxiv.org/abs/arXiv:1903.05890}{{\tt arXiv:1903.05890}}].

\bibitem{Pereniguez:2019eoq}
D.~Pere\~niguez, \emph{{$p$-brane Newton\textendash{}Cartan geometry}},
  \href{http://dx.doi.org/10.1063/1.5126184}{\emph{J. Math. Phys.} {\bf 60}
  (2019) 112501}, [\href{https://arxiv.org/abs/arXiv:1908.04801}{{\tt
  arXiv:1908.04801}}].

\bibitem{Kluson:2020rij}
J.~Kluso\v{n}, \emph{{Stable and unstable D$p$-branes in $p$-brane
  Newton\textendash{}Cartan background}},
  \href{http://dx.doi.org/10.1088/1751-8121/abf768}{\emph{J. Phys. A} {\bf 54}
  (2021) 215401}, [\href{https://arxiv.org/abs/arXiv:2003.14037}{{\tt
  arXiv:2003.14037}}].

\bibitem{Schwarz:1995dk}
J.~H. Schwarz, \emph{{An SL(2,Z) multiplet of type IIB superstrings}},
  \href{http://dx.doi.org/10.1016/0370-2693(95)01405-5}{\emph{Phys. Lett. B}
  {\bf 360} (1995) 13--18}, [\href{https://arxiv.org/abs/hep-th/9508143}{{\tt
  hep-th/9508143}}].

\bibitem{Aganagic:1997zq}
M.~Aganagic, J.~Park, C.~Popescu and J.~H. Schwarz, \emph{{World volume action
  of the M theory five-brane}},
  \href{http://dx.doi.org/10.1016/S0550-3213(97)00227-7}{\emph{Nucl. Phys. B}
  {\bf 496} (1997) 191--214}, [\href{https://arxiv.org/abs/hep-th/9701166}{{\tt
  hep-th/9701166}}].

\bibitem{Pasti:1997gx}
P.~Pasti, D.~P. Sorokin and M.~Tonin, \emph{{Covariant action for a D = 11
  five-brane with the chiral field}},
  \href{http://dx.doi.org/10.1016/S0370-2693(97)00188-3}{\emph{Phys. Lett. B}
  {\bf 398} (1997) 41--46}, [\href{https://arxiv.org/abs/hep-th/9701037}{{\tt
  hep-th/9701037}}].

\bibitem{Pasti:1995tn}
P.~Pasti, D.~P. Sorokin and M.~Tonin, \emph{{Duality symmetric actions with
  manifest space-time symmetries}},
  \href{http://dx.doi.org/10.1103/PhysRevD.52.R4277}{\emph{Phys. Rev. D} {\bf
  52} (1995) R4277--R4281}, [\href{https://arxiv.org/abs/hep-th/9506109}{{\tt
  hep-th/9506109}}].

\bibitem{Pasti:1996vs}
P.~Pasti, D.~P. Sorokin and M.~Tonin, \emph{{On Lorentz invariant actions for
  chiral p forms}},
  \href{http://dx.doi.org/10.1103/PhysRevD.55.6292}{\emph{Phys. Rev. D} {\bf
  55} (1997) 6292--6298}, [\href{https://arxiv.org/abs/hep-th/9611100}{{\tt
  hep-th/9611100}}].

\bibitem{Meessen:1998qm}
P.~Meessen and T.~Ortin, \emph{{An Sl(2,Z) multiplet of nine-dimensional type
  II supergravity theories}},
  \href{http://dx.doi.org/10.1016/S0550-3213(98)00780-9}{\emph{Nucl. Phys. B}
  {\bf 541} (1999) 195--245}, [\href{https://arxiv.org/abs/hep-th/9806120}{{\tt
  hep-th/9806120}}].

\bibitem{Simon:2011rw}
J.~Simon, \emph{{Brane Effective Actions, Kappa-Symmetry and Applications}},
  \href{http://dx.doi.org/10.12942/lrr-2012-3}{\emph{Living Rev. Rel.} {\bf 15}
  (2012) 3}, [\href{https://arxiv.org/abs/1110.2422}{{\tt 1110.2422}}].

\bibitem{Alvarez:1996up}
E.~Alvarez, J.~L.~F. Barbon and J.~Borlaf, \emph{{T duality for open strings}},
  \href{http://dx.doi.org/10.1016/0550-3213(96)00455-5}{\emph{Nucl. Phys. B}
  {\bf 479} (1996) 218--242}, [\href{https://arxiv.org/abs/hep-th/9603089}{{\tt
  hep-th/9603089}}].

\bibitem{Buscher:1987sk}
T.~H. Buscher, \emph{{A Symmetry of the String Background Field Equations}},
  \href{http://dx.doi.org/10.1016/0370-2693(87)90769-6}{\emph{Phys. Lett. B}
  {\bf 194} (1987) 59--62}.

\bibitem{Buscher:1987qj}
T.~H. Buscher, \emph{{Path Integral Derivation of Quantum Duality in Nonlinear
  Sigma Models}},
  \href{http://dx.doi.org/10.1016/0370-2693(88)90602-8}{\emph{Phys. Lett. B}
  {\bf 201} (1988) 466--472}.

\bibitem{Aganagic:1996pe}
M.~Aganagic, C.~Popescu and J.~H. Schwarz, \emph{{D-brane actions with local
  kappa symmetry}},
  \href{http://dx.doi.org/10.1016/S0370-2693(96)01643-7}{\emph{Phys. Lett. B}
  {\bf 393} (1997) 311--315}, [\href{https://arxiv.org/abs/hep-th/9610249}{{\tt
  hep-th/9610249}}].

\bibitem{Aganagic:1996nn}
M.~Aganagic, C.~Popescu and J.~H. Schwarz, \emph{{Gauge invariant and gauge
  fixed D-brane actions}},
  \href{http://dx.doi.org/10.1016/S0550-3213(97)00180-6}{\emph{Nucl. Phys. B}
  {\bf 495} (1997) 99--126}, [\href{https://arxiv.org/abs/hep-th/9612080}{{\tt
  hep-th/9612080}}].

\bibitem{Green:1996bh}
M.~B. Green, C.~M. Hull and P.~K. Townsend, \emph{{D-brane Wess-Zumino actions,
  T-duality and the cosmological constant}},
  \href{http://dx.doi.org/10.1016/0370-2693(96)00643-0}{\emph{Phys. Lett. B}
  {\bf 382} (1996) 65--72}, [\href{https://arxiv.org/abs/hep-th/9604119}{{\tt
  hep-th/9604119}}].

\bibitem{Hassan:1999bv}
S.~F. Hassan, \emph{{T duality, space-time spinors and RR fields in curved
  backgrounds}},
  \href{http://dx.doi.org/10.1016/S0550-3213(99)00684-7}{\emph{Nucl. Phys. B}
  {\bf 568} (2000) 145--161}, [\href{https://arxiv.org/abs/hep-th/9907152}{{\tt
  hep-th/9907152}}].

\bibitem{Witten:1985cc}
E.~Witten, \emph{{Noncommutative Geometry and String Field Theory}},
  \href{http://dx.doi.org/10.1016/0550-3213(86)90155-0}{\emph{Nucl. Phys. B}
  {\bf 268} (1986) 253--294}.

\end{thebibliography}\endgroup

\end{document}